\documentclass{ws-rv9x6}
\usepackage[square]{ws-rv-van}   
\usepackage{siunitx}
\usepackage[section]{placeins}

\newcommand{\sech}{\textrm{sech}}
\newcommand{\etal}{\textit{et al}.}

\begin{document}

\chapter[Low-frequency vibrational spectroscopy of glasses]{Low-frequency vibrational spectroscopy of glasses}\label{RHF_ch8}
\author[B. Ruffl\'e, M. Foret  \& B. Hehlen]{Benoit Ruffl\'e, Marie Foret and Bernard Hehlen}
\address{Laboratoire Charles Coulomb (L2C), University of Montpellier,\\ 
CNRS, Montpellier, France \\
benoit.ruffle@umontpellier.fr}

\begin{abstract}
Atomic vibrations in perfect, slightly defective or mixed crystals are to a large extent well understood since many decades. Theoretical descriptions are thus in excellent agreement with the experiments. As a consequence, phonon-related properties like specific heat, thermal conductivity or sound attenuation are also well explained in these solids. This is not yet the case in glasses where the lack of periodicity generates enormous difficulties in theoretical treatments as well as in experiments or in numerical simulations. Thanks to recent developments along all these lines, comprehensive studies have emerged in the last decades and several decisive advances have been made. This chapter is thus devoted to a discussion of the nature of the vibrational properties in glasses with particular emphasis on the low-frequency part of the vibrational density of states, including the acoustic excitations, and of the experimental techniques used to their study. 
\end{abstract}

\body

\section{Introduction}\label{RHF_sec1}
\subsection{Thermal properties}\label{RHF_sec1_1}

The lack of periodicity in glasses implies that no reciprocal lattice can be defined. As a consequence, the wavevector $\bf{q}$ is no longer a good quantum number for the description of the vibrational excitations and the phonon states cannot be characterized by dispersion curves $\omega(q)$. The quantity which can be safely used is the vibrational density of states $g(\omega)$ (vDOS), which is the distribution of the number of vibrational modes as a function of the frequency $\omega$. This quantity is of paramount importance as it also makes the connection between the microscopic and macroscopic aspects of atomic vibrations. The isochoric heat capacity $C_\textsc{v}(T)$ is for example directly derived from $g(\omega)$. 

The second macroscopic quantity of great interest is the thermal conductivity which relates heat transport and propagative elastic waves in dielectric crystals. Early measurements of these two thermal properties rapidly showed evidences of significant differences when compared to the crystalline counterpart. It is then natural to start the chapter with a very short review of what is usually called the low temperature thermal anomalies of glasses (see Chapter 2 and 3 for an extended discussion of the latter).

\subsubsection{Heat capacity}\label{RHF_sec1_1_1}

The temperature dependence of the heat capacity of crystalline dielectrics is well understood. Well below the Debye temperature $\theta_\textsc{d}$, the isochoric heat capacity $C_\textsc{v}$ follows the Debye law, $C_\textsc{v}=C_\textsc{d} T^3$, where $C_\textsc{d} \propto \theta_\textsc{d}^{-3}$ is given by the sound velocities and the atomic density. At higher $T$, $C_\textsc{v}$ shows a slower increase depending on the details of the atomic structure through the phonon dispersion relation, then flattens eventually to the classical Dulong--Petit limit at high temperature. 

\begin{figure}
\includegraphics[width=\textwidth]{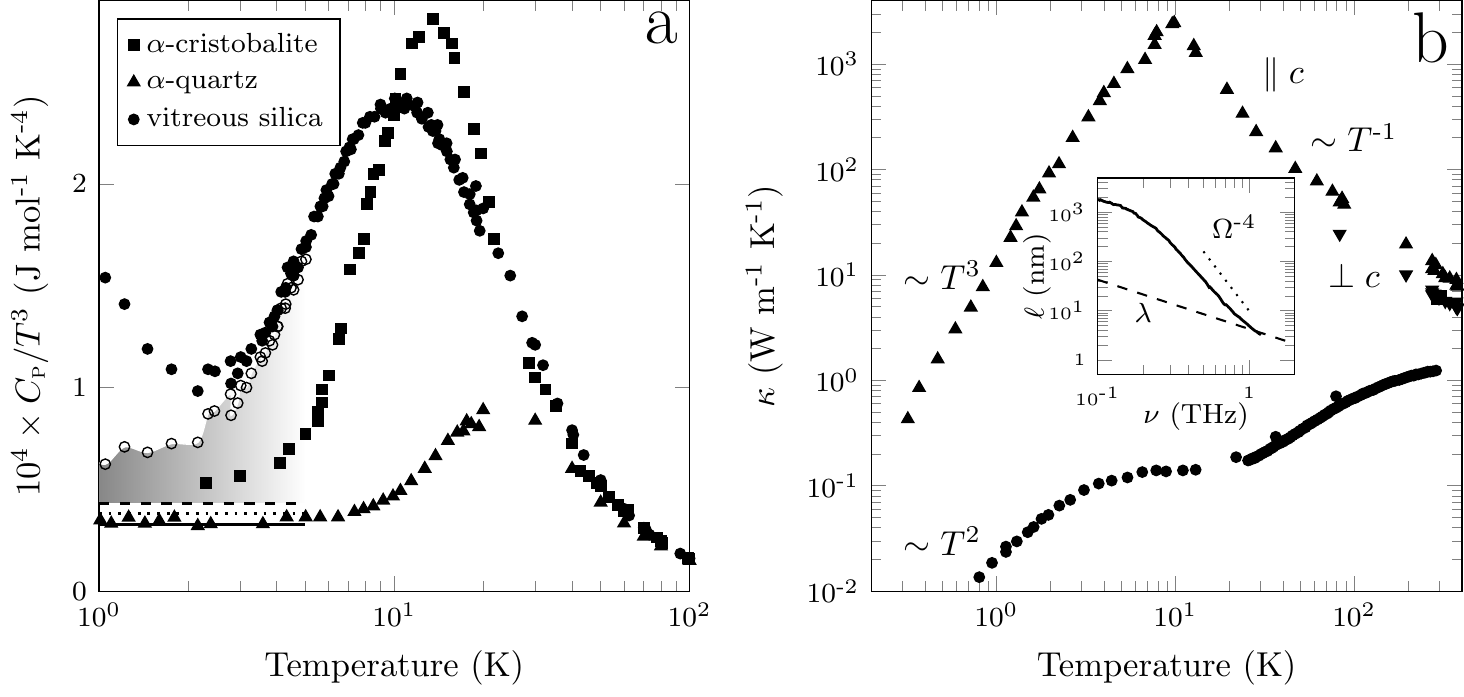}
\caption{
a) Low-temperature heat capacity of vitreous silica (dots~\cite{WIET1921, SIMO1922, SOSM1927, FLUB1959, ZELL1971, BUCH1986}), $\alpha$-cristobalite (squares~\cite{WIET1921, SIMO1922, BILI1975}) and $\alpha$-quartz (triangles~\cite{WIET1921, ZELL1971, BILI1975}) plotted as $C_\textsc{p}/T^3$ against $T$. The horizontal lines are the corresponding constant Debye values (dashed, dotted and solid line, respectively). The circles show the heat capacity of vitreous silica lessened by the two-level systems (TLS) contribution~\cite{RAMO1997a}. 
b) Temperature dependence of the thermal conductivity of vitreous silica (dots~\cite{ZELL1971, LASJ1975, CAHI1987}) and $\alpha$-quartz (up and down triangles for the two principal directions~\cite{EUCK1911, KAYE1926, BIRC1940, ZELL1971}). The square is a room $T$ measurement on $\alpha$-cristobalite~\cite{KUNU1972}. Inset: Frequency dependence of the mean free path of vitreous silica in the dominant phonon approximation.}
\label{fig:thermprop}
\end{figure}

Figure~\ref{fig:thermprop}a shows the reduced heat capacity $C_\textsc{p}/T^3$ at low temperature of three SiO$_2$ polymorphs: vitreous silica, $\alpha$-quartz and $\alpha$-cristobalite. Below \SI{5}{K}, $C_\textsc{p}/T^3$ of $\alpha$-quartz follows the expected constant Debye value whereas the bump at higher $T$ is fully described by the curvature of the acoustic phonon branches. According to their comparable densities and structure, it is however more appropriate to compare silica glass with $\alpha$-cristobalite. For that crystal, the constant Debye value of $C_\textsc{p}/T^3$ holds to about \SI{2}{K} while the large increase at higher $T$, giving rise to a large bump around \SI{13}{K}, originates essentially from the flattening of the low-velocity transverse acoustic branch in the $<$110$>$ direction~\cite{BILI1975,DOVE1997}. Hence both crystalline polymorphs nicely illustrate the general trends mentioned above with characteristic temperatures related to their local order and atomic packing. It is also worth recalling that optic modes contribute to $C_\textsc{p}$. The Raman active zone center mode around \SI{50}{\per\centi\meter}~\cite{SIGA1999} of $\alpha$-cristobalite is for example expected to strengthen $C_\textsc{p}$ just above the maximum of $C_\textsc{p}/T^3$.

Conversely, the heat capacity of vitreous silica is much higher than the Debye expectation at \SI{2}{K} and further never reaches it at low $T$, indicating low-frequency extra modes. At the lowest $T$, most of this excess originates from tunneling states giving rise to an almost linear contribution $C_\textsc{p}\propto T$, see Chapter 2. Subtracting this contribution to $C_\textsc{p}$ still leaves an excess over the Debye prediction as illustrated by the circles and the dashed region in Fig.~\ref{fig:thermprop}a. At higher $T$, $C_\textsc{p}/T^3$ of vitreous silica displays a peak around \SI{10}{K} analogous to that of  $\alpha$-cristobalite, in agreement with their similar structure and atomic density. 

Such common features indicated very early~\cite{BILI1975} that a peak in $C_\textsc{p}/T^3$ at low $T$ is not a peculiarity of glasses~\cite{SAFA2006,CHUM2014} although it does not preclude that part of the vibrational spectrum responsible for this peak in glasses may be very different in nature from that in the corresponding crystal. Oppositely to the case of $\alpha$-cristobalite, it is for example not possible to reconcile the low-$T$ part of the broad hump in $C_\textsc{p}/T^3$ of vitreous silica with the known linear dispersion of sound waves up to at least \SI{440}{\giga\hertz}~\cite{ROTH1983}, the latter giving a constant Debye value up to $\sim$\SI{5.5}{K}. This maximum above the Debye expectation in the $C_\textsc{p}/T^3$ of glasses is directly related to the peak that shows up in the \emph{reduced} vibrational density of states $g(\omega)/\omega^2$ (mostly not in $g(\omega)$ itself) and originally referred to as the \emph{boson peak} in Raman spectroscopy, see Sec.\ref{RHF_sec4_2}.

\subsubsection{Thermal conductivity}\label{RHF_sec1_1_2}

The temperature dependence of the thermal conductivity $\kappa$ of crystalline dielectrics is also well understood. Using the standard gas kinetic equation, $\kappa$ is approximately described in these solids by $\kappa = \frac{1}{3} C_\textsc{v} v \ell$, where $C_\textsc{v}$ is the specific heat per volume of the acoustic phonons providing the thermal transport, $v$ their velocity of propagation, and $\ell$ their mean free path. Starting from ambient temperature, $\kappa$ increases with decreasing temperatures as $\ell$ increases: phonons are scattered by anharmonic umklapp processes which become less frequent as there are less and less phonons. At a certain point, $\ell$ reaches sample dimensions and $\kappa$ decreases rapidly with the $T^3$ dependence of $C_\textsc{v}$. Figure~\ref{fig:thermprop}b illustrates such an ideal behavior in the case of $\alpha$-quartz in the $c$-axis direction. It also shows a room temperature measurement of the thermal conductivity of $\alpha$-cristobalite. Interestingly, the value $\sim$~\SI{6}{\watt\per\meter\per\kelvin} is almost identical to that obtained perpendicularly to the $c$-axis in $\alpha$-quartz, suggesting a similar behavior.

In striking contrast, dielectric glassy materials are known to exhibit much lower thermal conductivities with a markedly different temperature dependence. As an example, $\kappa$ of vitreous silica is plotted in Fig.~\ref{fig:thermprop}b. It follows an approximate $T^2$ law at very low $T$ instead of the $T^3$ dependence expected from the Debye approximation. This initial $T^2$ rise originates from interactions between phonons and tunneling states, reducing $\ell$ in glasses~\cite{PHIL1972,ANDE1972,RAMO1997a}. At higher $T$, $\kappa$ displays a remarkable plateau around \SI{10}{K}, a temperature close to the hump in $C_\textsc{p}/T^3$. It was early recognized that a very efficient phonon scattering mechanism was needed to generate such a plateau, at least with $\ell \propto \omega^{-4}$~\cite{GRAE1986}, suggesting a Rayleigh-type scattering mechanism arising from the glass disorder. 

Based on a dominant phonon approximation $\hbar\omega \sim 3.8\; k_\textsc{b}T$, the frequency dependence of the mean free path can be estimated from the above-mentioned kinetic equation of $\kappa$. The resulting $\ell(\omega)$ for vitreous silica is plotted in the inset of Fig.~\ref{fig:thermprop}b as a line. The mean free path exceeds \SI{1}{\micro\meter} below \SI{100}{\giga\hertz}, a value much larger than the acoustic wavelength $\lambda_a$ illustrating the propagative character of these acoustic phonons. In this frequency range, resonant relaxation by tunneling states dominates the acoustic attenuation at the corresponding temperatures, i.e., below \SI{1}{K} as seen on the top x-scale, and leads to $\ell\propto \omega^{-1}$. At higher frequencies, $\ell$ drops rapidly then follows an approximate $\omega^{-4}$ trend. Around \SI{1}{\tera\hertz}, $\ell$ is comparable to $\lambda_a$, indicating that the Ioffe--Regel criterion $\ell=\lambda_a/2$ is fulfilled in this frequency region ($\ell$ is here the energy mean free path). It corresponds to the propagation of the sound wave over a distance less than half of its wavelength. Above the corresponding frequency $\omega_\textsc{ir}$, sound waves do not propagate and cannot transfer energy anymore. The wave vector is no longer a good quantum number and the notion of phonon becomes ill-defined. At higher temperatures, $\kappa$ rises again and eventually saturates at values around $\sim$ \SI{1}{\watt\per\meter\per\kelvin}, following approximately $C_\textsc{p}(T)$. As the heat transport cannot be mediated anymore by propagating sound waves, it is generally admitted that $\kappa$ is governed here by diffusion mechanisms~\cite{SCHI2006,BELT2016}.

\subsection{Vibrations in glasses}\label{RHF_sec1_2}

In many aspects, the vibrational spectrum of glasses is similar to that of their crystalline counterparts. However, the anomalous thermal properties displayed by the disordered solids indicate that some fundamental differences must exist. Well-defined low-frequency sound waves also exist in glasses as their wavelength are much larger than the disorder length-scale. At higher frequencies, the rapid decrease of the acoustic phonon mean free path related to the plateau in $\kappa$ around \SI{10}{K} seems however specific to disordered solids. Whether and how this peculiar scattering of the acoustic phonons occurring at high frequencies is related to the excess above the Debye expectation in the low-temperature specific heat at temperature below the maximum in $C_\textsc{p}/T^3$ remains a central and still unsettled question. It is further interesting to note that while similar humps do exist in the low-temperature reduced specific heat both for vitreous silica and $\alpha$-cristobalite, indicating similar vibrational density of states in that region, thermal transport mechanisms at the nanometer scale seem very different in the two polymorphs.

Section 2 briefly describes the ($\omega,q$) range which must be addressed by the experimental techniques as well the quantities probed using scattering techniques. Section 3 is devoted to the spectroscopy of acoustic phonons whereas Section 4 discusses the different experimental techniques aiming at measuring the low-frequency part of the vibrational density of states, in particular in the boson peak region.

\section{Inelastic spectroscopy}\label{RHF_sec2}
\subsection{Dispersion diagram and experimental techniques}\label{RHF_sec2_1}

\begin{figure}
\begin{center}
\includegraphics[height=6cm]{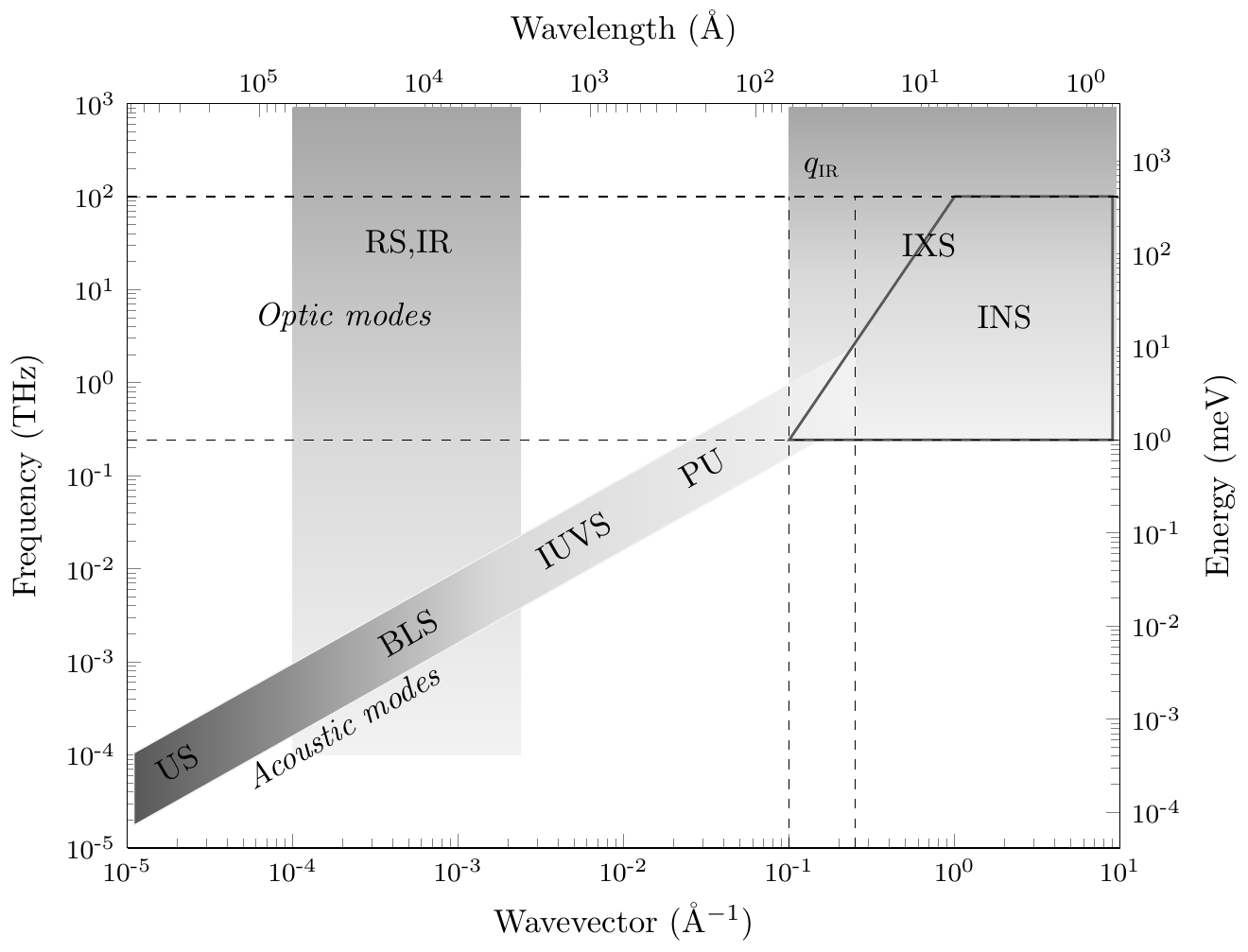}
\caption{Dispersion diagram and ($\omega$,$q$) windows of experimental techniques probing vibrational excitations in disordered materials. Ultrasonics (US), Brillouin light scattering (BLS), Raman scattering (RS), infrared absorption (IR), inelastic scattering with ultra-violet radiation (IUVS), picosecond acoustics (PU), inelastic neutron (INS) and x-ray scattering (IXS). The horizontal dashed lines mark the domain of network optic vibrations in glasses while the diagonal band indicates approximately the domain of acoustic excitations, merging with the optic ones at high $q$.}
\label{fig:OmQ}
\end{center}
\end{figure}

Figure~\ref{fig:OmQ} displays a ($\omega$,$q$) map of the main experimental techniques that can probe vibrational excitations in disordered solids. The diagonal region represents the linear dispersion relation for sound waves of velocity between \SI{1}{\kilo\meter\per\second} and \SI{6}{\kilo\meter\per\second}. Brillouin light scattering (BLS), inelastic scattering with ultra-violet radiation (IUVS), x-rays (IXS) or neutron (INS) probe an increasing well defined $q$ range associated with an increasing $\omega$ range according to the linear dispersion law. Conversely, ultrasonics (US) is not a scattering experiment but covers specific frequency ranges. Accordingly, no scattering vector $\bf{q}$ is really defined for this technique. Picosecond ultrasonics (PU) forms actually a family of optical techniques using pulsed lasers and belongs to the two categories depending on the setup. Concerning the optic modes, the ($\omega$,$q$) ranges covered by scattering techniques that can probe these excitations and measure the vibrational density of states are shown approximately as  vertical regions in Fig. \ref{fig:OmQ}. In increasing $q$ range order one finds Raman scattering (RS), IXS and INS. Due to kinematics constraints, the ($\omega$,$q$) range covered by INS is somewhat limited as compared to the one covered by IXS as shown by the approximate trapezoidal shape in Fig.~\ref{fig:OmQ}. Additionally, there also exists non-scattering techniques like infra-red (IR) spectroscopy or nuclear inelastic scattering (NIS) which give informations on the vibrational density  of states in a $q$ region related to the incoming radiation.

Even if much of the dispersion diagram seems nowadays accessible for studying acoustic excitations from the macroscopic length scale down to the atomic one, it is far from being the case for most glasses. Moreover, it is often feasible for the longitudinal acoustic mode only. As mentioned in the preceding section, of peculiar interest is the mean free path of the acoustic phonons, a quantity often difficult, if not impossible, to obtain with a useful accuracy. This is dramatically true for the transverse modes, in particular in the high $q$ range above about \SI{5e-3}{\per\angstrom}, approaching the Ioffe--Regel crossover region, defined by $q_\textsc{ir}$ in Fig. \ref{fig:OmQ}. This is all the more important as the transverse modes largely dominate the Debye density of states. Consequently, study of the longitudinal modes will give only incomplete informations.

A crucial point revealed by Fig. \ref{fig:OmQ} is that the acoustic linear dispersion curves merge with the low frequency optic vibrations in the Ioffe--Regel crossover region, where stands the boson peak feature around \SI{1}{\tera\hertz}. This explains why, concomitantly to the study of the acoustic excitations, a very large effort of the scientific community has been devoted to the understanding of the low-frequency part of the vibrational density of states of disordered materials.

\subsection{Inelastic scattering intensity}\label{RHF_sec2_2}

The incident radiation impinging on a material is inelastically scattered with the intensity $I({\bf q},\omega)$ corresponding to the space and time Fourier transforms of the autocorrelation function of the physical quantity $A$ which couples to the incoming radiation:
\begin{equation}
I({\bf q},\omega)\propto 
FT\left \{ \int_V\int_tA({\bf r},t)\,A({\bf r}+{\bf r'}, t+t')\,{\rm d}t{\rm d}{\bf r} \right \}=C(A') S({\bf q}, \omega),
\end{equation}
where $A'$ is the derivative of $A$ over the atomic displacements $U$ of the mode while the correlation function $C(A')$ contains the selection rules. The latter depend on the strength of the coupling of the incident radiation to a vibration and therefore modulate the intensity of its spectral shape $S({\bf q}, \omega)$. The physical quantity $A$ is the electronic density $\rho_e$ for X-ray scattering whereas it stands for the coherence length $b$ for neutron scattering and the dielectric susceptibility ${\bf \chi}$ for light scattering. In the latter case, the dynamical dielectric susceptibility $\chi'$ is usually expressed in terms of the polarizability tensor $\bar{\bar \alpha}'$ (Raman scattering), the hyper-polarizability tensor $\bar{\bar{\bar \beta}}'$ (hyper-Raman Scattering), and the photo-elastic tensor $\bar{\bar{\bar{\bar p}}}'$ (Brillouin scattering). All the atomic vibrations can be detected using neutron scattering, enabling the phonon dispersion curves to be measured over a large $({\bf q}, \omega)$  range, however limited by the kinematic conditions and the instrumental resolution. The normalized integral over all $\bf{q}$ values leads to the number of vibrations per unit frequency, also called the neutron vibrational density of states $g(\omega)$ (vDOS). The latter is however modulated by the coherent length of the atoms and remains an approximated quantity, in particular for polyatomic systems. 

As mentioned before, the vibrations are very well defined in crystals so are their spectral response functions. In that case, the group theory provides the mathematical support to compute the selection rules contained in $C(A')$ for all scattering experiments. The situation is again much more complicated in glasses owing to the structural disorder. Each elementary unit of the network is connected to neighboring units, inducing strong structural local distortions. The modes of a given symmetry may also involve different number of neighboring units. These effects modulate $C(A')$ and $S({\bf q}, \omega)$ from site to site inhibiting up to now the development of a general theory describing the vibrational spectra in glasses.

\subsection{Coherent and incoherent scattering}\label{RHF_sec2_3}

The relative spatial localization of extended waves in disordered media induces a loss of coherence of the scattering process due to the partly pointless vibration wavevector $\bf{Q}$. Acoustic or optic phonons in perfect crystals are plane waves with atomic displacements defined by:
\begin{equation}\label{eq:Uphon}
 U({\bf r},t)=u_0e^{\pm i\left ( {\bf Qr}-\Omega t\right )}.
\end{equation}
They are extended modes characterized by a spatial extension larger than the probed wavelength. The scattered radiation results from the interference integral over all the modes in the scattering volume $V$, which in case of light scattering reads:
\begin{equation}\label{eq:Escat}
{\bf E}_s=\int_V\delta {\bf P}e^{- i{\bf k}_s{\bf r}}{ d}V,
\end{equation}
where $\delta {\bf P}=\epsilon_0{\bf \chi}'U({\bf r},t){\bf E}_i$ is the modulation of the electrical polarization induced by the incident field  ${\bf E}_i={\bf E}_{0}e^{i({\bf k}_i{\bf r}-\omega t)}$. Putting Eq.~\ref{eq:Uphon} in Eq.~\ref{eq:Escat} leads to
\begin{equation}\label{eq:Escatf}
{\bf E}_s\propto{\bf E}_0\int_V {\bf \chi}'u_0 e^{i(({\bf k}_i-{\bf k}_s\pm{\bf Q}){\bf r} -(\omega_i\mp\Omega)t)}d{\bf r},
\end{equation} 
where $\bf{r}$ integrates over all orientations so that ${\bf E}_s$ is zero unless ${\bf k}_i-{\bf k}_s\pm {\bf Q}=0$, i.e., for momentum conservation. For a given scattering geometry, only the vibrations whose wavevector $\bf{Q}$ matches the one of the experiment ${\bf Q} = \pm ({\bf k}_i-{\bf k}_s) = {\bf q}$ will be active, denoting a coherent scattering process.
  
On the extreme opposite side, the atomic displacements of fully localized atomic motions are defined by $U(t)=u_0e^{\pm i\Omega t}$ as ${\bf Q}$ is not a relevant quantum number. ${\bf k}_s{\bf r}$ is constant over the very small spatial extension of the mode, and if ${\bf \chi}'$ is also supposed to be constant over the volume $V$ (gases and non-viscous liquids), Eq.~\ref{eq:Escatf} transforms into 
\begin{equation}
{\bf E}_s\propto{\bf \chi}'u_0{\bf E}_0e^{-i(\omega_i\mp\Omega)t},
\end{equation}
showing that the vibration scatters at all $\bf{q}$ values, indicating incoherent scattering.

For intermediate situations, i.e., damped plane waves or quasi-localized waves in disordered solids, the atomic displacements of these vibrations can be modeled by $U({\bf r},t)=u_0e^{-r/\ell_c}e^{\pm i\left ( {\bf Qr}-\Omega t\right )}$, where $\ell_c$ is the coherence length of the vibrational mode. For sufficiently small values of $l_c$, the spatial localization of the mode induces a wavevector spectral broadening $\Delta Q$. In that case, the contribution at the spectral frequency $\omega$ is the sum over all modes of frequency $\Omega=\omega$ having a wavevector spectral component ${\bf Q}$ matching the scattering wavevector $\bf{q}$ of the experiment. The above conclusion holds in glasses for optic modes as well as for short wavelength acoustic waves, since the latter progressively transform into quasi-localized vibrations  at \SI{}{\tera\hertz} frequencies. 

Alternatively, high frequency acoustic modes in amorphous solids can be treated by considering plane waves propagating in an inhomogeneous elastic medium \cite{MART1974} where ${\bf \chi}'$ is position-dependent. It leads to a spatial modulation of the photoelastic tensor which now reads $\bar{\bar{\bar{\bar p}}}'({\bf r})$.
The product $\bar{\bar{\bar{\bar p}}}'({\bf r})U({\bf r},t)$ in $\delta{\bf P}$ yields to distorted acoustic waves whose coherence length $\ell_c$ is limited by the local elastic inhomogeneities. The full treatment shows that the light scattering spectrum consists of the usual Brillouin peaks, i.e., coherent scattering, and a $\omega^2$ incoherent background induced by these local heterogeneities. 

\section{Spectroscopy of acoustic phonons}\label{RHF_sec3}
\subsection{Attenuation and sound velocity in dielectric crystals}\label{RHF_sec3_1}

In perfect dielectric crystals, it is well-known that the dominant mechanism for sound attenuation originates from the interaction between the acoustic wave and high frequency phonons constituting the thermal bath~\cite{MARI1971}. This interaction also leads to a significant variation of the sound velocity. Within the original Akhiezer relaxation model~\cite{AKHI1939}, it is assumed that the temperature is sufficiently high for the bath to be appreciably populated. The thermal phonon mode frequencies $\Omega_i$ in equilibrium are then perturbed by a sound wave of frequency $\Omega$ by $\gamma_i = \partial \ln \Omega_i/\partial e$, where $e$ is the strain developed by the sound wave. As a rough estimate, the $\gamma_i$ can be rationalized by using a mean-square average Gr{\"u}neisen parameter $\gamma^2 = \overline{\gamma_i^2}$. The perturbed thermal bath relaxes towards equilibrium with a characteristic thermal time $\tau_{\rm th}$ via anharmonic interactions and dissipates the energy of the sound wave with an internal friction
\begin{equation}\label{eq:Qm1ANH}
Q^{-1}(\Omega,T) = \frac{A\Omega\tau_{\rm th}}{1+\Omega^2\tau_{\rm th}^2},
\end{equation}
where $A = \gamma^2C_\textsc{v}Tv/2\rho v_\textsc{d}^3$~\cite{MARI1971}. In the latter equation, $C_\textsc{v}$ is the specific heat per unit volume, $\rho$ is the mass density, and $v_\textsc{d}$ is the Debye velocity. 

\begin{figure}
\includegraphics[width=\textwidth]{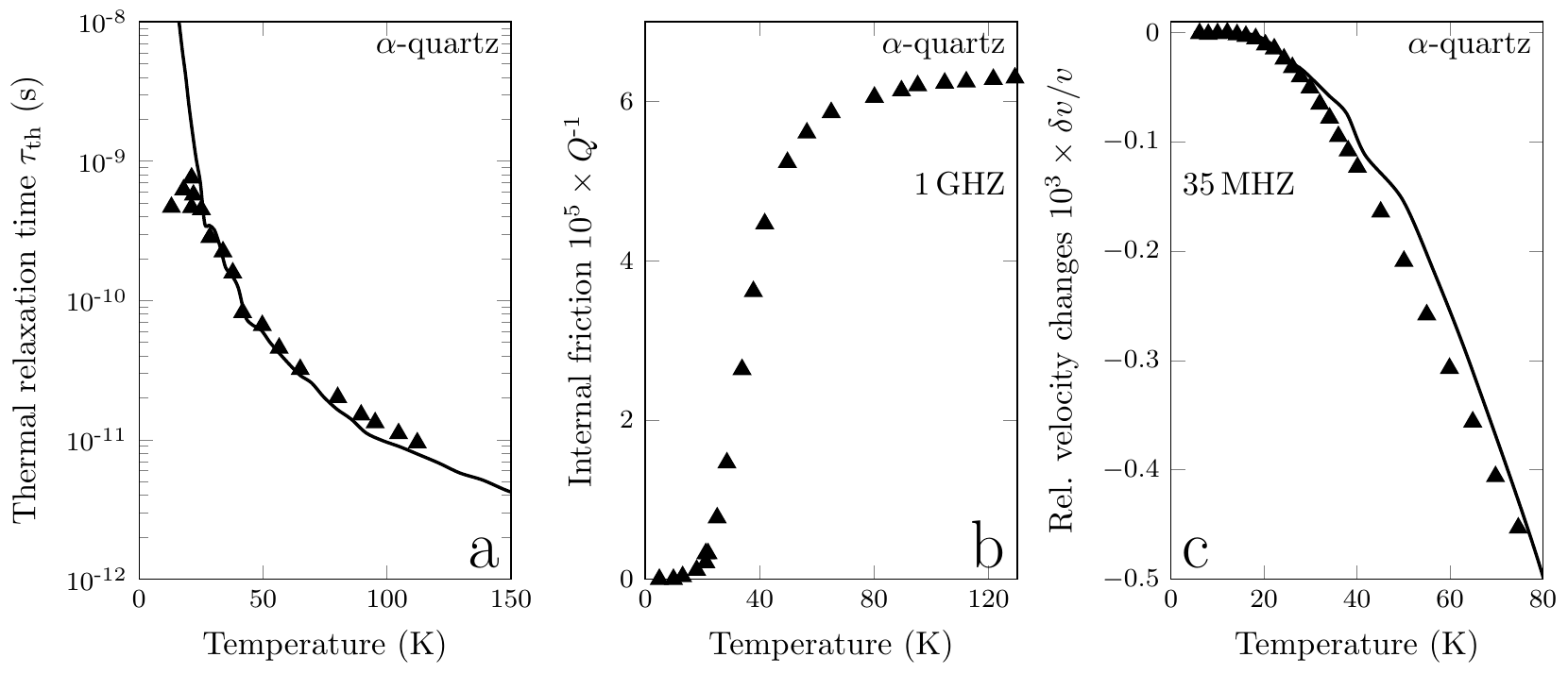}
\caption{a) Temperature dependence of the thermal relaxation time $\tau_{\rm th}$ in $\alpha$-quartz. The line is calculated from the kinetic equation $\kappa = \frac{1}{3}C_\textsc{v}v_\textsc{d}^2\tau_{\rm th}$ while the symbols are obtained from $Q^{-1}$ of panel b) using Eq.~\ref{eq:Qm1ANH} and $\gamma^2\simeq1$~\cite{VACH2005}.
b) Internal friction of \SI{1}{\giga\hertz} longitudinal waves in Z-cut $\alpha$-quartz~\cite{BOMM1960}. c) Fractional velocity changes of \SI{35}{\mega\hertz} shear waves in BC-cut $\alpha$-quartz~\cite{BLIN1970}. The line is obtained using $\tau_{\rm th}(T)$ from the kinetic equation and $\gamma^2\simeq 2.6$.}
\label{fig:quartz}
\end{figure}

The temperature dependence of the thermal phonon lifetime $\tau_{\rm th}$ can be roughly estimated from $\kappa$ using the kinetic equation $\kappa = \frac{1}{3}C_\textsc{v}v_\textsc{d}^2\tau_{\rm th}$. The corresponding values for crystal quartz are shown in Fig.~\ref{fig:quartz}a as a solid line. As $\tau_{\rm th}$ is typically short and further decreases rapidly with $T$, a significant damping of the sound wave is expected at high frequencies only. In the case of quartz, $(2\pi\tau_{\rm th})^{-1}\simeq$ \SI{1}{\giga\hertz} around \SI{35}{K}. The internal friction of \SI{1}{\giga\hertz} longitudinal waves in a Z-cut crystal quartz~\cite{BOMM1960} is plotted in Fig.~\ref{fig:quartz}b. At very low temperature, $\Omega\tau_{\rm th}\gg 1$ and the internal friction is very small. The rapid increase of $Q^{-1}$ with increasing $T$ is essentially driven by the temperature dependence of $C_\textsc{v}$, leading to $A\propto T^4$. At higher temperature, both $A$ and $\tau_{\rm th}$ evolve moderately so does $Q^{-1}$. In the latter region $\Omega\tau_{\rm th}\ll 1$ and $Q^{-1}\propto \Omega$. The acoustic damping $\alpha\propto\Omega^2$ is thus best observed using high frequency techniques. Another estimate of $\tau_{\rm th}$ can be extracted from the measured $Q^{-1}$ and Eq.~\ref{eq:Qm1ANH} using a temperature independent Gr{\"u}neisein parameter $\gamma^2\simeq1$ as demonstrated by the symbols in Fig.~\ref{fig:quartz}a. The agreement between the two sets of values emphasizes the physical significance of $\tau_{\rm th}$ as a mean lifetime of the vibrational excitations in the thermal bath. 

The frequency-dependent damping induces a frequency-dependent correction to the sound velocity \textit{via} the Kramers--Kronig transform
\begin{equation}\label{eq:dvsvANH}
\frac{\delta v}{v}(\Omega,T) = -\frac{1}{2}\frac{A}{1+\Omega^2\tau_{\rm th}^2},
\end{equation}
where $\delta v = v-v_\infty$. In a crystal, one expects that $v_\infty$, the high frequency limit of $v$, is constant in $T$. For $\Omega\tau_{\rm th}\ll 1$, which is the case in ultrasonics for crystal quartz, one simply has $\delta v/v = -A/2$. Such a decrease of the sound velocity with $T$ is illustrated in Fig.~\ref{fig:quartz}b for \SI{35}{\mega\hertz} shear waves in BC-cut quartz~\cite{BLIN1970}. It compares well with the predicted temperature dependence of $\delta v/v$ computed from Eq.~\ref{eq:dvsvANH} and $\tau_{\rm th}$ from $\kappa$ using $\gamma^2\simeq 2.6$ for the shear acoustic mode. All in all, this approach gives a simple and coherent picture of how anharmonicity affects sound waves in dielectric crystals.

\subsection{Attenuation and sound velocity in glasses}\label{RHF_sec3_2}

As mentioned before, sound waves are key properties to understand the low-temperature thermodynamic anomalies of disordered systems. In spite of its long history, the subject remains of much actuality and activity, particularly at high frequencies in relation with nanoscale thermal transport. Other additional mechanisms do affect sound velocity and attenuation in glasses. The direction of the present paragraph is to review progress during the last decades in the understanding of the frequency and temperature dependencies of acoustic properties in glasses and which experimental technique can be used to that purpose.

\subsubsection{Ultrasonics and mechanical techniques}\label{RHF_sec3_2_1}

\begin{figure}
\includegraphics[width=\textwidth]{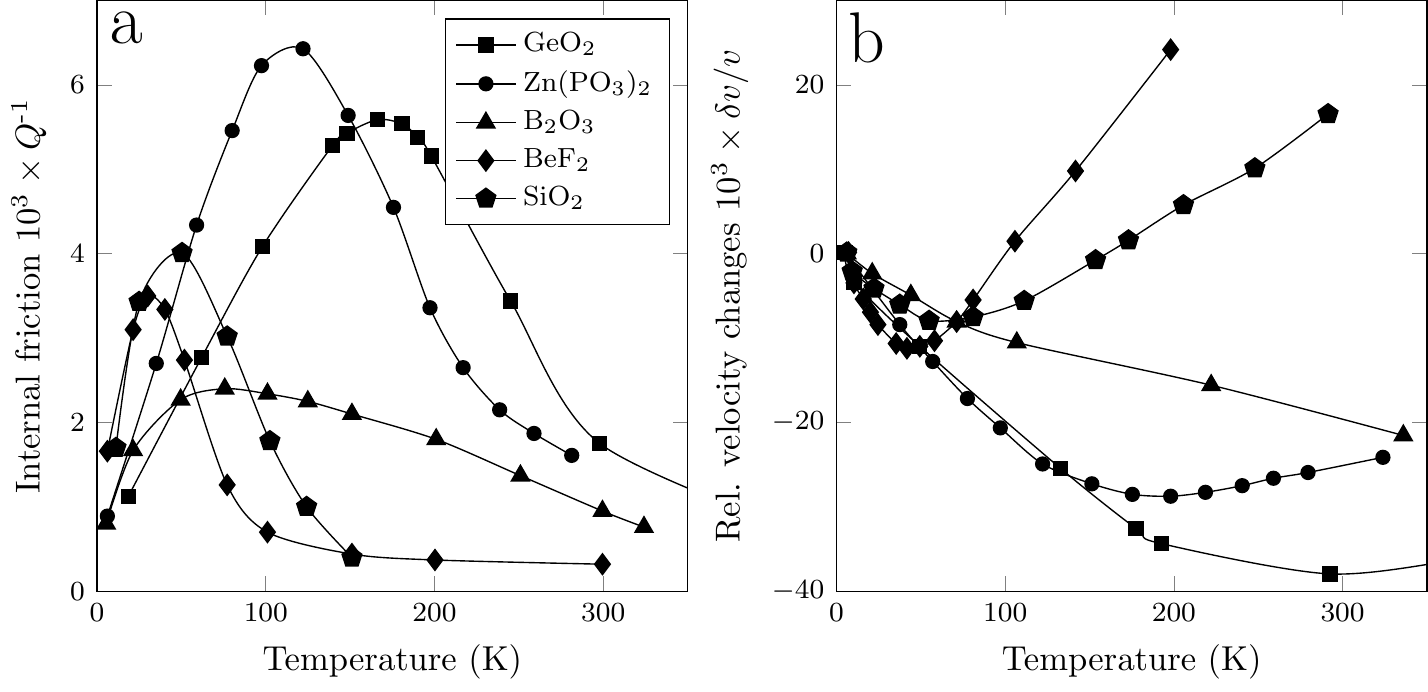}
\caption{a) Internal friction and b) fractional velocity changes of \SI{20}{\mega\hertz} longitudinal waves in different glass formers, adapted from~\cite{KRAU1968}.}
\label{fig:TARpeaks}
\end{figure}

At sonic and ultrasonic frequencies there are several well established methods to measure the sound velocity $v(\Omega,T)$ and the energy sound damping constant $\alpha(\Omega,T)$ related to the energy mean fee path $\ell = \alpha^{-1}$ or to the internal friction $Q^{-1} = \alpha v/\Omega$. Figure~\ref{fig:TARpeaks} shows examples of internal friction and velocity changes in dielectric glasses measured at ultrasonic frequencies as a function of temperature measured in the late 1960's~\cite{KRAU1968}. Oppositely to the case of crystalline quartz plotted in Fig.~\ref{fig:quartz}b the internal friction in these glasses increases rapidly at low temperature, $Q^{-1}\propto T$, evidencing a clear maximum which depends on the material. Moreover, the values of $Q^{-1}$ observed here are about two orders of magnitude larger than in quartz, while the frequency is 50 times lower. It is clear that the internal friction shown in Fig.~\ref{fig:TARpeaks}a must originate from another damping mechanism rather than anharmonic interactions.

There is also marked differences in the low-$T$ decay of the relative changes in the sound velocity illustrated in Fig.~\ref{fig:TARpeaks}b as compared to Fig.~\ref{fig:quartz}c for quartz. According to the Kramers--Kronig relation with the internal friction, the onset in $\delta v/v$ is much slower in crystalline quartz than in these glasses. Further, the velocity changes with $T$ show a clear minimum for four of the five glasses. This anomalous behavior is specific to tetrahedrally coordinated glasses~\cite{KRAU1968} and is related to progressive structural changes with increasing $T$. In contrast, the velocity changes in B$_2$O$_3$ at sufficiently high $T$ are similar to those observed in quartz. It was indeed soon identified that anharmonicity was governing the velocity changes in  glasses in this temperature range~\cite{CLAY1978,VACH1981}, except for the few that are tetrahedrally coordinated.

The ultrasonic absorption peaks observed in glasses at low-$T$ are usually described using a thermally activated structural relaxation process (TAR) represented by a broad distribution of asymmetric double-well potentials~\cite{PHIL1972,ANDE1972,JACK1972,JACK1976,GILR1981}. Owing to structural disorder, several atoms or groups of atoms can occupy two or more equilibrium positions. In that picture, their displacements correspond to these double well potentials. From site to site, the heights $V$ of the potential barrier and their asymmetries $\Delta$ are distributed according to a probability $P(\Delta,V)$. The sound wave couples to the defect site via a deformation potential $\gamma = \frac{1}{2}\frac{\partial\Delta}{\partial e}$, breaking the thermal equilibrium population which relaxes within the characteristic time
\begin{equation}\label{eq:tautar}
\tau = \tau_0 \exp\left(\frac{V}{T}\right) \sech\left(\frac{\Delta}{2T}\right),
\end{equation}
where $1/\tau_0$ is the attempt frequency. Integrating over all the thermally activated relaxations, the internal friction can be calculated following a similar approach used to described anharmonicity leading to~\cite{VACH2005, TIEL1992}
\begin{equation}\label{eq:Qm1TAR}
Q^{-1}_\textsc{tar} = \frac{\gamma^2}{\rho v^2 T}\int_{-\infty}^{\infty}\textrm{d}\Delta\int_{0}^{\infty}\textrm{d}V P(\Delta,V)\sech^2\frac{\Delta}{2T}\left(\frac{\Omega\tau}{1+\Omega^2\tau^2}\right).
\end{equation}

\begin{figure}
\includegraphics[width=\textwidth]{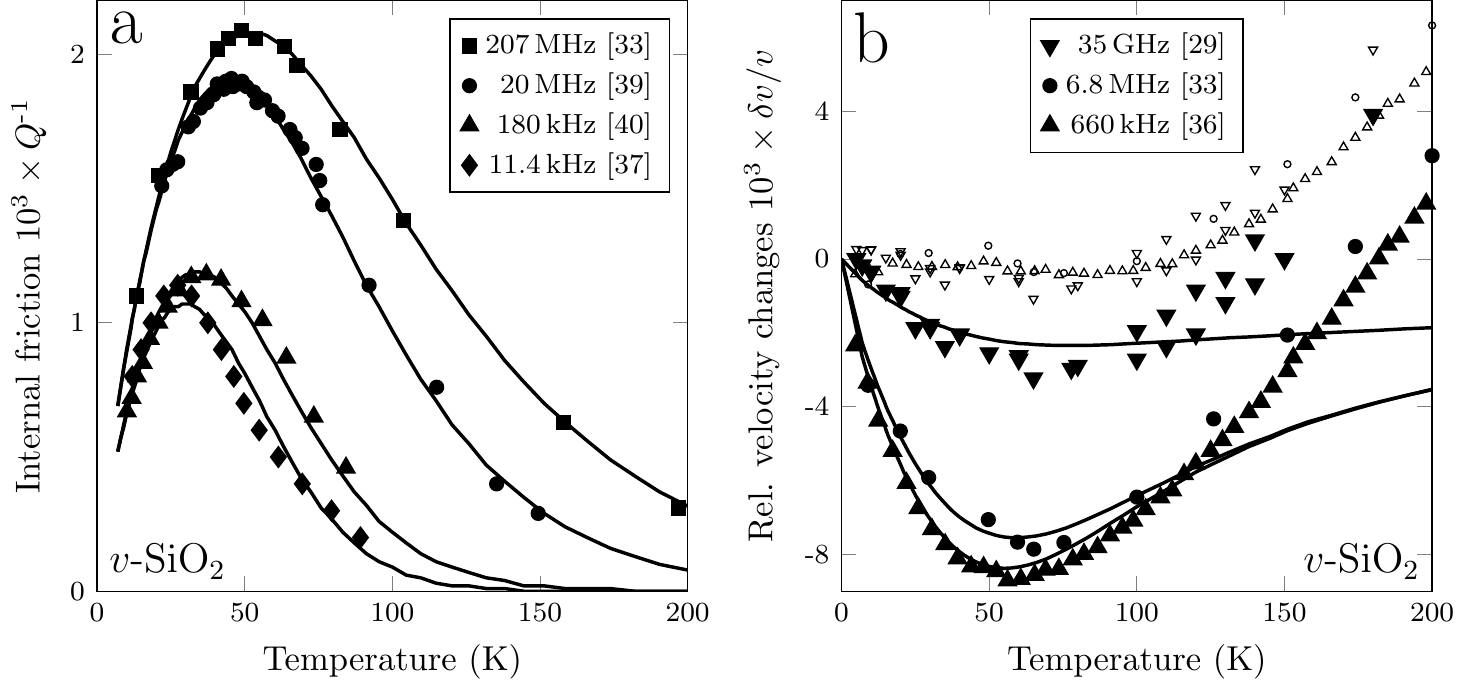}
\caption{a) Internal friction of $v$-SiO$_2$ at sonic and ultrasonic frequencies as a function of temperature taken from the literature~\cite{VACH1981, ANDE1955, KEIL1993, TIEL1992}. The lines show the adjustment of the data to Eq.~\ref{eq:Qm1TAR}. b) Fractional velocity changes of $v$-SiO$_2$ at sonic, ultrasonic and hypersonic frequencies~\cite{GILR1981, VACH1981, VACH2005} (solid symbols). The lines show the adjustment of low-$T$ part of the data to Eq.~\ref{eq:dvsvTAR} whereas the smaller hollow symbols are the experimental $\delta v/v$ corrected from the $\left(\delta v/v\right)_\textsc{tar}$ effect.}
\label{fig:TARSiO2}
\end{figure}

The most advanced analysis so far have been carried out on vitreous silica in the early 2000s~\cite{VACH2005}. This is illustrated in Fig.~\ref{fig:TARSiO2}a which shows the temperature dependence of the internal friction at several representative sonic and ultrasonic frequencies. The peak position depends logarithmically on $T$~\cite{HUNK1976}, supporting thermal activated relaxation as the main source of the internal friction. All the data set are simultaneously adjusted with Eq.~\ref{eq:Qm1TAR} using for $\Delta$ and $V$ independent gaussian distributions with cutoff $\Delta_c$ and $V_0$, respectively. The same set of parameters is also used to described the low-$T$ part of the sound velocity variations using the following equation,
\begin{equation}\label{eq:dvsvTAR}
\left(\frac{\delta v}{v}\right)_\textsc{tar} = -\frac{1}{2} \frac{\gamma^2}{\rho v^2 T}\int_{-\infty}^{\infty}\textrm{d}\Delta\int_{0}^{\infty}\textrm{d}V P(\Delta,V)\sech^2\frac{\Delta}{2T}\left(\frac{1}{1+\Omega^2\tau^2}\right),
\end{equation}
which is the Kramers--Kronig transform of Eq.~\ref{eq:Qm1TAR}. Interestingly, the three data set covering nearly five decades, corrected from this relaxational effect, collapse on a single curve at low temperature revealing that the dip around \SI{50}{K} originates from dynamical effects. Oppositely the anomalous continuous increase of the sound velocity at high temperatures is specific to the tetrahedral networks, thus of structural origin.

The parameters used to draw the lines in Fig.~\ref{fig:TARSiO2} are $V_0\simeq$ \SI{660}{K}, $\Delta_c\simeq$ \SI{80}{K}, $\tau_0\simeq$ \SI{0.6}{\pico\second} and a dimensionless constant strength $C = 1.4 \times 10^{-3}$ related to $\gamma^2$ and the density of relaxing defects~\cite{VACH2005}, the latter being estimated at $N\simeq$ \SI{0.1}{\per\nano\meter\cubed}. This phenomenological description of internal friction at sonic and ultrasonic frequencies recently found a strong support from atomic-scale modeling~\cite{DAMA2018}. In that work, statistical properties of the two-level systems responsible for sound absorption above \SI{10}{K} are obtained for a model of vitreous silica. The agreement with the above analysis is very good, both qualitatively and quantitatively. Additionally, the microscopic nature of the relaxing defects is revealed. It involves the cooperative rotation of Si$-$O$-$Si bonds, see Fig. 6 in~\cite{DAMA2018}. Similar results have been obtained later for some other glasses like GeO$_2$~\cite{GUIM2012}, B$_2$O$_3$~\cite{CARI2012, CARI2014}, where the boroxol rings were identified as the units subjected to thermally activated relaxations or binary borate glasses~\cite{CARI2009,CARI2008}.

\subsubsection{Brillouin light scattering}\label{RHF_sec3_2_2}

Increasing frequency up to the hypersonic regime, it was soon anticipated that anharmonicity should contribute significantly to the sound attenuation in glasses as it is the case in dielectric crystals~(see Sect.~\ref{RHF_sec3_1})~\cite{VACH1981}. In this frequency domain, typically in the range of several tens of \SI{}{\giga\hertz}, Brillouin light scattering (BLS) is a method of choice for sufficiently transparent glasses. BLS deals with light inelastically scattered by the density fluctuations exiting in the probed material which are related to the thermally activated acoustic modes. The scattering vector $\bf{q}$ is fixed by the geometry, the laser wavelength $\lambda$ and the refractive index $n$ of the material, $q = 4\pi n \sin(\theta)/\lambda$, where $\theta$ is the angle between the incident and the scattered light directions. It also corresponds to the wavevector $\bf{Q}$ of the observed acoustic mode $Q=2\pi/\lambda_a$, where $\lambda_a$ is the acoustic wavelength. BLS technique measures a Brillouin shift frequency, which is actually the frequency $\Omega$ of the probed sound wave. $\Omega$ relates to the acoustic velocity $v$ through $\Omega=v/q$. The linewidth $\Gamma$ of the Brillouin spectrum quantifies the acoustic energy attenuation $\alpha$ at the frequency $\Omega$ or the inverse energy mean free path $\ell^\text{-1}$ by $\alpha = \ell^\text{-1} = \Gamma/v$ where $\Gamma$ is expressed in angular frequencies, just like $\Omega$. The internal friction is actually a direct output of a BLS experiment as $Q^\text{-1} = \Gamma/\Omega$. The ratio $\Omega/\Gamma$ is thus the quality factor of the damped harmonic oscillator (DHO) used to adjust the Brillouin spectrum.

Typical BLS experiments are nowadays carried out using a tandem multi-pass plane Fabry--Perot interferometer (T-FPI) developed 40 years ago by J. R. Sandercock~\cite{LIND1981}. This spectrometer is  characterized by a high contrast, better than $10^{10}$ thanks to the six Fabry--Perot (FP) etalons in series, and a high flexibility due to the tandem arrangement. It gives a frequency resolving power of about $2\times10^6$ corresponding to a spectral bandwidth of \SI{250}{\mega\hertz} (HWHM) for a Brillouin frequency shift of \SI{30}{\mega\hertz}. Compared to the very abundant ultrasonic results, there exist relatively few damping data at frequencies above \SI{1}{\giga\hertz} in glasses. This is essentially due to the fact that the frequency resolving power of the T-FPI hardly meets the demanding needs for proper sound attenuation measurements in glasses well below the glass transition temperature.

This experimental difficulty can be overcome by using a Brillouin spectrometer consisting in a tandem arrangement of a multi-pass plane FP etalon followed by a confocal one~\cite{VACH1976}. The plane FP acts here as a prefilter effectively eliminating other spectral components and reducing considerably the elastic signal strength. It is maintained at a fixed spacing which is dynamically adjusted to one of the Brillouin lines with the help of an electro-optically modulated signal at the Brillouin frequency. The confocal FP is piezo-electrically scanned for the spectra analysis. Different confocal FPs can be used to adjust the spectrometer resolution which is ranging from \SIrange{10}{100}{\mega\hertz}. In this way a high contrast (better than $10^{10}$) and a high-frequency resolving power ($10^7-10^8$) is obtained.

As far as sound damping is concerned, most BLS experiments relate actually to the longitudinal acoustic (LA) mode. This is due to the fact that experiments are generally performed in the backscattering geometry offering several advantages: (i) higher Brillouin frequency shift; (ii) lower uncertainty of velocity measurements; (iii) lower geometrical broadening of the Brillouin line due to the finite aperture. Selection rules for Brillouin scattering indicate however that transverse acoustic modes are not active in the backscattering geometry for isotropic materials. One concludes that high-resolution BLS gives accurate and absolute measurements of $Q^\text{-1}$ as both the Brillouin frequency shift $\Omega$ and the Brillouin linewidth $\Gamma$ can be obtained simultaneously. 

\begin{figure}
\includegraphics[width=\textwidth]{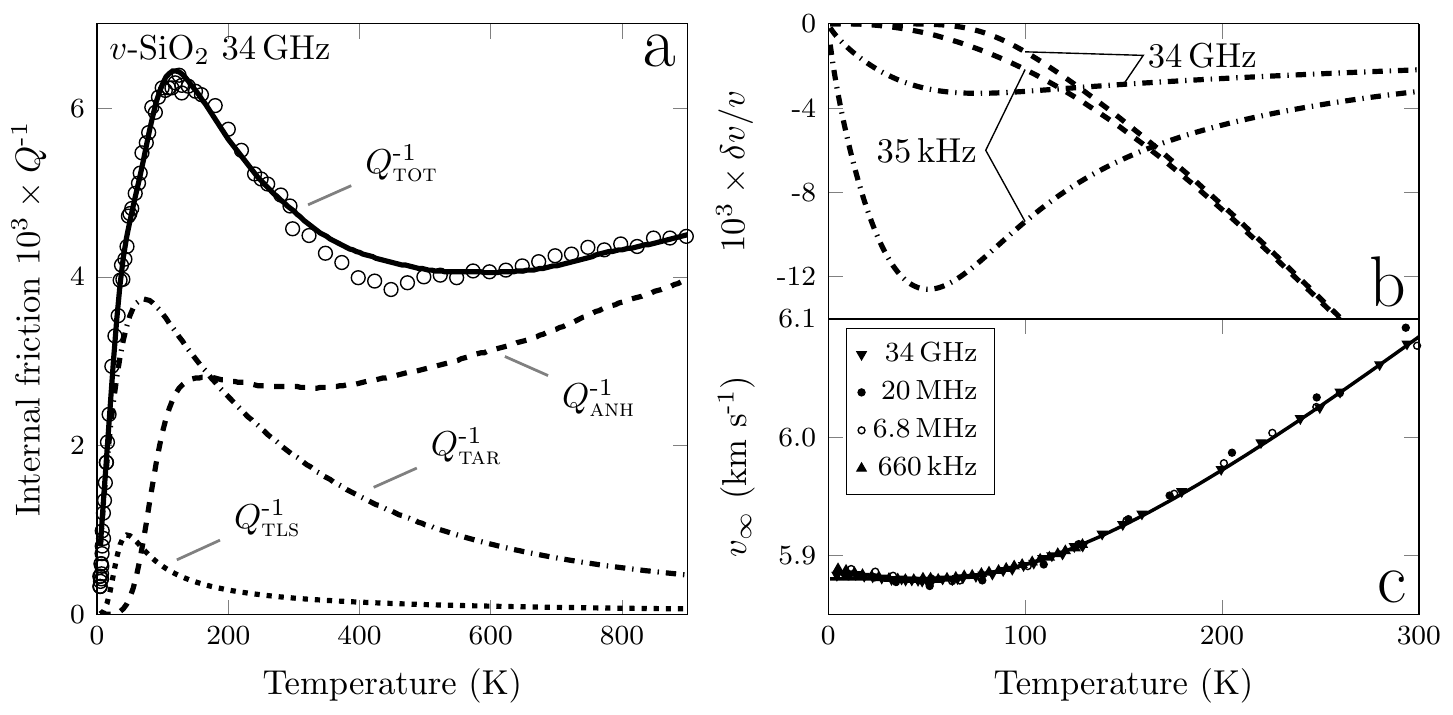}
\caption{a) Internal friction of $v$-SiO$_2$ at hypersonic frequencies as a function of temperature~\cite{VACH1981, VACH2005, RUFF2010}. The solid line shows the adjustment of the curve to the sum of the three damping processes $Q^\text{-1}_\textsc{tot} = Q^\text{-1}_\textsc{tls} + Q^\text{-1}_\textsc{tar} + Q^\text{-1}_\textsc{anh}$. b) Calculated fractional velocity changes at sonic and hypersonic frequencies resulting from TAR and anharmonic processes. c) Unrelaxed velocities $v_\infty$ after correction for TAR and anharmonic processes at four selected frequencies. The solid line is a Wachtman's adjustment of the data.}
\label{fig:ANHSiO2}
\end{figure}

We pursue with the case of vitreous silica which is by far the most studied system. Figure~\ref{fig:ANHSiO2}a shows the internal friction of $v$-SiO$_2$ at hypersonic frequencies as a function of the temperature. A maximum of the damping coefficient is observed around \SI{150}{K}, similar to what is shown in Fig.~\ref{fig:TARSiO2}a. However, $Q^\text{-1}$ does not decrease rapidly to zero above that maximum but remains at a high level up to the highest temperatures, indicating the appearance of another damping mechanism in this frequency range. From Eq.~\ref{eq:Qm1TAR}, we can anticipate that $Q^\text{-1}_\textsc{tar}$ cannot account for the entire $Q^\text{-1}$. At the peak position corresponding to $\omega\tau\simeq1$, $Q^\text{-1}_\textsc{tar}$ is indeed nearly independent of $\Omega$. Second, anharmonicity is still in the regime  $\Omega\tau_{\rm th}\ll 1$ in that temperature/frequency range, i.e., $Q^\text{-1}_\textsc{anh}\propto\Omega$. This is illustrated in Fig.~\ref{fig:ANHSiO2}a where the dash-dotted line is the $Q^\text{-1}_\textsc{tar}$ values calculated with the parameters given in the preceding section for the TAR damping mechanism in $v$-SiO$_2$ to which a small contribution from the incoherent scattering of the two-level systems is added ($Q^\text{-1}_\textsc{tls}$ dotted line, see Chapter 4). The difference, $Q^\text{-1}_\textsc{anh}$, is the contribution due to the anharmonic coupling of the LA waves with the thermally excited vibrational modes. It clearly appears that at sufficiently high temperature and frequency, $Q^\text{-1}_\textsc{anh}$ takes over $Q^\text{-1}_\textsc{tar}$. The analysis of the anharmonic damping has been done using Eq.~\ref{eq:Qm1ANH} with $A(T) = \gamma^2C_\textsc{v}Tv/2\rho v_\textsc{d}^3$~\cite{VACH2005}. The amplitude of $Q^\text{-1}_\textsc{tar}$ determines the value of the Gr{\"u}neisen parameter $\gamma^2 = 3.8$, the unique unknown parameter entering $A(T)$ while the temperature dependence of the thermal time $\tau_{\rm th}$ is extracted from the shape of $Q^\text{-1}_\textsc{tar}(T)$. The values of $\tau_{\rm th}$ are proportional to $T^\text{-2}$ in the approximate range \SIrange{100}{300}{K}, evolving towards a $T^\text{-1}$ behavior at higher temperatures in a similar way to those of $\alpha$-quartz plotted in Fig.~\ref{fig:quartz} but three time shorter. It is important to stress again at that point that contrarily to the case of $\alpha$-quartz, it is not possible to obtain $\tau_{\rm th}(T)$ in that temperature range from the thermal conductivity. The kinetic equation $\kappa=\frac{1}{3}C_v v \ell$ is indeed valid only for propagating modes, a condition that is presumably no more fulfilled beyond the plateau in $\kappa(T)$ located around \SI{10}{K} in $v$-SiO$_2$. Finally, the validity of the analysis of the internal friction solely based on the TAR contribution at ultrasonic frequencies illustrated in Fig.~\ref{fig:TARSiO2} is verified \textit{a posteriori}.

The effect of anharmonicity on the velocity dispersion can be calculated following Eq.~\ref{eq:dvsvANH}. This is illustrated in Fig.~\ref{fig:ANHSiO2}b where the relative sound velocity variations $\delta v/v$ induced by anharmonicity are plotted for two different frequencies and compared to those due to the thermally activated relaxation processes. While $(\delta v/v)_\textsc{tar}$ strongly depends on the frequency below \SI{100}{K}, $(\delta v/v)_\textsc{anh}$ remains essentially unaffected by changing $\nu$, in particular above \SI{100}{K}. At higher temperatures, the magnitude of $(\delta v/v)_\textsc{anh}$ increases almost linearly with temperature. Linear decrease of the sound velocity have indeed been observed in a large number of glasses~\cite{KRAU1968,CLAY1978,CARI2005,CARI2014}. Finally, Fig.~\ref{fig:ANHSiO2}c shows the unrelaxed values $v_\infty$ of the frequency dependent measured sound velocity $v$ that are corrected for the two damping processes, $v_\infty = v - [\delta v_\textsc{tar}+\delta v_\textsc{anh}]$. One observes that the four series of data covering nearly five decades in frequency nicely collapse on a single curve $v_\infty(T)$. The latter is not a constant but rather increases rapidly with $T$ and can be adjusted to a Wachtman's equation~\cite{WACH1961}. This is a peculiarity of tetrahedrally-coordinated glasses and associated to a progressive structural change occurring in the network when the temperature is raised~\cite{AYRI2011a}.

\begin{figure}
\includegraphics[width=\textwidth]{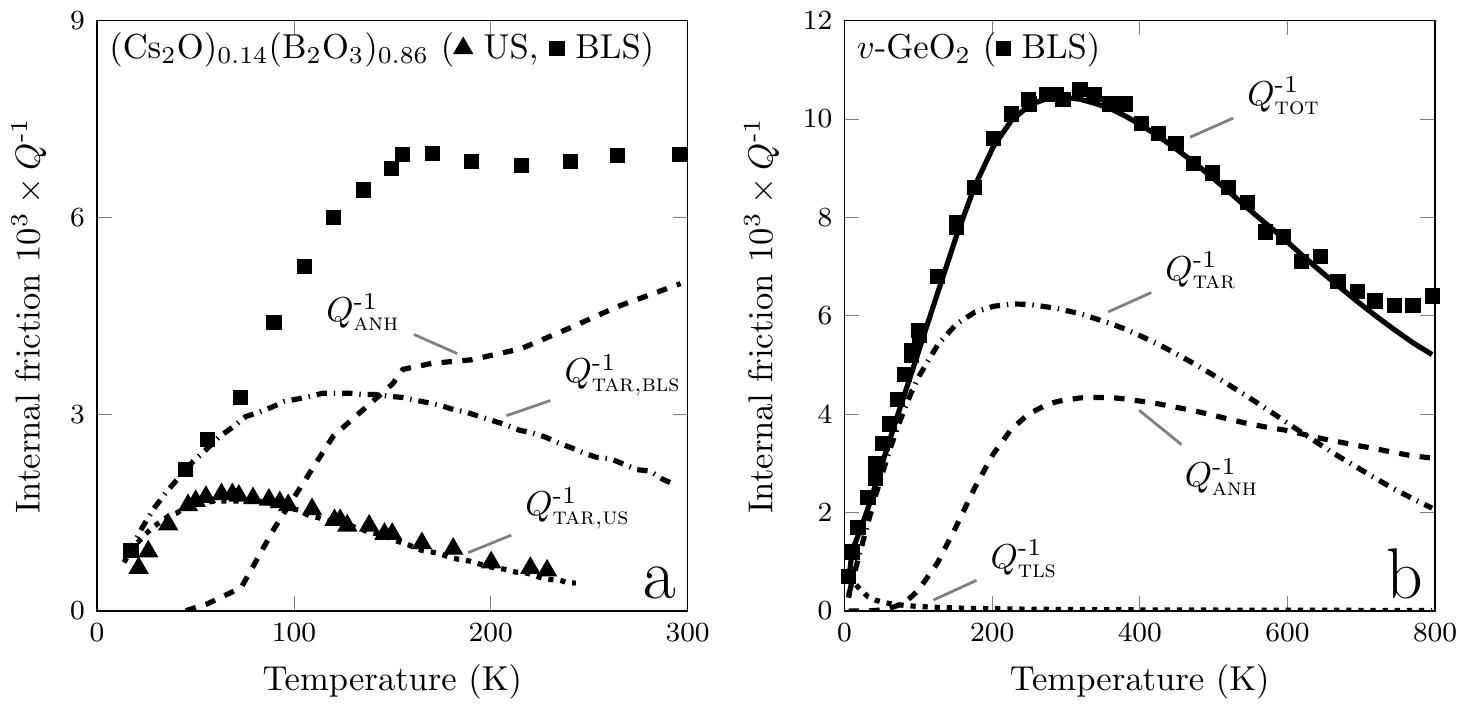}
\caption{a) Internal friction of (Cs$_2$O)$_{0.14}$(B$_2$O$_3$)$_{0.86}$ at ultrasonic~\cite{CARI2006} and hypersonic~\cite{CARI2009} frequencies. The dotted line is the extrapolation at hypersonic frequencies of the $Q^\text{-1}_\textsc{tar}$ solid line adjusted to the ultrasonic data. The dash-dotted line is the anharmonic contribution to the damping resulting from the subtraction of $Q^\text{-1}_\textsc{tar}$ from the BLS data~\cite{CARI2009}. b) Internal friction of $v$-GeO$_2$ at hypersonic frequencies. The solid line shows the adjustment of the BLS data to the sum of the three damping processes $Q^\text{-1}_\textsc{tot} = Q^\text{-1}_\textsc{tls} + Q^\text{-1}_\textsc{tar} + Q^\text{-1}_\textsc{anh}$~\cite{GUIM2012}.}
\label{fig:AnhBLS}
\end{figure}

It is expected that the above-mentioned competing damping processes are not unique to vitreous silica. Even if accurate sound attenuation data are scarce at hypersonic frequencies some attempts have been made in that direction. Figure~\ref{fig:AnhBLS} illustrates the case for two other systems. In Fig.~\ref{fig:AnhBLS}a the internal friction of a cesium borate glass at ultrasonic frequencies had been adjusted to TAR alone following Eq.~\ref{eq:Qm1TAR}~\cite{CARI2006}. Its extrapolation at hypersonic frequencies, shown by the dotted line, largely fails to reproduce the Brillouin light scattering data. Clearly, the BLS damping data must contain an appreciable contribution from anharmonicity, shown by the dash-dotted line. The latter has been analyzed using Eq.~\ref{eq:Qm1ANH}~\cite{CARI2009} leading to the determination of $\tau_{\rm th}$ and of the mean Gr{\"u}neisen parameter for that glass as well. Figure~\ref{fig:AnhBLS}b shows the case of vitreous germania, $v$-GeO$_2$, for which several sound absorption experiments exist at low frequencies ensuring a fair estimate of $Q^\text{-1}_\textsc{tar}$. In a nutshell, an analysis similar to that done for $v$-SiO$_2$ lead to comparable conclusions~\cite{GUIM2012}.

\begin{figure}
\includegraphics[width=\textwidth]{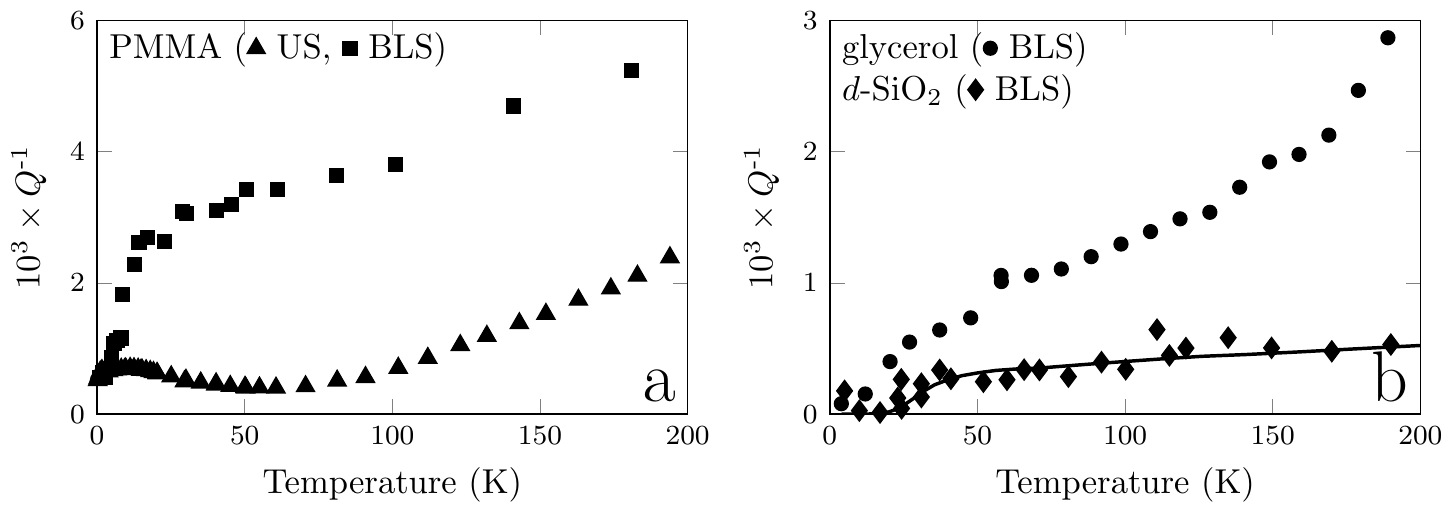}
\caption{b) Internal friction of PMMA at ultrasonic~\cite{FEDE1982,FEDE1982a} and hypersonic~\cite{VACH1976,SCHM1982} frequencies. b) Internal friction obtained using BLS in densified silica~\cite{RAT2005} and in glycerol~\cite{VACH1985,RUFF2007a}. Note the different y-scale of the plots compared to Fig.~\ref{fig:AnhBLS}.}
\label{fig:AnhBLS2}
\end{figure}

This does not seem to be restricted to oxide glasses as illustrated by the internal friction of poly(methyl methacrylate) (PMMA) obtained at ultrasonic~\cite{FEDE1982,FEDE1982a} and hypersonic~\cite{VACH1976,SCHM1982} frequencies and plotted in Fig.~\ref{fig:AnhBLS2}a. It has been shown that the TAR model alone fails to explain consistently the internal friction in the \SI{}{\hertz}--\SI{}{\giga\hertz} frequency domain above \SI{10}{K}~\cite{CALI2001}. The authors concluded that an additional relaxational mechanism seems to appear at high frequency, whose strength increases with $T$. They invoked a constant loss spectrum ($Q^\text{-1}_\textsc{loss}\propto\omega$) whose exact nature is not clear. Alternatively, anharmonicity could explain the missing contribution to the sound damping at high frequency, a possibility actually suggested by the authors themselves. Permanently densified glasses seems to show a much lower contribution from the TAR, likely related to the loss of the free volume~\cite{WEIS1996}. It offers unique possibilities to study the sound attenuation originating from anharmonicity at hypersonic frequencies. Fig.~\ref{fig:AnhBLS2}b illustrates this point with densified silica glass having the density of crystal quartz. In that case, the TAR are almost completely suppressed~\cite{WEIS1996}. The internal friction measured at \SI{}{\giga\hertz} frequencies is indeed quite small as compared to that of normal silica~\cite{RAT2005}.
Hence, adjusting the measured $Q^\text{-1}$ with Eq.~\ref{eq:Qm1ANH} gives the solid line in Fig.~\ref{fig:AnhBLS2}b. The suppressing of $Q^\text{-1}_\textsc{tar}$ with densification seems also to happen in vitreous boron oxide but accurate high resolution BLS data are still missing to properly estimate the anharmonicity in $d$-B$_2$O$_3$~\cite{CARI2014}. Finally, Fig.~\ref{fig:AnhBLS}2b also shows internal friction data for glycerol obtained using BLS. The damping coefficient is surprisingly low for a molecular glass and no well defined peak, characteristic of TAR processes is observed. No attempt has been made yet to described these data using Eq.~\ref{eq:Qm1ANH}.

\subsubsection{Inelastic scattering of UV radiation}\label{RHF_sec3_2_3}

\begin{figure}
\includegraphics[width=\textwidth]{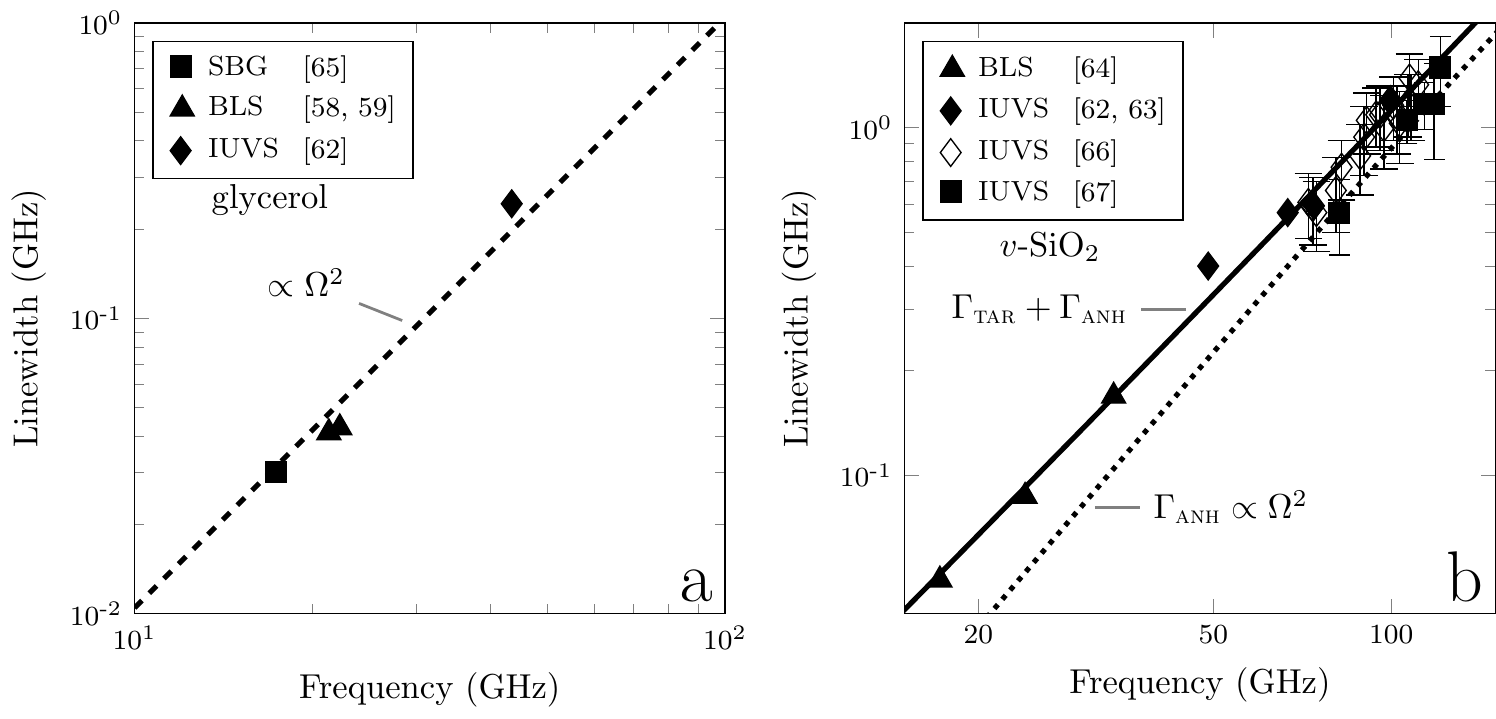}
\caption{a) Frequency dependence of the Brillouin linewidth in glycerol using SBG~\cite{GRUB1994}, BLS~\cite{VACH1985,RUFF2007a} and IUVS~\cite{MASC2004} at $T=$ \SI{150}{K}. The dashed line is the $\Gamma\propto\Omega^2$ adjustment of the data. b) Frequency dependence of the Brillouin linewidth in vitreous silica using BLS~\cite{VACH2006} and IUVS~\cite{MASC2004,BENA2005,MASC2006,RUFF2011} at $T=$ \SI{300}{K}. The line is the theoretical expectation based on Eqs.~\ref{eq:Qm1ANH}~and~\ref{eq:Qm1TAR} used to describe the internal friction at lower frequency in $v$-SiO$_2$  as explained in the preceding paragraphs. The dotted line is $\Gamma_\textsc{anh}$ only, showing a quadratic behavior.}
\label{fig:IUVS}
\end{figure}

Brillouin scattering at shorter wavelength can be used to increase the Brillouin shift frequency. This can be done using excitation with UV radiation~\cite{MASC2004,BENA2005}, as long as the samples are sufficiently transparent~\cite{VACH2006}. This has been attempted for example in glycerol, using the second harmonic of a visible laser operating at \SI{488}{\nano\meter} and the IUVS spectrometer located at the Elettra synchrotron light source in Trieste, Italy~\cite{MASC2004}. In this operation mode, the total resolving power is $1.1\times 10^6$. This setup allows an increase of $Q$ by a factor of 2. Figure~\ref{fig:IUVS}a shows the Brillouin linewidth at $T=$ \SI{150}{K} ($T_\text{g}=$ \SI{187}{K}) extracted from the inelastic spectrum obtained with IUVS at $\simeq$ \SI{43}{\giga\hertz}. It is compared to the values obtained at lower frequencies using stimulated Brillouin gain spectroscopy (SBG)~\cite{GRUB1994} and BLS~\cite{VACH1985,RUFF2007a}. The data show that, in this frequency range, the sound attenuation coefficient closely follows a $\Omega^2$ dependence, as expected for anharmonicity, and in agreement with the temperature dependence discussed just before.

Vitreous silica is known to have a rather high absorption edge at $E_\text{i}\approx$ \SI{8}{\electronvolt}, allowing the opportunity to access Brillouin frequencies above \SI{100}{\giga\hertz} and a momentum transfer region of \SI{0.14}{\per\nano\meter}. In that case, the IUVS spectrometer uses the UV beam produced by the synchrotron and the total resolving power is $5\times 10^5$. Figure~\ref{fig:IUVS}b shows part of the first IUVS results obtained at two wavelengths~\cite{MASC2004} or at two angles~\cite{BENA2005}, plotted as solid symbols. The graph also shows high resolution BLS values obtained with visible light at three scattering angles~\cite{VACH2006}. The IUVS data remarkably align with the expected damping values calculated from Eqs.~\ref{eq:Qm1ANH}~and~\ref{eq:Qm1TAR} with the parameters used to described internal friction of vitreous silica at lower frequencies. One observes that the power law frequency exponent of $\Gamma(\Omega)$ is about 1.7 at \SI{20}{\giga\hertz}, still slightly smaller than the quadratic behavior of the growing anharmonic contribution $\Gamma_\textsc{anh}$. 

The understanding of the sound attenuation in vitreous silica and more generally in glasses was challenged by deep UV Brillouin scattering experiments, i.e., close to the absorption gap, claiming an unexpected dramatic increase of the linewidth beyond \SI{110}{\giga\hertz} in $v$-SiO$_2$~\cite{MASC2006}. Below this frequency, these new data agreed very well with the preceding IUVS ones. At higher frequencies, a much steeper increase was clearly observed that was interpreted as a crossover associated with elastic disorder. The latter was however not detected by a second experiment carried out using a high grade superpolished silica sample up to about \SI{125}{\giga\hertz}~\cite{RUFF2011}. As we will see in the next section devoted to picosecond ultrasonics, several other experiments covering this frequency range also disproved the deep UV results. As a consequence, only the deep UV Brillouin data which truly relate to acoustic damping are plotted in Fig.~\ref{fig:IUVS}b as hollow symbols. Brillouin scattering close to the absorption gap is clearly a delicate experiment. Approaching the gap from below, the collected Brillouin signal considerably decreases, owing to rapidly increasing sample absorption of both the incident and the scattered light. At first sight, the unexpected increase of the Brillouin linewidth observed in the deep UV data above \SI{110}{\giga\hertz} could be related to finite-size effects induced by the dramatic increase of the imaginary part of the refractive index approaching the absorption gap~\cite{SAND1972,VACH2006}. However, this effect is still negligible in the investigated region~\cite{RUFF2011}. If not an artifact, the anomalous increase in damping could originate from growing refractive index fluctuations close to the gap, producing an enhanced uncertainty broadening in Brillouin scattering~\cite{INAM2010}.

\subsubsection{Picosecond ultrasonics}\label{RHF_sec3_2_4}

There is, then, a limit set by the optical absorption edge to the proper determination of acoustic attenuation in dielectric samples using Brillouin light scattering. It corresponds to acoustic waves around \SI{100}{\giga\hertz} in the favorable case of vitreous silica. The samples remain then opaque to electromagnetic radiation up to the x-ray region where they become again transparent. Brillouin spectroscopy known as inelastic x-ray scattering allows then observation of sound properties in the \SI{}{\tera\hertz} frequency range, see Sec.~\ref{RHF_sec3_2_5}.

In the crucial frequency region between \SI{50}{\giga\hertz} and \SI{1}{\tera\hertz}, there exists but a single spectroscopy, known as picosecond ultrasonics (PU). This delicate but very promising technique is essentially based on the generation and the detection of \SI{}{\giga\hertz}$--$\SI{}{\tera\hertz} acoustic waves propagating through a thin glass film by ultrashort optical pulses, typically shorter than \SI{1}{\pico\second}. This optical technique has been pioneered in the 1980s~\cite{THOM1984} but first attenuation measurements on a glass have been reported in the 1990s only~\cite{ZHU1991}. Generally, an ultrashort light pulse is focused onto a small area of a thin metallic film deposited on the dielectric glass sample, which is itself often a film deposited on a substrate. This light pulse, the pump, is absorbed by the metallic layer which expands, generating a longitudinal strain pulse into the glass. Typically, for a \SI{12}{\nano\meter} thick aluminum film on vitreous silica, the launched acoustic pulse is a plane wave packet of about \SI{20}{\nano\meter} spatial extent whose Fourier spectrum peaks around \SI{250}{\giga\hertz}~\cite{THOM1986}. The propagation of the acoustic pulse in the media and its reflection at the different interfaces can be monitored using a second time-delayed light pulse focused on the same spot. The main part of this probe light is reflected at the fixed interfaces whereas a weaker component is reflected by the propagating strain pulse. All these reflections interfere leading to oscillations of the measured reflectivity with the time delay due to the strain pulse bouncing back and forth in the sample. 

\begin{figure}
\begin{center}
\includegraphics[height=6cm]{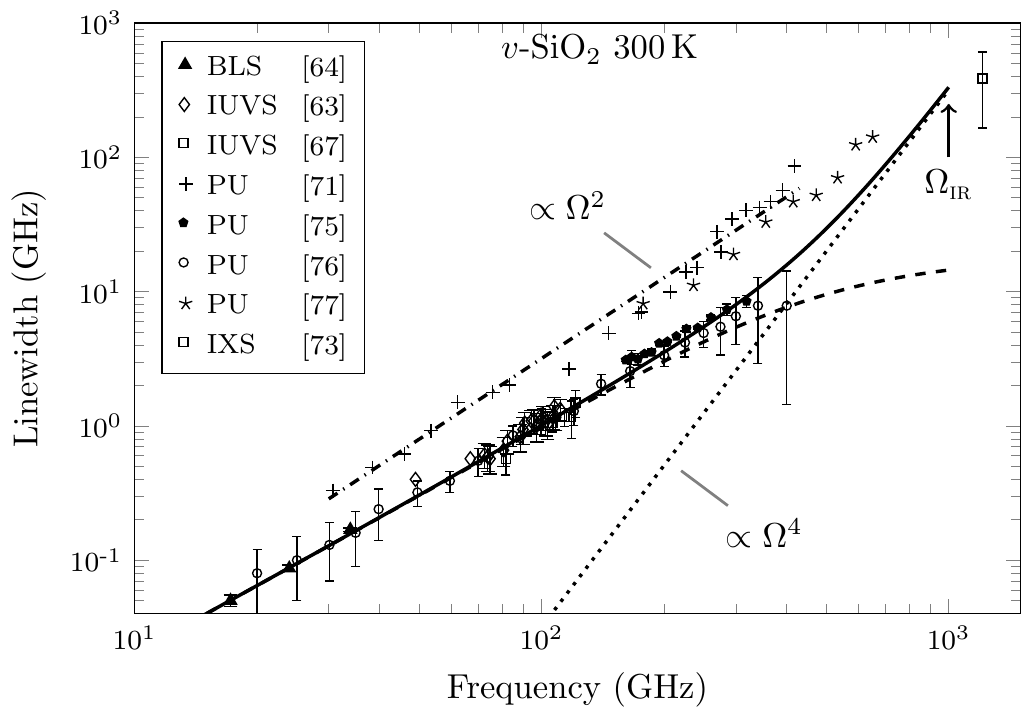}
\caption{Frequency dependence of the Brillouin linewidth in vitreous silica at room temperature using BLS~\cite{VACH2006}, IUVS~\cite{MASC2004, BENA2005, RUFF2011},  different PU schemes~\cite{ZHU1991, DEVO2008, AYRI2011, KLIE2011, WEN2011} and IXS~\cite{BENA1996}. The solid line is the sum $\Gamma_\textsc{tar}+\Gamma_\textsc{anh}$ discussed in the preceding section plus the expected $\Omega^4$ law leading to the Ioffe--Regel crossover at $\Omega_\textsc{ir}$.}
\label{fig:POT}
\end{center}
\end{figure}

In the early work by Zhu \etal~\cite{ZHU1991}, the frequency dependence of the sound attenuation in vitreous silica layers was determined from the decay of successive echoes reaching the Al transducer after a round trip in the glass. A single-crystal tungsten substrate was used to produce numerous and strong echoes thanks to the large acoustic mismatch between $v$-SiO$_2$ and tungsten. Based on the modification of the shape of the acoustic pulse reflecting the frequency dependent attenuation coefficient of its spectral components, this method allows accessing frequencies well above the Brillouin frequency. These early data obtained at room temperature are plotted as stars in Fig.~\ref{fig:POT} showing an approximate quadratic behavior. It is interesting to note that extrapolating this $\Omega^2$ behavior (dashdotted line in Fig.~\ref{fig:POT}) to the \SI{}{\tera\hertz} region leads to values comparable to those reported using IXS (hollow square in Fig.~\ref{fig:POT})~\cite{BENA1996}. We will come back to this point in the next section. Up to about \SI{200}{\giga\hertz}, the picosecond ultrasonic data roughly follow the expected frequency dependence shown by the dashed line discussed in the preceding section as the sum $\Gamma_\textsc{tar}+\Gamma_\textsc{anh}$. However, it became clear with time that the sound attenuation was systematically overestimated by a factor between 2 and 3. It is now well admitted that incomplete corrections for losses at the interfaces were the likely source of errors.

Those difficulties could be alleviated with a different scheme using the oscillations produced by the interference of the probe partly reflected at the sample surface with its reflection by the moving acoustic pulse. In that case, the probe light interacts with one of the Fourier components of the acoustic pulse according to momentum conservation. This \textit{modus operandi} is actually the time-domain analog of the usual Brillouin light scattering and as such suffers in principle from the same limits in the accessible frequency domain~\cite{LIN1991}. If the dielectric sample film is deposited on a substrate with significantly higher refractive index and acoustic velocity, like crystalline silicon, the oscillations detected in the substrate correspond then to much higher frequency acoustic waves. A study of the amplitude of these higher frequency acoustic oscillations in the silicon substrate as a function of the glass film thickness allows then determining hypersound attenuation in the amorphous sample at a frequency beyond the usual BLS limit. This scheme has been used for silica films on crystalline silicon~\cite{DEVO2008,AYRI2011} and the more recent data covering the frequency range \SIrange{165}{320}{\giga\hertz} are plotted in Fig.~\ref{fig:POT} as solid pentagons. They nicely superimpose with the theoretical expectations (dashed line) based on Eqs.~\ref{eq:Qm1ANH} and~\ref{eq:Qm1TAR} up to about \SI{250}{\giga\hertz} in a domain where the anharmonic contribution $\Gamma_\textsc{anh}\propto\Omega^2$ becomes dominant as described in Fig.~\ref{fig:IUVS}b. At higher frequencies, the latter tends to saturates while the measured sound attenuation still increases. This is in fair agreement with the solid line which takes into account the expected $\Omega^4$ law leading to the Ioffe--Regel crossover at $\Omega_\textsc{ir}$. The latter is located around \SI{1}{\tera\hertz} in vitreous silica as shown in Fig.~\ref{fig:POT} together with the lowest point obtained using IXS~\cite{BENA1996}. This was not yet a clear observation of the $\Omega^4$ law in that frequency range but thanks to the very high quality of the data the onset seems to have been detected here.

Similar results have been obtained later on using a narrowband ultrafast photoacoustic approach that allows covering more than a decade in frequency~\cite{KLIE2011}. The attenuation coefficients, converted to linewidth, are shown in Fig.~\ref{fig:POT} as hollow disks. As expected for a multiband technique, the most precise values are obtained in the mid-range from 50 to \SI{270}{\giga\hertz} while data in the frequency wings are rapidly approximate. It is particularly true for the data above \SI{300}{\giga\hertz} which cannot allow the observation of the anticipated $\Omega^4$ law. However, the PU technique is continuously improving and other works reported ultra-broadband PU data reaching \SI{650}{\giga\hertz} in $v$-SiO$_2$~\cite{WEN2011}, shown by stars in Fig.~\ref{fig:POT}. In that case, high frequency Fourier components in the acoustic pulse are obtained replacing the metallic transducer with a piezoelectric nanolayer one. Unfortunately, the reported sound attenuation extracted from the decaying echoes is still clearly overestimated as in~\cite{ZHU1991}. Nevertheless, with a proper analysis of the interface contribution one can envision in the near future to bridge the gap between BLS and IXS so as to firmly detect the onset of the expected $\Omega^4$ law up to the IR crossover at \SI{}{\tera\hertz} frequencies. 

\begin{figure}
\includegraphics[width=\textwidth]{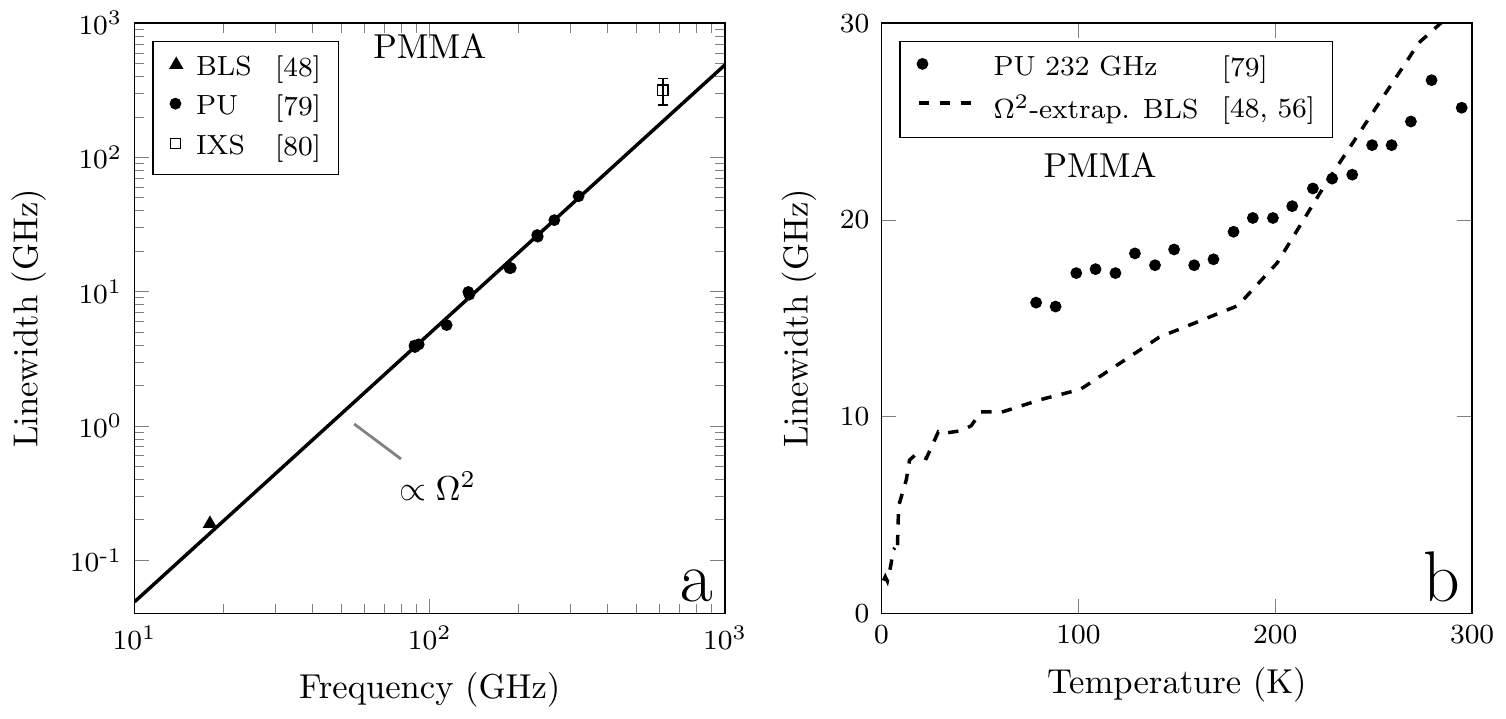}
\caption{a) Frequency dependence of the Brillouin linewidth in PMMA at room temperature using BLS~\cite{VACH1976}, PU~\cite{MORA1996} and IXS~\cite{MERM1998}. The solid line is the $\Gamma\propto\Omega^2$ best fit of the PU data. b) Temperature dependence of the Brillouin linewidth in PMMA at \SI{232}{\giga\hertz}~\cite{MORA1996} compared to $\Omega^2$-extrapolated \SI{18}{\giga\hertz} BLS data (dashed line).}
\label{fig:POTPMMA}
\end{figure}

One must note however that most of the PU data focus on vitreous silica due to the tremendous requirements for making decent glass films having properties comparable to bulk samples. Nevertheless, similar measurements for other amorphous materials have been sparsely collected to investigate the universality of the attenuation at high frequency~\cite{MORA1996}. In the latter work, a general $\Gamma\sim\Omega^2$ law is evidenced up to about \SI{320}{\giga\hertz} for the amorphous polymers PMMA, Polyethyl methacrylate (PEMA) and polystyrene (PS) as well as for the metallig glass TiNi, all at room temperature. Concerning PMMA, the attenuation coefficient measured at low frequencies using BLS nicely extrapolate the obtained $\Omega^2$ law at ambient $T$ as shown in Fig.~\ref{fig:POTPMMA}a. This does not seem to be the case on the high frequency side where a single measurement at about \SI{600}{\giga\hertz} using IXS shows an attenuation coefficient about 70\% larger~\cite{MERM1998}, leaving room for the observation of a $\Omega^4$ law in PMMA. Besides, the temperature dependence of the sound attenuation can be studied using PU as well. Both the transient and steady-state heating of the sample by the laser pump must be controlled, in particular at low-$T$ where heat capacity is  rapidly decreasing. A typical result is displayed in Fig.~\ref{fig:POTPMMA}b for PMMA at \SI{232}{\giga\hertz}~\cite{MORA1996}. The phonon linewidth is continuously decreasing with decreasing temperature, in agreement with a possibly anharmonicity-dominated sound absorption. The temperature dependence seems however different from that obtained using BLS around \SI{18}{\giga\hertz} as shown in Fig.~\ref{fig:POTPMMA}b where the dashed line is the measured BLS linewidth~\cite{VACH1976} extrapolated to \SI{232}{\giga\hertz} with a $\Gamma\propto\Omega^2$ law. This probably means that several competing sound damping mechanisms, each with its own frequency and temperature dependence, are present in PMMA as well. More recently, the temperature dependence of the sound absorption up to about \SI{200}{\giga\hertz} has been studied in vitreous silica showing fair agreement with the theoretical expectations based on Eqs.~\ref{eq:Qm1ANH} and~\ref{eq:Qm1TAR}~\cite{HUYN2017}. 

An attractive perspective given by the rapidly developing picosecond ultrasonic technique is finally the possibility to observe the predicted $\Omega^4$ law for sound attenuation in a ($\omega,T$) region where this process would dominate over relaxation mechanisms. The first direction is to push the technique towards higher frequencies using more complex transducers as acoustic pulse generator/detector. The second path is to sufficiently reduce the temperature to suppress the relaxations, leaving the disorder-induced process dominate the acoustic attenuation.

\subsubsection{Inelastic neutron and x-ray scattering}\label{RHF_sec3_2_5}

Glass samples become again transparent in the soft x-ray region. The latter allows then the observation of sound properties in the \SI{}{\tera\hertz} range using a Brillouin scattering technique known as inelastic x-ray scattering (IXS). Due to the large decrease of the wavelength compared to visible light, the experiment must be done at relatively small angles to achieve the lowest momentum transfer~\cite{MASC1996}. Hence, to observe Brillouin scattering of \SI{}{\tera\hertz} acoustic excitations in amorphous materials, IXS experiments are made in forward scattering in transmission while conventional BLS uses backscattering to increase the sensitivity. The instrumental resolution of the IXS spectrometer is now around \SI{1}{\milli\electronvolt} $\simeq$ \SI{0.25}{\tera\hertz} while the lowest achievable $q$ value is in practice about \SI{1}{\per\nano\meter}. On the other hand, inelastic neutron scattering (INS) is not a technique of choice for such experiments, at least in amorphous materials with rather high sound velocities. Close to $q=0$, it is indeed not possible to measure sound waves whose velocities are larger than those of the incident neutrons due to kinematic conditions. Amorphous selenium is one of the few examples where LA phonons are sufficiently slow to be accessible by INS~\cite{FORE1998}. Both techniques however share the same drawback which is their inability in probing transverse acoustic modes at low $q$. 


\FloatBarrier
\vspace{10pt}
\noindent \textit{a) Two competing scenario}

Measuring the acoustic modes in glasses in the ($\omega,q$) region of the expected IR crossover together with the search for a damping $\Gamma\propto\Omega^4$ has been a strongly debated issue emerging in the late nineties when IXS experiments became operational. The very first IXS experiments in glasses reported on glycerol and LiCl:6H$_2$O in the $q$ region 2--\SI{8}{\per\nano\meter}~\cite{MASC1996}. The measured spectra, which are proportional to the dynamical structure factor $S(q,\omega)$, consist of a strong elastic peak plus a weaker LA Brillouin doublet modeled as a damped harmonic oscillator (DHO) both convoluted with the instrumental resolution. In the classical approximation and neglecting Debye--Waller effects, $S(q,\omega)$ can be written as
\begin{equation}
S(q,\omega) = f_q S(q) \delta(\omega) + (1-f_q)S(q) \frac{1}{\pi} \frac{\Omega_q^2\Gamma_q}{\left[\omega^2-\Omega_q^2\right]^2+\left[\omega\Gamma_q\right]^2}
\label{eq:LA_DHO_IXS}
\end{equation}
where $S(q)$ is the structure factor and $f_q$ the elastic fraction of the scattering at this $q$. $\Omega_q$ is the Brillouin frequency shift, related to the phase velocity of sound waves $v_q =\Omega_q/q$ while $\Gamma_q$ is the full width at half maximum of the Brillouin peak, associated to the sound attenuation. Parameters with the index $q$ depend on the magnitude of the scattering vector $q$ but not on the frequency $\omega$. 

\begin{figure}
\includegraphics[width=\textwidth]{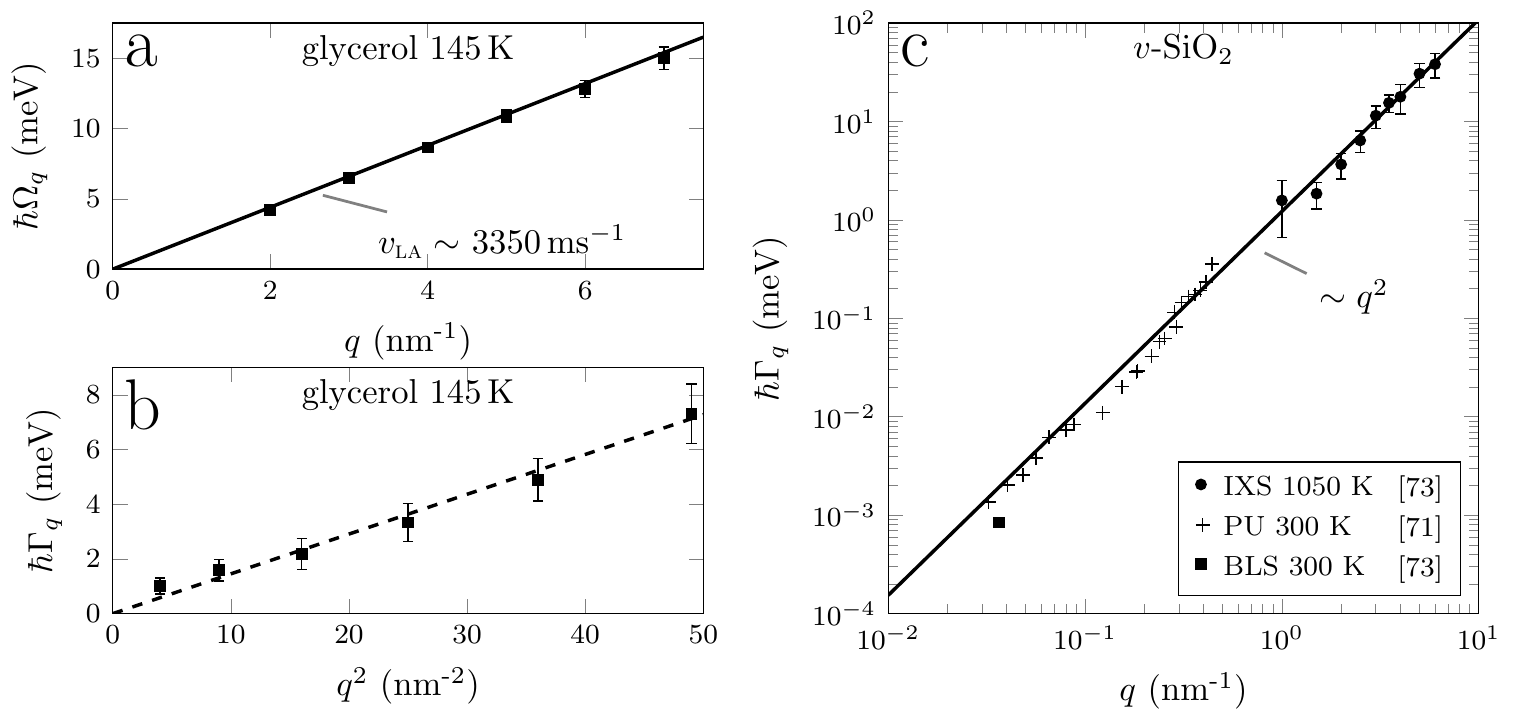}
\caption{Wavenumber dependence of a) the Brillouin frequency shift $\Omega_q$ and b) the Brillouin linewidth $\Gamma_q$ in glycerol at \SI{145}{K} measured using IXS~\cite{MASC1996}. The solid line in a) is the $\Omega_q=v_\textsc{la}q$ best fit giving a sound velocity $v_\textsc{la}\simeq$ \SI{3350}{\meter\per\second} while the dashed line in b) is the $\Gamma_q\propto q^2$ best fit. Adapted from ~\cite{MASC1996}. c) Wavenumber dependence of the Brillouin linewidth $\Gamma_q$ in vitreous silica at \SI{1050}{K} measured using IXS~\cite{BENA1996}, PU~\cite{ZHU1991} and BLS~\cite{BENA1996} at \SI{300}{K}. The solid line is the $\Gamma_q\propto q^\alpha$ best fit of the IXS data giving $\alpha\simeq 1.95$. Adapted from ~\cite{BENA1996}.}
\label{fig:1stIXS}
\end{figure}

The pioneering values obtained in glassy glycerol at $T=$ \SI{145}{K} are displayed in Figs.~\ref{fig:1stIXS}a and~\ref{fig:1stIXS}b showing an almost linear dependence of $\Omega_q$ with $q$ and an approximate quadratic $q$ dependence of $\Gamma_q$. The linear dispersion was corresponding to a sound velocity $v_\textsc{la}\simeq$ \SI{3350}{\meter\per\second}, lower but close to the known longitudinal sound velocity measured at much larger acoustic wavelength. Very similar results were obtained for the second glass, LiCl:6H$_2$O, leading the authors to conclude that propagating sound waves were existing at \SI{}{\tera\hertz} frequencies in those two glasses. This picture was reinforced when first IXS data measured in the archetypal glass $v$-SiO$_2$ also revealed linear dispersion with a longitudinal sound velocity comparable to the macroscopic one together with a quadratic $q$ dependence of the linewidths. The latter behavior was further shown to be reasonably consistent with the sound attenuation coefficient known at lower frequencies at that time\cite{ZHU1991,BENA1996} as illustrated in Fig.~\ref{fig:1stIXS}c where the solid line is the adjustment of the sole IXS data to a $\Gamma_q\propto q^\alpha$ law with $\alpha\simeq 2$. 

Interestingly, a distinct combined analysis of room temperature INS and IXS data in vitreous silica proposing a very different scenario appeared simultaneously~\cite{FORE1996}. In that work, the inelastic part of the spectra was analyzed with a $S_{\rm in}(q,\omega)$ function derived from an effective-medium approximation (EMA) for the force constants of a percolating network that reads~\cite{POLA1988}:
\begin{equation}
S_{\rm in_\textsc{ema}}(q,\omega) = \frac{k_\textsc{b}T q^2}{\pi m\omega}
\frac{q^2 W''(\omega)}{\left[\omega^2-q^2 W'(\omega)\right]^2+\left[q^2 W''(\omega)\right]^2}
\label{eq:SinEMA}
\end{equation}
Eq.~\ref{eq:SinEMA} is a generalization of the second term in Eq.~\ref{eq:LA_DHO_IXS}, appearing clearly when the second moment rule giving $(1-f_q)S(q)\Omega_q^2 = k_\textsc{b}Tq^2/m$ is used in the later~\cite{BUCH2014}:
\begin{equation}
S_{\rm in_\textsc{dho}}(q,\omega) = \frac{k_\textsc{b}T q^2}{\pi m\omega} \frac{\omega\Gamma_q}{\left[\omega^2-\Omega_q^2\right]^2+\omega^2\Gamma_q^2}
\label{eq:SinDHO}
\end{equation}

In the EMA, the disorder-induced inhomogeneous force constants are replaced by a complex, frequency-dependent, uniform function $W=W'-iW''$. $W'(\omega)$ and $W''(\omega)$ result in effective sound velocity $c_\textsc{ema}(\omega)$ and scattering length $\ell_\textsc{ema}$ which depend on the frequency~\cite{POLA1988}
\begin{equation}
c_\textsc{ema}(\omega) = \frac{\left| W(\omega) \right|}{\operatorname{Re}\sqrt{W(\omega)}}
\hspace{2cm}
\ell_\textsc{ema}(\omega) = \frac{1}{\omega} \frac{\left| W(\omega) \right|}{\operatorname{Im}\sqrt{W(\omega)}} 
\label{eq:cdewGdewEMA}
\end{equation}
The physical picture is that, as $\omega$ increases towards $\omega_\textsc{ir}$, the definition in $q$ of the vibrations becomes less and less precise due to the disorder. Thus vibrations of different frequencies have an appreciable Fourier component at the scattering vector $q$. This leads to a dependence in frequency (rather than in wavevector, which is not well defined for the vibrations) of the parameters entering $S_{\rm in}(q,\omega)$. Although by themselves measurements of $S(q,\omega)$ cannot demonstrate phonon propagation or localization, it seemed relevant to investigate the applicability of the two models. 

\begin{figure}
\includegraphics[width=\textwidth]{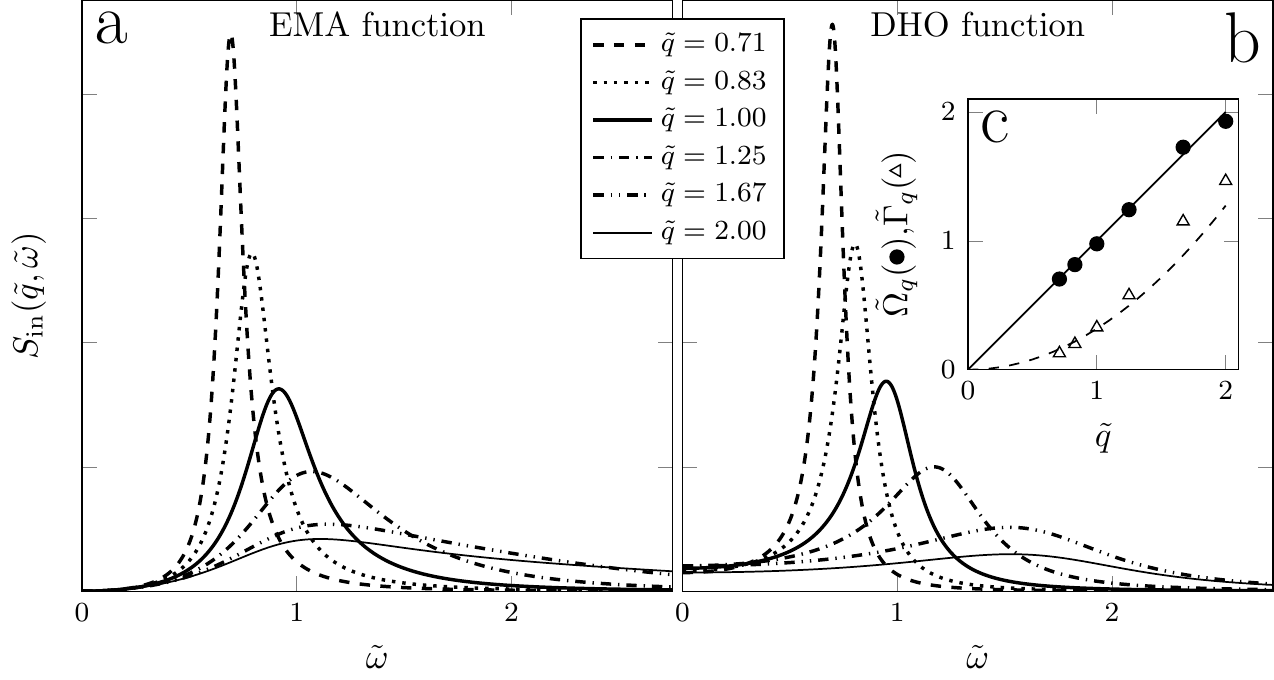}
\caption{Spectral shapes of a) the EMA function (Eq.~\ref{eq:SinEMA}) at several $q$ values around $q_\textsc{ir}$~\cite{FORE1996}, b) of the DHO function (Eq.~\ref{eq:SinDHO}) obtained using the best $\Omega_q$ and $\Gamma_q$ values that fit the EMA profiles. Inset: $q$-dependence of  $\Omega_q$ and $\Gamma_q$. The line is $\Omega_q = c_0 q$ whereas the dashed line is $\Gamma_q \propto  q^2$. The scales are normalized by $\omega_\textsc{ir}$ or $q_\textsc{ir}$.}
\label{fig:DHOEMA}
\end{figure}

To illustrate the difference between Eqs~\ref{eq:SinEMA} and~\ref{eq:SinDHO}, typical spectral shapes are shown in Fig.~\ref{fig:DHOEMA}. The frequency scale is normalized by the hypothetical Ioffe--Regel frequency $\omega_\textsc{ir}$ occurring when the linewidth $\Gamma_q$ reaches $\Omega_q/\pi$ at the wavenumber $q_\textsc{ir} = \omega_\textsc{ir}/c_0$ where $c_0$ is the constant low-frequency sound velocity. Figure~\ref{fig:DHOEMA}a shows EMA spectral shapes at several $q$ values around $q_\textsc{ir}$ obtained from a phonon-fracton crossover model~\cite{COUR1988} which was also used to describe IXS data in
$v$-SiO$_2$~\cite{FORE1996}. In that case the effective sound velocity $c_\textsc{ema}(\omega)$ is $c_0$ well below $\omega_\textsc{ir}$ and increases as $\sqrt{\omega}$ above while the sound damping coefficient $\Gamma_\textsc{ema}(\omega)=c_\textsc{ema}(\omega)/\ell_\textsc{ema}(\omega)$ rapidly increases as $\omega^4$ below $\omega_\textsc{ir}$, then flattens to a linear frequency dependence above. Figure.~\ref{fig:DHOEMA}b displays the corresponding DHO spectral shapes obtained using the best $\Omega_q$ and $\Gamma_q$ values that adjust the EMA profiles. At small $q$ values, below $q_\textsc{ir}$, the spectral shapes in Figs.~\ref{fig:DHOEMA}a and~\ref{fig:DHOEMA}b are quite similar showing a sharp Brillouin peak. The main difference arises at larger $q$ values when the linewidth is large. The homogeneous DHO spectrum shows an increasing high intensity for $\omega$ tending to 0 compared to the peak intensity. Conversely, the EMA spectrum, which is intrinsically inhomogeneous, shows an intensity going to zero together with $\omega$. Furthermore, the EMA spectrum for $q \gg q_\textsc{ir}$ becomes practically $q$ independent, with a peak around $\omega_\textsc{ir}$. 

The inset of Fig.~\ref{fig:DHOEMA}b shows the $q$-dependence of the normalized $\Omega_q$ and $\Gamma_q$ values. An almost linear dispersion of the former is observed even for $q>q_\textsc{ir}$ while the sound damping fairly compares to a quadratic behavior. These schematic graphs illustrate how contradictory conclusions could be drawn from similar IXS experiments: one supporting the existence of propagating sound waves up to high $q$ values~\cite{MASC1996,BENA1996} characterized by linear dispersion and quadratic sound damping, the other one implying a strong disorder-induced scattering of these excitations leading to a Ioffe--Regel crossover~\cite{FORE1996}. However, the limited statistics and energy resolution of the early IXS experiments clearly prevented any definite conclusions at that time. 

\FloatBarrier
\vspace{10pt}
\noindent \textit{b) Search for a quartic attenuation law}

The two competing views have been extensively discussed for many years, thereby including better experiments, progresses in numerical modeling and  theoretical advances. On the experimental side, the linear dispersion of the longitudinal acoustic excitation and the quadratic trend of the linewidth using IXS were reported for many different amorphous systems~\cite{MASC1997, MASC1998, SETT1998, MONA1998, FIOR1999, PILL2000, MATI2001, MATT2003, SCOP2004, MATI2004, BOVE2005}. However, spectroscopic evidences for the alternative scenario were slowly showing up~\cite{FORE1998, RAT1999, MATI2001a, FORE2002}. The latter were  concomitantly strengthened by numerical simulations, in particular of $v$-SiO$_2$ models. IR crossovers for acoustic excitations were observed both for LA and TA modes at $\sim$\SI{1}{\tera\hertz}~\cite{TARA1999, TARA2000a}, in very good agreement with the value obtained from the analysis of INS data~\cite{FORE1996} and ideas developed in~\cite{FORE1998, RAT1999, MATI2001a, FORE2002}. Molecular dynamics further demonstrated that the IR crossover was actually characterizing a continuous transition from propagative acoustic waves to diffusive excitations, and not localized vibrations as suggested by the phonon-fracton approach~\cite{TARA2000a}. Localized modes exist only at much higher frequencies. The disorder-induced quartic behavior of the acoustic damping was nevertheless not observed, due to the low-$q$ box-size limitation of the simulation.
 
Similarly, a direct observation of the $\omega^4$ law using x-rays was still a very demanding experiment. From the thermal conductivity, it was indeed anticipated that the expected $\omega_\textsc{ir}$ hardly exceed 1--\SI{2}{\tera\hertz} in most materials. Combined with the typical sound velocities found in amorphous solids, it leads to crossover wave vectors $q_\textsc{ir}$ in the range 1--\SI{2}{\per\nano\meter}. This happened to be the lowest accessible limit in IXS. Further, with a Lorentzian-like instrumental resolution profile of about \SI{1.7}{\milli\electronvolt} (FWHM), features below \SI{1}{\tera\hertz} were easily buried into the wings of the intense elastic signal. However, studying selected materials with IR crossovers expected at high values appeared to be the right strategy for further progresses. 

\begin{figure}
\includegraphics[width=\textwidth]{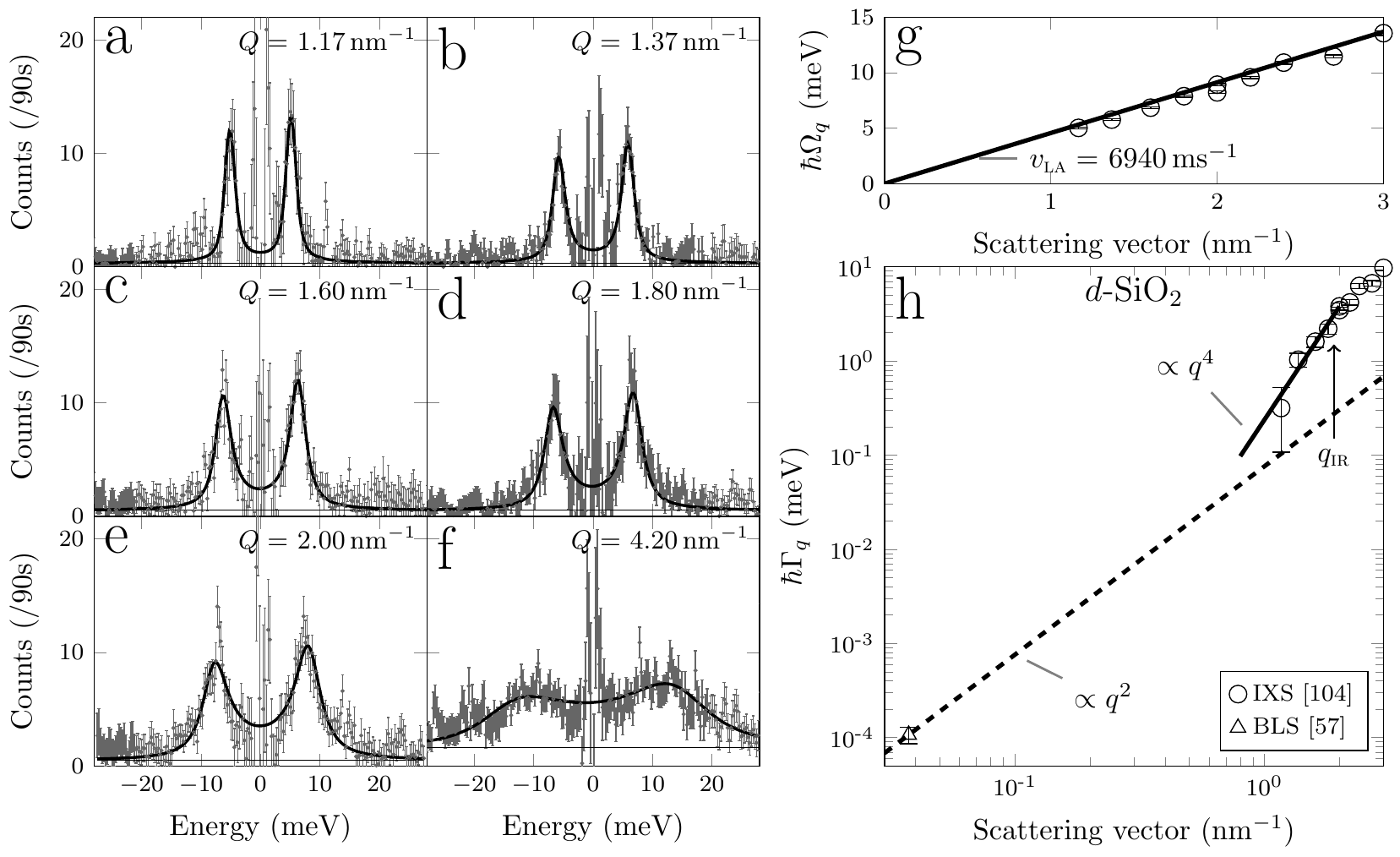}
\caption{a-f) IXS spectra of $d$-SiO$_2$ at several $q$ values and their DHO fits of the inelastic part after subtraction of a central peak freely adjusted in the fitting process~\cite{RUFF2003}. The background shown by the thin baselines is fixed to the detector noise, measured independently. g) Evolution of $\Omega_q$ with $q$ where the line extrapolates the BLS result. h) Evolution of $\Gamma_q$ with $q$ where the dashed line extrapolates the quadratic behavior of the sound attenuation from BLS results~\cite{RAT2005}. The solid line is a $\Gamma_q\propto q^4$ law interpolating the five lowest points. Adapted from~\cite{RUFF2003}.} 
\label{fig:IXSdSiO2}
\end{figure}

The first example is densified silica which is about 20\% denser than normal vitreous silica. The expected crossover is then pushed to higher values, $\omega_\textsc{ir}\sim$ \SI{2.2}{\tera\hertz} and $q_\textsc{ir} \sim$ \SI{2.2}{\per\nano\meter}~\cite{RAT1999, FORE2002}, as opposed to the estimates $\omega_\textsc{ir} \sim$ \SI{1}{\tera\hertz} and $q_\textsc{ir} \sim$ \SI{1}{\per\nano\meter} for $v$-SiO$_2$~\cite{FORE1996}. $d$-SiO$_2$ is also more homogeneous so that the elastic intensity is significantly reduced. For these reasons, it was anticipated that the crucial region below and near $\omega_\textsc{ir}$ might be accessible to spectroscopy in $d$-SiO$_2$ using existing IXS capabilities~\cite{RAT1999,FORE2002}. Figure~\ref{fig:IXSdSiO2} shows typical inelastic spectra obtained at $T =$ \SI{565}{K}~\cite{RUFF2003}. The measured spectra were fitted to a DHO plus a delta function for the elastic part, convoluted with the Lorentzian instrumental function. In Fig.~\ref{fig:IXSdSiO2}a the width of the Brillouin peak is mostly instrumental, while in \ref{fig:IXSdSiO2}e it is due in large part to a real broadening of the Brillouin signal. 

The values obtained for $\Omega_q$ are reported in Fig.\ref{fig:IXSdSiO2}g and show an almost linear $q$-dependence as in other materials and in agreement with the discussion of Fig.~\ref{fig:DHOEMA}. The $q$ dependence of the resulting $\Gamma_q$ values is shown in Fig.\ref{fig:IXSdSiO2}h. $\Gamma_q$ increases very rapidly with $q$ and is significantly above the values that can be extrapolated from the precise BLS measurement. As discussed before for $d$-SiO$_2$, see Fig.~\ref{fig:AnhBLS2}, the sound attenuation at long wavelength originates from anharmonicity in this temperature range and it seems reasonable to assume $\Gamma_q\propto q^2$ in the sub-\SI{}{\tera\hertz} range as shown as the dashed line in Fig.\ref{fig:IXSdSiO2}h. This homogeneous broadening contributes only to 10\% of the linewidth measured using IXS at $q=$ \SI{1.37}{\per\nano\meter}. Below about \SI{2}{\per\nano\meter}, the four points can be well described with a $\Gamma_q\propto q^4$ expression, giving the solid line in Fig.\ref{fig:IXSdSiO2}h. At larger $q$ values, $\Gamma_q$ extracted from the DHO fits seems to increase as $q^2$, in agreement with observations in other materials. It was strongly suggesting  that a \emph{new damping mechanism} should exist in between. Clearly, the latter must lead to a rapid increase of the linewidth, faster than $q^2$, as two different $q^2$ laws could not even crossover. 

In addition, the DHO parameters fulfill the Ioffe--Regel criterion ($\Gamma_q =  \Omega_q/\pi$) at $q\simeq$ \SI{2}{\per\nano\meter}, corresponding approximately to  $\hbar\omega_\textsc{ir}\simeq\hbar\omega_\textsc{bp}$. Finally, another piece of the puzzle came from the shape of the inelastic spectra. It seems clear from Fig.\ref{fig:IXSdSiO2} that with increasing $q$, the DHO profile demands a larger and larger signal at small $\omega$ which is not measured, as already noted in the preceding experiments in $d$-SiO$_2$~\cite{RAT1999,FORE2002}. This is particularly clear in Fig.~\ref{fig:IXSdSiO2}f at $q\simeq 2q_\textsc{ir}$, an averaged spectrum obtained after a week of measurement. All these spectra were also adjusted to an EMA function using Eq.~\ref{eq:SinEMA}, giving similar results at low $q$ values but much better fits approaching $q_\textsc{ir}$ and above, as expected. The sole usefulness of the DHO profile was then to evidence the existence of a new sound damping mechanism for the LA waves which approximately behaves as $q^4$ or $\omega^4$ up to a IR crossover~\cite{RUFF2003}.

\FloatBarrier
\vspace{10pt}
\noindent \textit{c) Fate of the high-frequency acoustic waves}

It was crucial to examine to what extent the observations reported for $d$-SiO$_2$ were common to glasses. To that effect, a second glass, lithium diborate Li$_2$O$-$2B$_2$O$_3$, or LB2, was carefully studied using IXS upon approaching the expected IR crossover at two temperatures. Again, the feeling that the IR frequency should be related to the boson peak guided the choice of a material with a boson peak located at a high energy to facilitate the spectroscopy. Further, some preliminary results already suggested that such a IR crossover should exist in LB2~\cite{MATI2001}. The same procedure as the one previously described for analyzing $d$-SiO$_2$ IXS data was applied to LB2. 

\begin{figure}
\includegraphics[width=\textwidth]{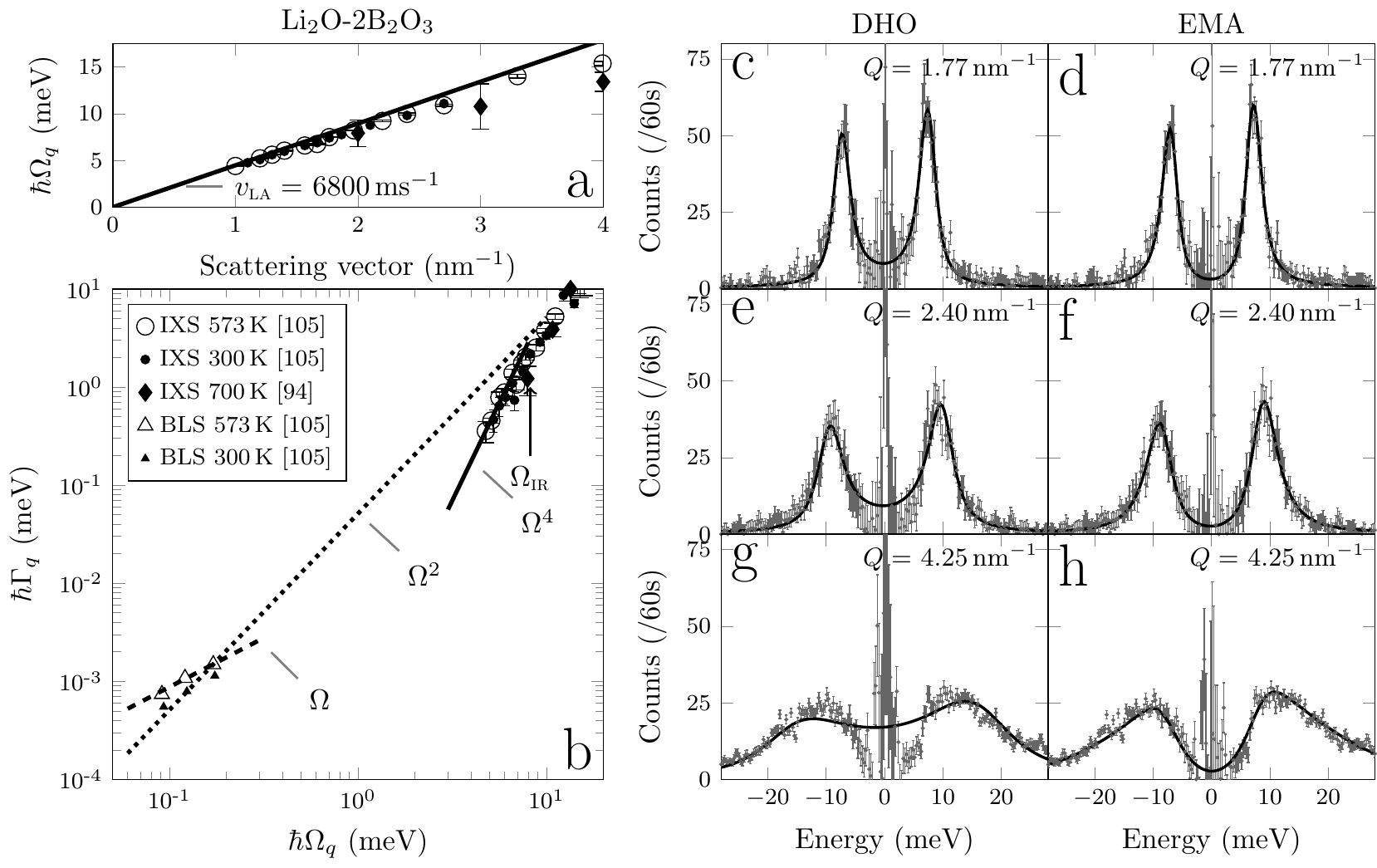}
\caption{a) Evolution of $\Omega_q$ with $q$ where the line extrapolates the BLS result for LB2 glass. b) Evolution of $\Gamma_q$ with $\Omega_q$ where the dashed line at low energy extrapolates the linear behavior of the sound attenuation from BLS results~\cite{VACH2008}. The solid line is a $\Gamma_q\propto q^4$ law interpolating the lowest $q$ points of IXS data~\cite{RUFF2006}. The dotted line is a quadratic behavior extrapolating IXS data obtained above the IR crossover. c-h) Comparison between IXS spectra adjusted to a DHO lineshape  or to an EMA lineshape showing the failure of the former approaching $q_\textsc{ir}$. Adapted from~\cite{RUFF2006}.} 
\label{fig:IXSLB2}
\end{figure}

The parameters extracted from the DHO fits of the IXS data are shown in Fig.~\ref{fig:IXSLB2}a-b~\cite{RUFF2006}. They exhibit very little dependence on $T$, if any. Below about \SI{2}{\per\nano\meter} there is only a slight departure from linear dispersion as seen in Fig.~\ref{fig:IXSLB2}a where the slope of the line is the BLS sound velocity at \SI{573}{K}~\cite{VACH2008}. There is however a very rapid increase of the full width $\Gamma_q$ in this $q$ region, as shown in Fig.~\ref{fig:IXSLB2}b. The latter is definitely faster than a $q^2$ behavior and is consistent with a $q^4$ law as can be inferred from the solid and the dotted lines. Above about \SI{2}{\per\nano\meter}, $\Gamma_q$ increases slower with an exponent close to 2, in agreement with numerous reports. All these results are fully consistent with those which have been obtained previously for densified silica, including the occurrence of the Ioffe--Regel crossover $q_\textsc{ir}$ below \SI{2}{\per\nano\meter} which corresponds to a IR frequency $\omega_\textsc{ir}$ close to the frequency of the maximum of the boson peak $\omega_\textsc{bp}$. The two series of points reported in Fig.~\ref{fig:IXSLB2}b at lower energy have been obtained using BLS analyzed with a high-resolution tandem spectrometer~\cite{VACH2008} for three different scattering angles. They show that contrarily to the $d$-SiO$_2$ case the sound attenuation depends linearly on the frequency in the hypersonic range, pointing to TAR processes as the dominant broadening mechanism in this ($\omega,T$) region. The anharmonic contribution should take over at higher frequencies. 

As discussed before, the experimental evidence for a disorder-induced $q^4$ sound attenuation mechanism leading to a Ioffe--Regel crossover indicates that the DHO cannot be the adequate representation of the spectral lineshape approaching $q_\textsc{ir}$. This is illustrated in Figs.~\ref{fig:IXSLB2}c-h which compare inelastic IXS spectra just below and above $q_\textsc{ir}$ fitted to the DHO lineshape on the left and to an EMA lineshape on the right~\cite{RUFF2006, COUR2006}. Well below $q_\textsc{ir}$, the experimentally selected scattering vector $q$ matches  acoustic modes of well defined $Q$ number, the linewidth $\Gamma_q$ truly reflects an inverse lifetime. In this case, the DHO model gives an excellent representation of the spectral shape. In the opposite limit, where $Q$ is ill-defined at a given mode frequency $\Omega$, a spectrum $S(q,\omega)$ at constant $\Omega=\omega$ just reflects an average spatial profile of the excitation packets at $\omega$~\cite{VACH1999}. Conversely, a spectrum $S(q,\omega)$ at constant $q$ is then the sum of contributions of all the modes with $\Omega=\omega$ at this particular scattering $q$ value. This is inhomogeneous broadening. There is no basis for using the DHO in this case. The failure of the DHO above $q_\textsc{ir}$ is clearly illustrated in Figs.~\ref{fig:IXSLB2}e,g and can be observed in Figs.~\ref{fig:IXSdSiO2}e,f as well.

\begin{figure}
\includegraphics[width=\textwidth]{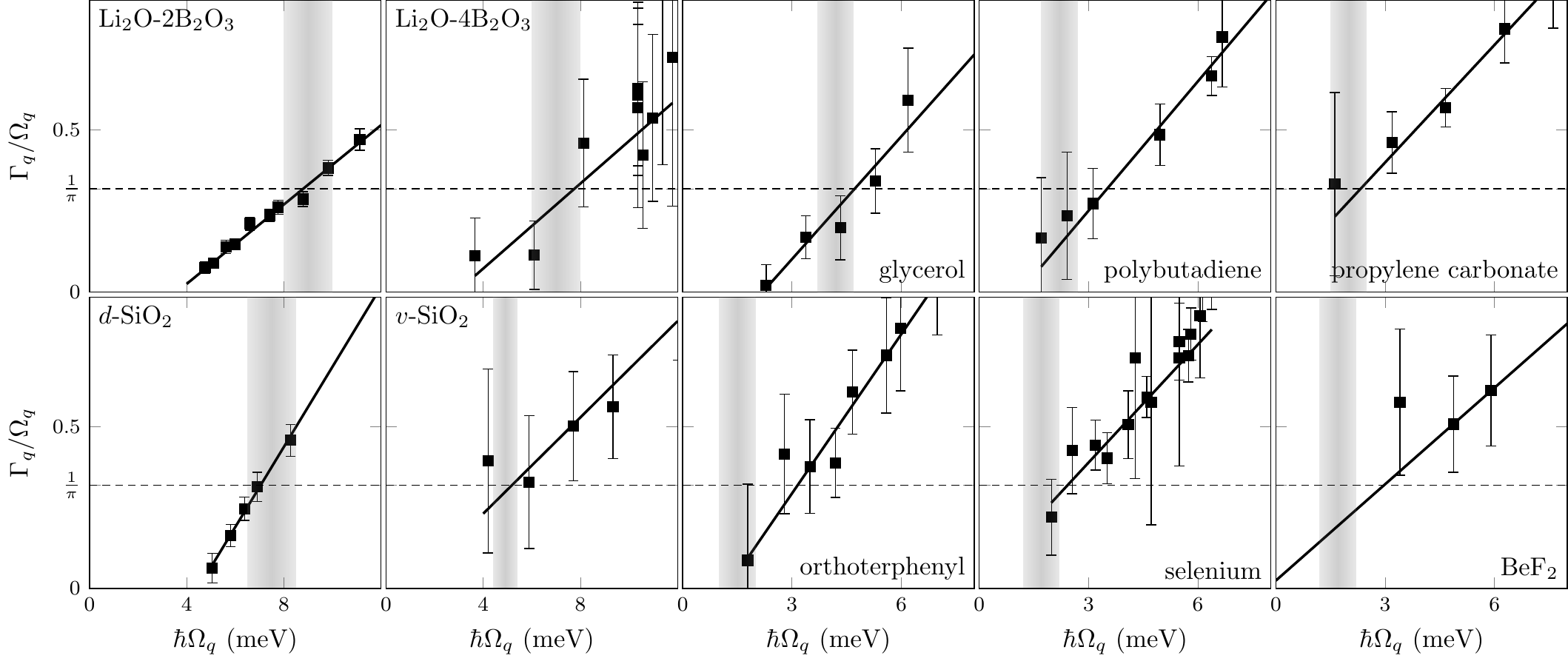}
\caption{Determination of the Ioffe--Regel energy $\hbar\omega_\textsc{ir}$ (intercept with the dashed line) for different glasses and comparison with the energy region of the maximum of the boson peak $\hbar\omega_\textsc{bp}$ (hatched region). References for the experimental points can be found in~\cite{RUFF2006}. The positions and ranges of the BPs are discussed in~\cite{COUR2006, RUFF2007}.} 
\label{fig:IRcrossover}
\end{figure}

The question of why such a rapid increase of the Brillouin linewidth was not detected in all the other glasses investigated so far was elucidated by considering the values of the ratio $\Gamma_q/\Omega_q$ obtained in these IXS experiments. Figure~\ref{fig:IRcrossover} shows values of that ratio vs $\Omega_q$ for ten different glasses. For each glass, the points were arbitrarily interpolated with a straight line in the IR crossover region. Its intercept with the dashed horizontal line $\Gamma_q/\Omega_q = 1/\pi$ gave an estimate for $\omega_\textsc{ir,la}$. One observes that only for LB2 and $d$-SiO$_2$ are high-quality data available below the crossover. In the other cases, this region where the $q^4$ was expected was simply not investigated at that time. For all these glasses, the estimated $\omega_\textsc{ir,la}$ values were clearly in the same frequency range than the maximum of the boson peak $\omega_\textsc{bp}$, marked by the hatched region, and not a decade above~\cite{SCHI2007}. All in all, after a decade of vivid scientific discussions, it appeared that a large part of the picture originally proposed in~\cite{FORE1996} was experimentally verified. It must be noted that the latter observation made on the LA modes gave an incomplete view. The low-frequency region of the vibrational density of states is indeed largely dominated by the transverse acoustic excitations, i.e. as $2c_\textsc{l}^3/c_\textsc{t}^3$. Even if no experimental evidence exists yet, it is largely admitted that $\omega_\textsc{bp}$ is closed to $\omega_\textsc{ir,ta}$ for all glasses while the ratio $\omega_\textsc{ir,la}/\omega_\textsc{ir,ta}$ is system- dependent. The case of orthoterphenyl in Fig.~\ref{fig:IRcrossover} already illustrated this point with $\omega_\textsc{ir,la}/\omega_\textsc{bp}>2$ whereas computer simulation of a soft sphere glass even showed a larger different behavior of the two branches with well defined LA modes up to the pseudo zone boundary and IR crossover of the TA modes at the boson peak frequency~\cite{SCHO2004}. Similarly, well-defined LA modes have been observed later at frequencies well above $\omega_\textsc{bp}$ in metallic glasses~\cite{SCOP2006, COUR2007, BRUN2011}.

\FloatBarrier
\vspace{10pt}
\noindent \textit{d) Interaction of theory and experiment}

Concomitantly with the latter experimental evidences, theoretical approaches were developed to describe vibrations in glasses. They can be grossly separated in two main categories. The first one, the so-called Soft Potential Model (SPM), assumes that there exists in glasses, non-acoustic, soft quasi-local vibrations (QLVs) at low frequencies~\cite{KARP1983}. The latter form an excess $g_\textsc{loc}$ of the vibrational density of states above the Debye contribution, $g_\textsc{loc}(\omega) \propto \omega^4$~\cite{KARP1985}. Their mutual interactions, mediated by the acoustic phonons, lead to the universal shape of the boson peak~\cite{PARS2007}. In return, the acoustic phonons are strongly scattered by the QLVs leading to $\Gamma\propto\omega^4$ up to a IR crossover at $\omega_\textsc{ir} \simeq \omega_\textsc{bp}$~\cite{BUCH1992,RAMO1997a,PARS2001}. As the SPM is an extension of the standard tunneling model~\cite{ANDE1972,PHIL1972}, it offers a unified approach and a comprehensive understanding of the thermo-mechanical anomalies observed in glasses and discussed in Sec.~\ref{RHF_sec1_1}. For more details, the reader is referred to Chapter 9. 

The second category considers harmonic elastic disorder. A self consistent effective medium approach based on fluctuating elastic constants has been worked out in great detail and is now often termed Fluctuating Elasticity Theory (FET)~\cite{SCHI1993,SCHI1998}. The elastic disorder induces an excess of the reduced vibrational density of states over the Debye theory, which is interpreted as the boson peak located at a fraction of the Debye frequency. Similarly to the SPM results, the acoustic waves are  strongly scattered in the boson peak frequency range~\cite{SCHI2006}. In this region, a crossover from an acoustic damping proportional to $q^4$ at low frequency to $q^2$ at high frequency is obtained, as observed experimentally~\cite{SCHI2007}. For more details, the reader is referred to Chapter 10. 

Both theories give very similar predictions, for which increasing the disorder leads to a downshift of the boson peak, eventually driving the system to an instability. The boson peak anomaly marks the crossover from plane-wave like vibrational states to disorder-dominated states with increasing frequency. The inelastic structure factor related to the scattering of the acoustic modes takes the EMA form given by Eq.~\ref{eq:SinEMA} with distinct self-energy functions $W(\omega)$ for the two theoretical treatments~\cite{SCHO2011,SCHI2006}, permitting quantitative comparisons with experiments and computational simulations. It must be stressed than conversely to the FET approach which leads to the vDOS of the eigenmodes of the disordered system, the SPM starts from the coexistence at low frequency of pure propagating acoustic modes and boson peak modes with $g_\textsc{exc}(\omega)$ which are not the eigenmodes~\cite{RUFF2006, COUR2006, RUOC2007, RUFF2007}. 

Numerical simulation has become an increasing powerful tool to shed light on the vibrational properties of glasses. It can provide very relevant physical quantities that are not yet accessible to experiments and thus are certainly useful for testing theories. On one hand, elastic heterogeneity at a typical length scale $\xi\simeq 10-15 \sigma$, where $\sigma$ is the atomic scale, is well documented in glasses. It leads to the breakdown of the continuum elastic description and thus of the Debye approximation at $\xi$ which corresponds to the acoustic wavelength of frequencies close to $\omega_\textsc{ir}$~\cite{WITT2002, TANG2002, YOSH2004, LEON2005, MIZU2014}. On the other hand, there is also substantial evidence from simulation that QLVs exist in disordered systems~\cite{LAIR1991, HAFN1994, TARA1999a, SCHO2004, BELT2016}. More recently, large-scaled simulations gave new insights into the low-frequency continuum limit, evidencing the $\omega^4$ law for the QLVs thus supporting the SPM approach~\cite{LERN2016, MIZU2017, MIZU2018, SHIM2018, SHIM2018a, KAPT2018, BONF2020, RICH2020}.

As far as experiment is concerned, first comparisons using the IXS Brillouin linewidths of the LA modes in several glasses showed a much better agreement of the experimental data with the SPM expectations~\cite{RUFF2008}. The strength of the $q^4$ behavior derived from FET fell systematically well below the observed widths. This can be partly attributed to the fact that the model used was not able to produce sufficiently intense boson peaks above the Debye level. This was particularly true for vitreous silica~\cite{RUFF2008}. It was latter shown that in principle the agreement between experiments and FET calculations could be improved. Introducing spatial correlations in the elastic disorder~\cite{SCHI2008}, or replacing the original Gaussian distribution of the elastic constants by some non-Gaussian ones~\cite{KOHL2013}, leads indeed to a significant enhancement of the boson peak close to the instability. High quality new IXS data were later obtained on the H-bonded glass glycerol which, as guessed from Fig.~\ref{fig:IRcrossover}, was the next system to study. IXS spectroscopy have been performed in great details across the expected crossover $q_\textsc{ir}\simeq 2$--\SI{2.5}{\milli\electronvolt}. All the features previously discussed for $d$-SiO$_2$ and LB2 were confirmed, extending the phenomenology to molecular glasses~\cite{MONA2009}. Additionally, a marked softening of the sound velocity was evidenced below the IR crossover, thus exactly in the region of the rapid increase of the acoustic damping. A negative dispersion is actually the Kramers--Kronig counterpart of the $q^4$ behavior of the sound attenuation and as such was a further evidence of its existence. Both the SPM and FET approaches produce a softening of the apparent sound velocity for frequency below the boson peak~\cite{SCHO2011,SCHI2013}. Similar IXS results were obtained afterward in silica melt~\cite{BALD2010}, glassy sorbitol~\cite{RUTA2010,RUTA2012} and 2Na$_2$O$-$3SiO$_2$~\cite{BALD2014}. Further comparisons between the original FET model~\cite{SCHI2006,SCHI2007} and IXS data for the case of silica melt and glassy sorbitol concluded that FET was able to reproduce qualitatively the experimental data but not quantitatively all the features~\cite{BALD2011,RUTA2012}. For more details, the reader is referred to Chapter 7. 

The relationship between the boson peak spectrum and the Debye density of states, characteristic of the continuous elastic medium was also studied. A successful scaling was supposed to promote the idea that vibrational modes forming the boson peak were only acoustic, as developed by the FET. Small changes of the boson peak spectrum following the evolution of the Debye vDOS were indeed observed in some glasses~\cite{MONA2006,MONA2006a,BALD2009}, while the Debye scaling clearly failed in others~\cite{NISS2007,HONG2008,ORSI2012,RUFF2010}. In the vitreous silica case, it was found that the temperature dependence of the boson peak spectrum rather scales to the sound velocity based on the unrelaxed bulk modulus with scaling exponents in line with the SPM. All in all, the Debye scaling approach was not really successful to discriminate between both theories. 

The importance of coordination number or fluctuations of the particle contact number and applied stress on elastic properties of disordered systems has been underlined recently~\cite{DEGI2014}. An effective medium theory based on these two quantities has been developed which reproduced all the features obtained using the FET approach including Eq.~\ref{eq:SinEMA}. Interestingly, this work emphasizes the relevance of the effective sound velocity $c_\textsc{ema}(\omega)$ and scattering length $\ell_\textsc{ema}(\omega)$ of Eqs.~\ref{eq:cdewGdewEMA} as the physical quantities characterizing the total response of the system. Above the IR crossover, these quantities differ significantly from those extracted from a simple analysis of the IXS data using Eq.~\ref{eq:SinDHO} which leads to:
\begin{equation}
c_q(\omega) = \sqrt{\operatorname{Re}W(\omega)}
\hspace{2cm}
\ell_q(\omega) = \frac{1}{\omega} \frac{\operatorname{Re}W(\omega)}{\operatorname{Im}\sqrt{W(\omega)}} 
\label{eq:cdewGdewDHO}
\end{equation}
with $c_q=\Omega_q/q$ and $\ell_q = c_q/\Gamma_q$. In addition, analysis in term of constant-energy scans using Eq.~\ref{eq:SinEMA} instead of usual constant-wavenumber scans using Eq.~\ref{eq:SinDHO} provide the opportunity to easily include the diffuse Umklapp scattering contribution as illustrated by U. Buchenau~\cite{BUCH2014}. Depending on the system, this contribution can also greatly affects the obtained parameter values. Hence, a promising way to test the several theories is to directly use Eq.~\ref{eq:SinEMA} on constant-energy spectra to find the frequency dependence of the complex self-energy $W(\omega)$, taking into account the quadratic Umklapp scattering contribution.


\section{Spectroscopy of the low-frequency region of the vDOS}\label{RHF_sec4}

This section aims at describing the nature of the modes constituting the low frequency region of the vibrational density of states and to what extend this information can be obtained from the experiments. 
For example, inelastic neutron scattering and inelastic x-ray scattering are sensitive to all vibrations but the $q$-dependence of the scattering allows separating different type of excitations. 
The vibrational contrast in light experiments is more pronounced, even though the well established crystalline or molecular selections rules are lifted out by the structural disorder. 
In glasses those are gathered in a coupling-to-light function, $C(\omega)$, whose expression (including $C(\omega)=0$ for inactive vibrations) depends on the mode-symmetry (atomic displacements) as well as on the experimental technique. 
Comparing the vibrational responses of infrared spectroscopy, Raman scattering, and hyper-Raman scattering, can therefore be very informative. It is particularly true in simple glasses where selections rules are more pronounced.

\subsection{Inelastic neutron scattering}\label{RHF_sec4_1}

\begin{figure}
\includegraphics[width=\textwidth]{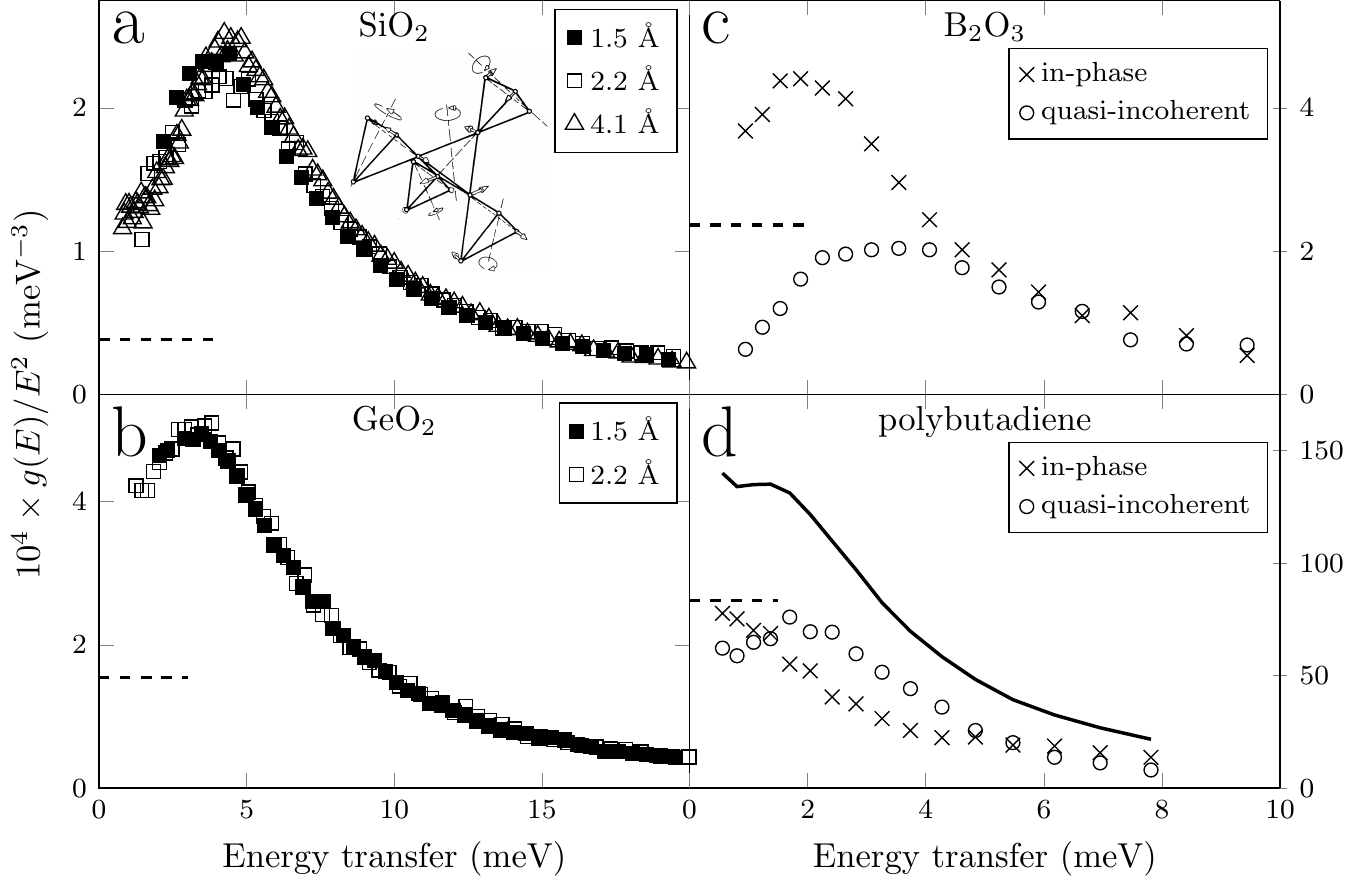}
\caption{bosons peak of (a) SiO$_2$ and (b) GeO$_2$ at different wave vectors~\cite{FABI2008} and schematic representation of coupled SiO$_4$ librations in an undistorted structural model containing five units~\cite{BUCH1986}. bosons peak of (c) B$_2$O$_3$~\cite{ENGB1999} and (d) polybutadiene~\cite{BUCH1996} decomposed into in-phase and quasi-incoherent excitations. The dashed lines indicate the Debye level.}
\label{NS_BP}
\end{figure}

One of the first experimental breakthrough in the understanding of the boson peak modes occurred in the 80's using inelastic neutron scattering (INS). 
Atomic vibrations are all actives in INS and hence it is a technique of choice to access the density of vibrational states $g(\omega)$ of a material (vDOS). 
One drawback however is that it is impossible (except by doing a series of experiments with atomic isotopic substitutions) to normalize the experimental data by the scattering length of each type of atom constituting the material. 
This possibly distorts the experimental result when one wants to compare it, e.g., with atomistic simulations. 
Within the Debye model, the boson peak is an excess of modes over the $\omega^2$ dependence of $g(\omega)$ at low frequency and hence it is observed in a $g(\omega)/\omega^2$ plot, see Fig.~\ref{NS_BP}a. 
Despites the above-mentioned limitations, at the frequencies of the boson peak, the specific heat $C_\textsc{p}(T)$ calculated from the density of vibrational states compares well with the experimental data, at least in vitreous silica~\cite{BUCH1986, FABI2008}. 

In the pioneering work of Buchenau \etal~\cite{BUCH1986}, a quantitative analysis of the low-frequency vDOS of $v$-SiO$_2$ is performed by decomposing the inelastic signal into two components, one proportional to $q^2S(q)$ and another one to $q^2e^{-2W}$, where $S(q)$ is the elastic structure factor and $e^{-2W}$ the Debye--Waller factor. 
This implies two kinds of low-frequency vibrations: a \emph{coherent} contribution corresponding to \emph{in-phase} motions of neighboring atoms or molecular units whose inelastic structure factor reproduces $S(q)$ and hence the first sharp diffraction peak (FSDP), and a \emph{quasi-incoherent} part corresponding to uncorrelated atomic motions with its own oscillations as a function of the scattering wavevector $q$.   
Long wavelength acoustic modes are collective excitations and belong to the first family, while local vibrations (at least to the extent of a few structural units) or optic-like modes, to the second. 
These two contributions add to construct the total vDOS. 
In vitreous silica the quasi-incoherent excitations were interpreted in terms of coupled librations of SiO$_4$ units.
A simplified schematic representation is given in Fig.~\ref{NS_BP}a.
Interestingly, these motions also define the soft mode of the $\beta$-$\alpha$ transition in quartz~\cite{TEZU1991} and 
are known to be instable in quartz and in cristobalite. 
As such, they naturally vibrate at low frequency~\cite{DOVE1997}. 
Similar decompositions were made in $v$-B$_2$O$_3$~\cite{ENGB1999} and polybutadiene~\cite{BUCH1996}, and the two contributions are well separated, as shown in Figs.~\ref{NS_BP}c,d. 
Differently, they superimpose in silica, suggesting a strong hybridization of the modes. 
The possibility that librations of coupled BO$_3$ triangles and B$_3$O$_6$ rings participate to the boson peak of $v$-B$_2$O$_3$ was proposed later~\cite{SIMO2006}, providing thereby a possible explanation for the quasi-incoherent contribution in boron oxide as well.   

\begin{figure}
\includegraphics[width=\textwidth]{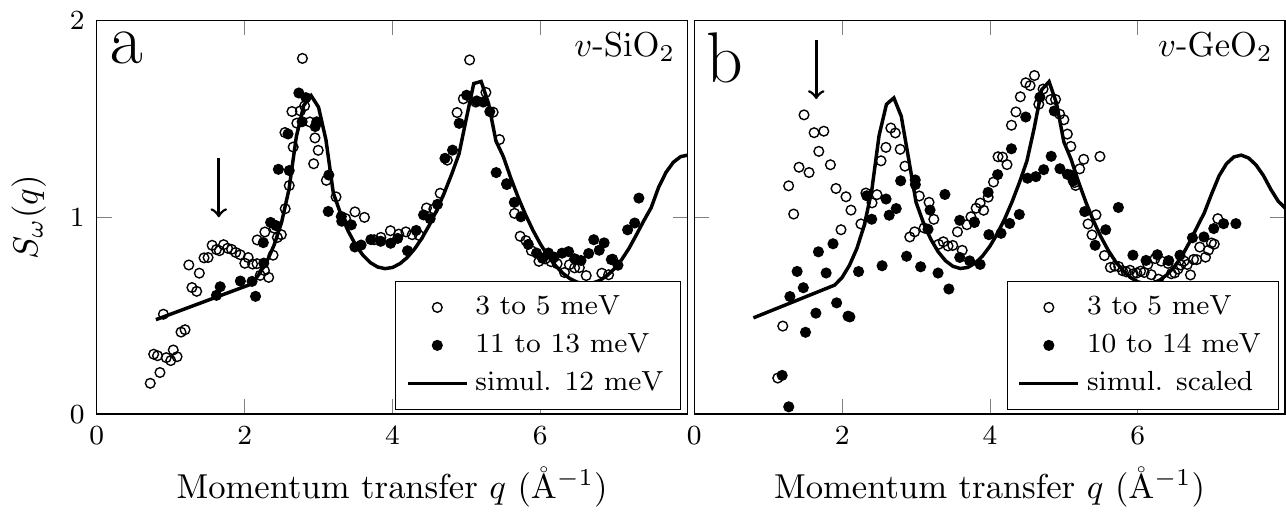}
\caption{Oscillation function of the inelastic structure factor integrated over two frequency domains for a) $v$-SiO$_2$ and b) $v$-GeO$_2$. The arrows mark the position of the first sharp diffraction peak. (adapted from~\cite{FABI2008}).}  
\label{NS_SQW}  
\end{figure}

A second campaign of INS experiments was performed about 10 years after using the high flux of the reactor at the Institut Laue--Langevin at Grenoble. 
The high quality of the data allowed improving the conclusions of the first campaign as well as to compare the situation in $v$-SiO$_2$ and $v$-GeO$_2$ (see Fig.~\ref{NS_BP})~\cite{FABI2008}. 
The spectroscopy allowed defining at each frequency $\omega$, a function $S_\omega(q)$ corresponding to the oscillations of the inelastic structure factor over a purely incoherent scattering (Fig.~\ref{NS_SQW}).
The appearance of an additional oscillation at the position of the FSDP ($\sim$\SI{1.6}{\per\angstrom}), solely in the low-frequency response (3--\SI{5}{\milli\electronvolt}), shows that in-phase atomic motions contribute mostly at frequency close or below the BP maximum. 
For frequency above \SI{10}{\milli\electronvolt}, typically, $S_\omega(q)$ only reproduces the oscillations around 3 and \SI{5}{\per\angstrom} characteristic of coupled librations of SiO$_4$ or GeO$_4$ tetrahedra (solid lines).    
The full analysis reveals that in vitreous silica, about 25\% of in-phase motions and 75\% of quasi-incoherent excitations contribute to the BP below its maximum (40\% and 60\%, respectively, in vitreous germania). 
Above the maximum, the in-phase part decreases so do the long wavelength excitations.  

Among all these excitations, all but the long wavelength acoustic modes can be identified as boson peak modes, the former defining the Debye level (dashed lines in  Figs.~\ref{NS_BP}). 
In silica, germania, and boron oxide, coupled librations of rigid elementary units (SiO$_4$ and GeO$_4$ tetrahedra, BO$_3$ triangles, boroxol rings) contribute partly to the excess of modes, but it is important to notice that the sole contribution of \emph{in-phase} motions also exceeds the Debye level (see, e.g., Fig.~\ref{NS_BP}c). 
This shows that in addition to acoustic \emph{plane waves}, the in-phase contribution contains modes that also built-up the reservoir of boson peak modes.
When considering the connected networks of silica, germania, and boron oxide, it is natural that local librations of elementary structural units (SiO$_4$ tetrahedra, BO$_3$ triangles, B$_3$O$_6$ rings) induce slightly delocalized atomic translations, providing thereby in-phase-type contributions to the boson peak. 

\begin{figure}
\begin{center}
\includegraphics[width=\textwidth]{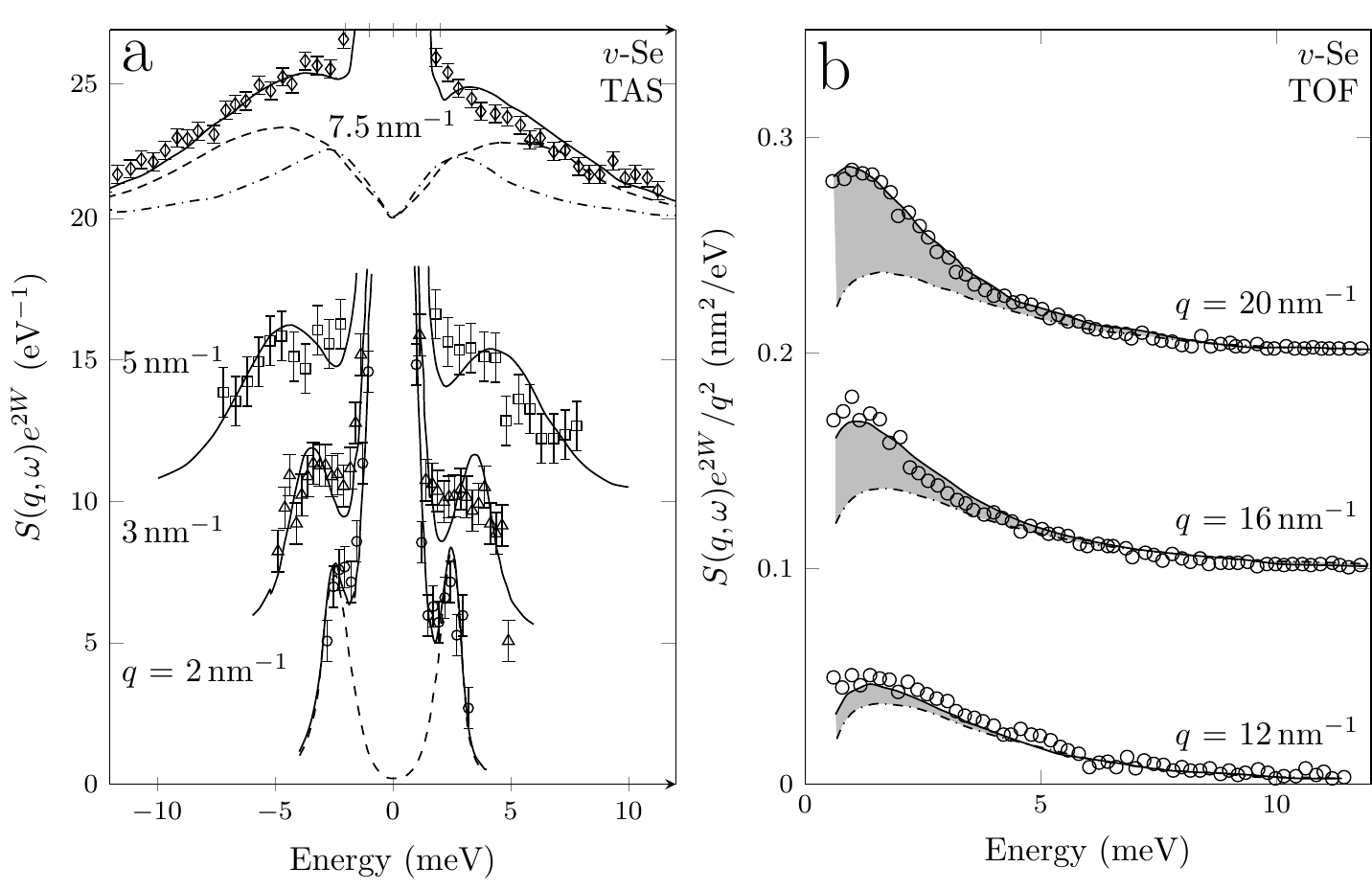}
\caption{INS triple axis experiment in vitreous selenium. LA plane-waves are seen at the smallest $q$ values. Dashed lines in a) and d) illustrate the Brillouin component alone.  The solid lines represent the overall fits (see~\cite{FORE1998} for details). Note the unusual abscissa scale, $\omega/q$ in meV.nm.}  
\label{NS_Se}  
\end{center}
\end{figure}

Attempts to measure the dispersion curves of long wavelength acoustic phonons in glasses with neutrons have also been made.
The main issue in using INS is that in addition to working in near forward scattering ($\theta\simeq$ 1$-$\ang{2} with good instrumental resolution, the experiment must also fulfill drastic kinematic conditions (conservation of momentum and energy). 
The latter basically require that the probed sound velocities must be slower than the speed of the incident neutrons. 
Fulfilling all these conditions at once is very challenging but still, experiments using the thermal neutrons of a triple axis spectrometer have been performed in vitreous selenium, a glass with a low longitudinal sound velocity, $v_\textsc{la}\simeq$ \SI{1800}{\meter\per\second}~\cite{FORE1998}. 
Spectra were recorded at wavevectors as low as \SI{0.2}{\per\angstrom}, as shown in Fig.~\ref{NS_Se}a. 
The cutoff at high frequency is due to the kinematic conditions and limits the analysis. However, one clearly observes the narrow lineshapes of the longitudinal phonons at low wavevector which rapidly broaden with increasing $q$. Above about \SI{1}{\per\angstrom}, the measured spectral shape starts mirroring the BP measured by time-of-flight experiment, see Fig.~\ref{NS_Se}b.
This qualitatively demonstrates the Ioffe--Regel crossover of LA phonons as discusses in Sec~\ref{RHF_sec3_2_5}. 

An analysis of the $q$-dependence of the signal, similar to that performed in ~\cite{BUCH1986, FABI2008}, highlighted three spectral components: 
the Brillouin LA response at low $q$ (dashed line in Fig.~\ref{NS_Se}a), the quasi-incoherent or local-mode component (dot-dashed line in Fig.~\ref{NS_Se}a,b), and the in-phase component developing at high $q$ (gray region in Fig.~\ref{NS_Se}b). 
The latter was called acoustic umklapp scattering in~\cite{FORE1998} as it corresponds to $q$-conserving inelastic processes occurring by exchanging part of the neutron momentum with the peaks of the elastic structure factor. 
It is worth noting that transverse modes dominate in this scattering process and the analysis therefore provides a unique way for obtaining high-frequency spectroscopic information on TA modes in glasses.   
Similarly to INS in silica and germania, its structure factor mimics that of $S(q)$, i.e., $\propto q^2S(q)$, and we recall that for the quasi-incoherent part one simply has a signal $\propto q^2$.  
LA and quasi-incoherent contributions as well as the total scattering response have been fitted with EMA functions derived from the effective-medium approximation (Eq.~\ref{eq:SinEMA}), and the in-phase component corresponds to the difference signal.
The analysis shows that as $q$ increases, the quasi-incoherent contribution progressively overwhelms that of acoustic phonons. 
At high $q$, the in-phase component (umklapp scattering) adds to these local modes to construct the boson peak. 
One also observes in Fig.~\ref{NS_Se}b that the in-phase contribution disappears around 3 to \SI{5}{\milli\electronvolt} revealing that TA phonons ceases to propagate at very low frequency in $v$-Se.

\subsection{Inelastic spectroscopies of light}\label{RHF_sec4_2}
\begin{figure}
\begin{center}
\includegraphics[height=6cm]{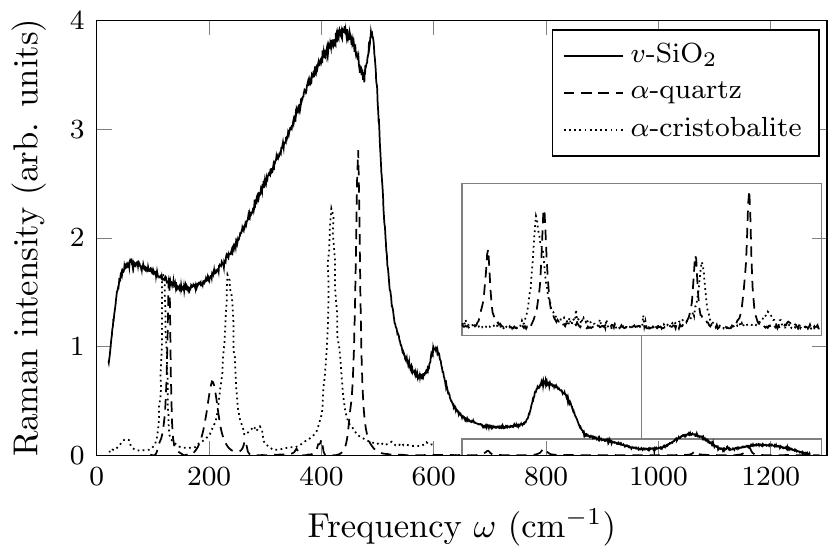}
\caption{Raman spectra of vitreous silica, $\alpha$-quartz, and  $\alpha$-cristobalite \cite{HEHL2020}.}  
\label{RS_XtalVer}  
\end{center}
\end{figure}

In a light scattering experiment the dipole $\bf{p}$ induced by an incident electric field $\bf{E}$ can be expanded in terms of $\bf{E}$, 
\begin{equation}
{\bf p}=\alpha.{\bf E}+\frac{1}{2}\beta:{\bf EE}+...
\label{Dip}
\end{equation}
where $\alpha$ and $\beta$ are the polarizability and hyper-polarizability tensors, respectively.  
The first order term gives the Raman signal whose spectrum $I(\omega)$ for a mode $\sigma$ corresponds to the space and time Fourier transform of the modulation of $\alpha$ by the mode amplitude $W_\sigma$, $\alpha'_\sigma=(\partial\alpha/\partial W_\sigma$) (see Sec.~\ref{RHF_sec2_2}).
 Contrary to INS where all modes participate to the scattering, a vibration will be active or not in light spectroscopy depending on the symmetry properties of the atomic displacements $W_\sigma$ (eigenvectors).
In Raman, the density of vibrationnal states $g_\sigma(\omega)$ of the mode $\sigma$ is observed via the coupling-to-light coefficient $C^\textsc{rs}_\sigma$ so that $I(\omega)$ reads:
\begin{equation}
I(\omega)=C^\textsc{rs}_\sigma g_\sigma(\omega)\frac{n(\omega)+1}{\omega}
\label{RS}
\end{equation}
where  $n(\omega)$ is the Bose population factor and a sum over all the modes $\sigma$ is implicit.  
As the boson peak appears in a $g(\omega)/\omega^2$ plot it is worthwhile defining the "reduced intensity" $I_\textrm{red}(\omega)$ which can also be related to the imaginary part of the Raman susceptibility $\chi"$ 
\begin{equation}
I_\textrm{red}(\omega) = \frac{I(\omega)}{\omega(n(\omega)+1)} = C^\textsc{rs}_\sigma\frac{g_\sigma(\omega)}{\omega^2} \propto \frac{\chi"(\omega)}{\omega}
\label{RS2}
\end{equation}
$C_\sigma^\textsc{rs}$ is known for every space groups and normal modes in crystals and molecules. 
These selections rules are lifted in amorphous materials owing to the structural disorder and $C_\sigma^\textsc{rs}$ transforms into a frequency-dependent function $C_\sigma^\textsc{rs}(\omega)$.
However, for simple network glasses with elementary structural units similar to these of the crystalline polymorphs, the vibrational spectrum resembles to a smeared-out version of the crystal one. 
For example, the similarities between the Raman spectrum of vitreous silica and that of cristobalite or $\alpha$-quartz are striking as illustrated in Fig.~\ref{RS_XtalVer}~\cite{HEHL2020}. 
One major difference, however, is the broad boson peak response appearing at low frequency in all vitreous materials.  
The knowledge of $C^\textsc{rs}_\textsc{bp}(\omega)$ (we will call it $C^\textsc{rs}(\omega)$ for ease of writing) and its frequency dependence is therefore of crucial importance for understanding the nature of the vibrations at BP frequencies.

Within classical expectations, $C^\textsc{rs}(\omega)$ is a constant for local or quasi-local vibrations~\cite{SHUK1970}. For acoustic waves propagating in a mechanically inhomogeneous medium, Martin and Brenig~\cite{MART1974} predicted a response composed by the usual sharp Brillouin peaks at \SI{}{\giga\hertz} frequencies and a background rising as $\omega^2$ originating from the incoherent contribution of the local heterogeneities, so that $C^\textsc{rs}(\omega) \propto \omega^2$.
This points out a fundamental difference between Raman scattering in crystals and glasses at low frequency. 
In the former the signal in null while in the latter the disorder opens up a new scattering channel for  short wavelength acoustics phonons based on incoherent processes (see Sec.~\ref{RHF_sec2_3}). 
This density of states of acoustic phonons builts the low frequency part of the boson peak (typically up to $\omega_{BP}$) to which may adds a coupling with low-lying optic modes.  

\begin{figure}
\includegraphics[width=\textwidth]{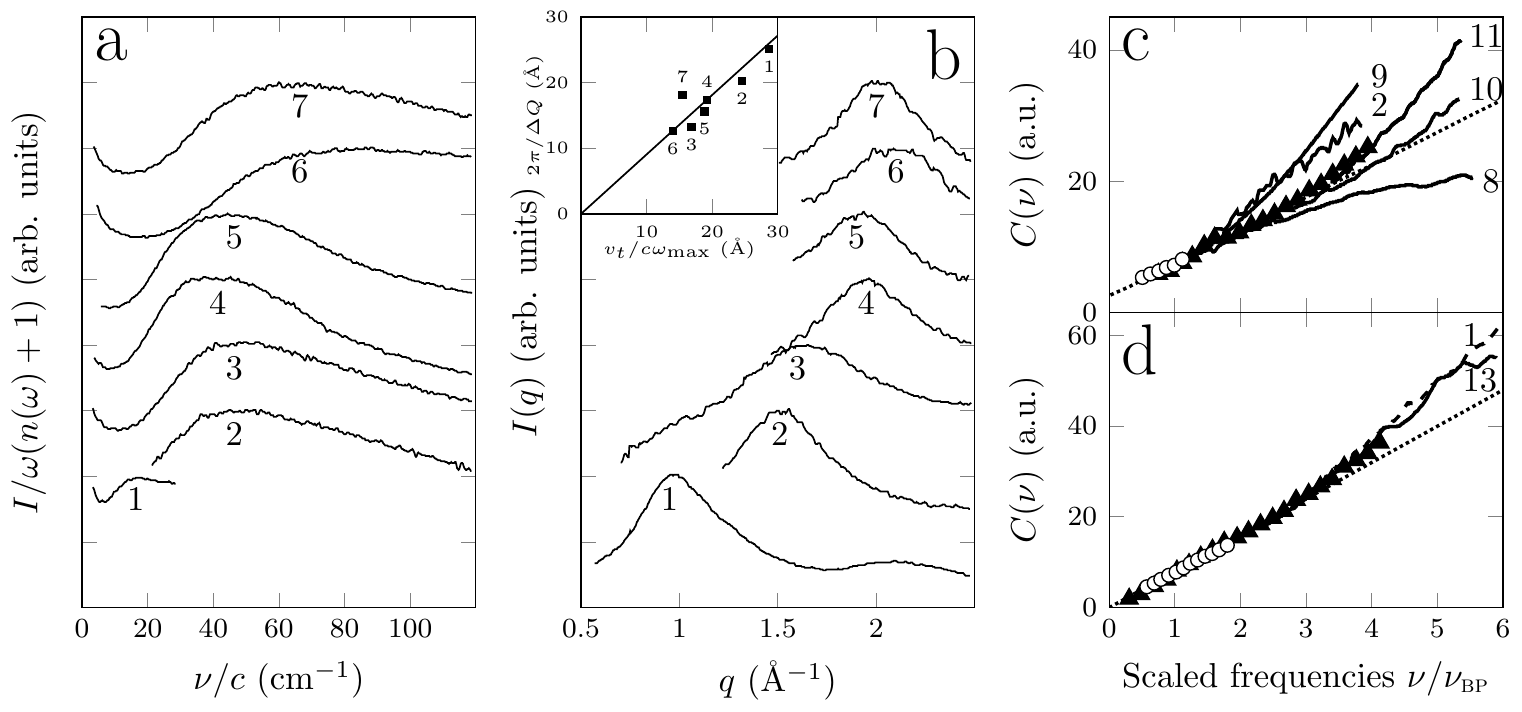}
\caption{a) boson peak and b) FSDP of different glasses (number 1 to 7 in Tab.~\ref{tabRam}).  The inset in b) shows the correlation between characteristic lengths derived from the glass structure and vibrations~\cite{SOKO1992}. c) Frequency dependence of the coupling coefficient $C^\textsc{rs}(\omega)$ for glasses 2, 8--11 in Tab.~\ref{tabRam}, CKN (circles), and Ag$_2$O$-$6B$_2$O$_3$ (triangles) vs $\nu/\nu_\textsc{bp}$. The dotted line in c) is  $C^\textsc{rs}(\nu)\propto\nu/\nu_\textsc{bp}+0.5$. 
d) same as in c) for glasses 1, 13 in Tab.~\ref{tabRam},  As$_2$O$_3$  (triangles), and GeO$_2$ (circles). The dotted line is  $C^\textsc{rs}(\nu)\propto \nu/\nu_\textsc{bp}$ \cite{SURO2002}.}  
\label{RS_BP1}  
\end{figure}

These predictions have been the subject of extensive experimental and theoretical investigations to unravel the nature of the boson peak modes.  
The comparison between the boson peak position $\omega_\textsc{bp}$ in Raman and the spectral broadening and position of the first sharp diffraction peak measured by x-ray diffraction has highlighted the close link between this vibrational feature and the medium range order~\cite{SOKO1992} (Figure~\ref{RS_BP1}a,b). Based on these observations it was proposed that the position of the BP relates to a characteristic length scale in the glass, typically in the nanometer range.
The microscopic origin of the nanometric heterogeneities, either elastic, structural, or dynamical, is still controversial, but it is likely that all of these effects coexist with a glass specific weighting.
An early qualitative model developed by Duval \etal~\cite{DUVA1990} proposed that the frequency of the Raman BP is inversely proportional to the size of the nanodomains, $\xi \simeq v/\omega_\textsc{bp}$ where $v$ is the transverse or longitudinal acoustic velocity for torsional or spherical modes, respectively. 
For most glasses $\xi$, $v$, and $\omega_\textsc{bp}$ vary by a factor about two only, giving thus similar numbers which limit the predictive power of the model.  

\begin{table}
\tbl{Frequency (in \SI{}{\per\centi\meter}) of the boson peak maximum $\omega_\textsc{bp}$ in a Raman reduced spectra and in a $g(\omega)/\omega^2$ plot in INS or specific heat measurement compiled from~\cite{SOKO1992,SURO2002} and additional data (when indicated).}
{\begin{tabular}{@{}cccc||cccc@{}}
\toprule
 & Glass & $\omega_{\rm max}^\textsc{rs}$ & $\omega_{\rm max}^\textsc{ins}$ & & Glass & $\omega_{\rm max}^\textsc{rs}$ & $\omega_{\rm max}^\textsc{ins}$\\
\colrule
1 & PMMA & 16.5 & 12.5 & 9 & Se & 23$^{(2)}$ & 16$^{(3)}$ \\
2 & SiO$_2$ & 51.5 & 33.5 & 10 & PC & 12--15 & 11.0 \\
3 &Na$_2$O$-$18GeO$_2$& 48.5 &  &  11 & PS & 14--18 & 11.5 \\ 
4 & SF1 & 38.5 & & 12 & CKN &  & 20.5 \\
5 & F1 & 43.5 & & 13 & GeSe$_2$ & 10.0  & \\
6 & BaSF64 & 83.5  & & 14 & GeO$_2$ & 41$^{(4)}$ & 27.0\\
7 & LaSF9 & 64.0 & &15 & Ag$_2$O$-$6B$_2$O$_3$& 22.5 &  \\
8 & B$_2$O$_3$ & 23$^{(1)}$ & 18.0 &16 & As$_2$S$_3$ &  & 16.5 \\ \botrule
\end{tabular}}
\begin{tabnote}
$^{1}$from\cite{SIMO2007}, $^{2}$from\cite{PROT2008}, $^3$from \cite{FORE1998}, $^{4}$from \cite{FONT2006}
\end{tabnote}\label{tabRam}
\end{table}

Since each spectroscopy is sensitive to particular atomic motions, another way to probe the boson peak modes is to extract the frequency dependence of its coupling-to-light coefficient. 
According to Eq.~\ref{RS2}, $C^\textsc{rs}(\omega)$ is extracted from the comparison between the Raman intensity with $g(\omega)$ obtained using either INS or specific heat.    
The results reveal that below $\sim 2\omega_\textsc{bp}$ the coupling coefficient exhibits a universal behavior $C^\textsc{rs}(\omega) = A + B\omega/\omega_\textsc{bp}$ \cite{SURO2002} with $A=0$ or $0.5$ depending on the glass (Figure \ref{RS_BP1}c,d). 
For frequencies above $\sim2\omega_\textsc{bp}$ no universal character is found and the BP shape becomes fully material-dependent. 
The progressive transition from a universal to a non-universal behavior of $C^\textsc{rs}(\omega)$ correlates with the frequency of the Ioffe--Regel crossover, providing further evidence of the close link between the strong scattering regime of acoustic branches and the boson peak maximum.
 
Until recently, infrared spectroscopy has been barely used to study the BP modes due to the technical difficulty of reaching low frequencies.
The recent development of \SI{}{\tera\hertz} light sources (far IR) in laboratories or of soft x-rays spectrometer at synchrotrons enabled to overcome this issue. 
Time-domain or \SI{}{\tera\hertz}-spectroscopies (\SI{}{\tera\hertz}-TDS) are now mature for measuring the absorption coefficient $\alpha(\omega)$ in the far infrared, i.e., down to a few wavenumbers~\cite{SIBI2014}.
Within the linear response theory $\alpha(\omega)$ relates to the density of vibrational states $g(\omega)$ by $\alpha(\omega)=C^\textsc{ir}(\omega)g(\omega$), where $C^\textsc{ir}(\omega)$ is the infrared coupling coefficient. Since the BP is seen in a $g(\omega)/\omega^2$ plot one simply has
\begin{equation}
\frac{\alpha(\omega)}{\omega^2}=C^\textsc{ir}(\omega)\frac{g(\omega)}{\omega^2}
\label{IR1}
\end{equation}
In glasses, Taraskin \etal~\cite{TARA2006} proposed a model where $C^\textsc{ir}(\omega)=A+B\omega^\beta$ for frequencies below the Ioffe--Regel crossover. 
$A$ and $B$ are material-dependent constants related to long-range uncorrelated charge fluctuations and correlated short-range (inter-atomic) charge fluctuations, respectively.  $\beta=2$ in the model of Ref.~\cite{TARA2006} but exponents as low as $\beta=1$ has been measured in molecular glasses, e.g., glucose~\cite{KABE2016}. 
This departure has been interpreted in terms of acousto-optic mode couplings and bending of the acoustic branches which are not taking into account in Taraskin and Elliott model.  

Hyper-Raman scattering (HRS) is a non-linear spectroscopy where two incident photons scatter one photon after interaction with a vibration in the medium. 
It corresponds to the first non-linear term in Eq.~\ref{Dip} and hence the signal originates from the fluctuations of the hyper-polarizability tensor $\beta$~\cite{DENI1987}. 
$\beta'$ has different symmetry properties than $\alpha'$ yielding to different selection rules.  
For example, polar vibrations are always active in HRS whatever the crystalline or molecular symmetry is. Conversely, there exists modes that are active in HRS but inactive in both RS and IR, improperly named \emph{silent modes}. 
The relation between the measured intensity $I_\textsc{hrs}$ and the density of state $g(\omega)$ is similar to that of Eq.~\ref{RS} and~\ref{RS2} replacing $C^\textsc{rs}_\sigma(\omega)$ by $C^\textsc{hr}_\sigma(\omega)$, the latter mirroring the properties of the hyper-polarizability tensor.  
\begin{figure}
\includegraphics[width=\textwidth]{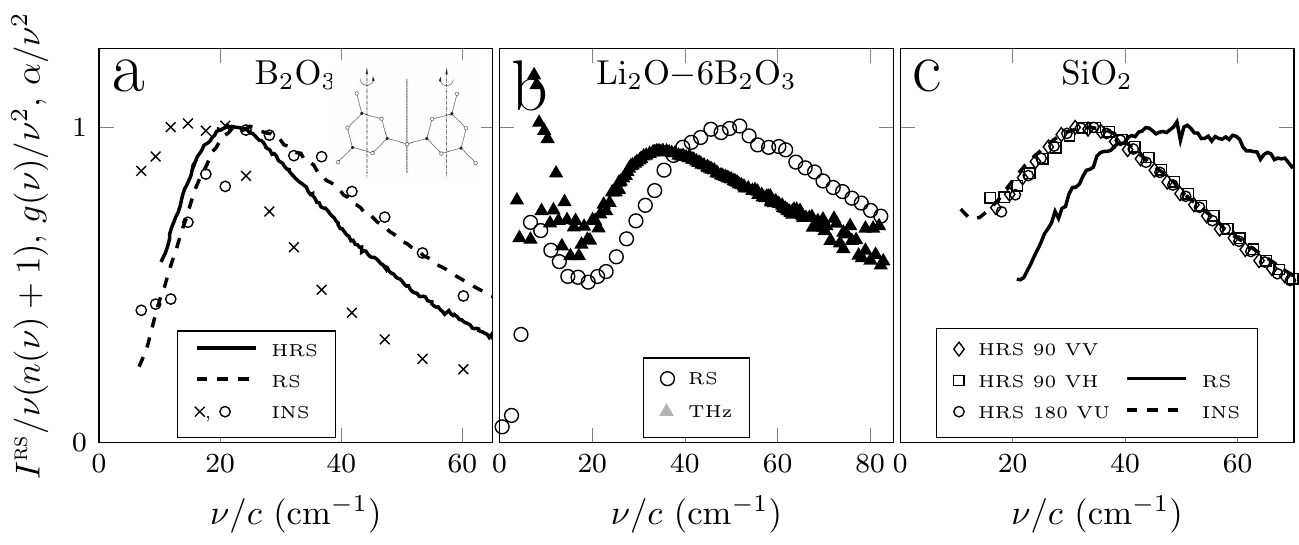}
\caption{a) boson peak of $v$-B$_2$O$_3$ : HRS (solid line), RS (dashed line), out-of-phase INS component (circles), and in-phase INS component (crosses) \cite{SIMO2006}. A schematic representation of $E''$-symmetry librations of boroxol rings is also shown. b) boson peak in a lithium borate glass with \SI{14}{\mol\percent} Li$_2$O content in RS and \SI{}{\tera\hertz}-TDS spectroscopy \cite{IIJI2018}. c) boson peak of $v$-SiO$_2$ : HRS (symbols), RS (solid  line), and INS (dashed line) \cite{HEHL2012}. }  
\label{RS_BP2}  
\end{figure} 

Figure~\ref{RS_BP2} compiles the boson peaks obtained in vitreous silica and vitreous borate glasses using neutron scattering, Raman scattering, hyper-Raman scattering, and \SI{}{\tera\hertz}-TDS spectroscopy.
The different spectral shapes arise from the different frequency dependence of the coupling coefficients combined to the sensitivity of each technique to specific vibrations.
In $v$-B$_2$O$_3$, see Fig. \ref{RS_BP2}a, the Raman and hyper-Raman BP are rather similar and coincide with the quasi-incoherent INS contribution shown in  Fig.\ref{NS_BP}c. 
The combined analysis revealed that the latter could correspond to local vibrations involving out-of-plane librations of rigid BO$_3$ triangles and B$_3$O$_6$ rings whose schematic representation is also shown in Fig.~\ref{RS_BP2}a. 
Within the $D_\textrm{3h}$ point group of these molecular units, this corresponds to $E''$-symmetry modes which are indeed active in both RS and HRS~\cite{SIMO2006, SIMO2007}. 
\SI{}{\tera\hertz}-TDS in pure $v$-B$_2$O$_3$ is very challenging, but data obtained in a lightly doped lithium borate glass Li$_2$O$-6$B$_2$O$_3$~\cite{IIJI2018} highlights a BP at significantly lower frequency than the RS one (see Fig.~\ref{RS_BP2}b), and suggests that the IR response rather associates with the in-phase INS contribution.  
In silica, the HRS BP has been associated to local or quasi-local libration motions involving rigid SiO$_4$ units~\cite{HEHL2020,HEHL2012} introduced by Buchenau \etal~(see Fig.~\ref{NS_BP}c).  
Within the T$_\textrm{d}$ point group symmetry of SiO$_4$ tetrahedra, this vibration is inactive in both IR (thus non-polar) and RS, and only active in HRS (\emph{silent mode} of $F_1$-symmetry).  
This is likely the reason why RS and HRS responses are so different. 
The former has been ascribed to a leakage of these modes due to the structural disorder and dipolar-type contributions~\cite{HEHL2012}. 
Moreover, since the $F_1$-symmetry modes are non-polar, they do not carry an electric charge and therefore do not participate to the local contribution of charge fluctuations ($B$ term). 
This is consistent with the observed  constant behavior of $C^\textsc{ir}$ in vitreous silica~\cite{TARA2006}. 
Finally, the fact that the HRS-BP perfectly matches the INS one supports INS assumption that these modes play an important role in the excess of modes at low frequency in vitreous silica. 
Besides, libration motion of SiO$_4$ tetrahedra is the soft mode of the structural $\alpha$-$\beta$ instability at \SI{846}{K} in quartz~\cite{TEZU1991}. 
Its frequency extrapolates to $\sim$ \SI{36}{\per\centi\meter} at $T_g\sim$ \SI{1600}{K}~\cite{DOLI1992}, i.e., to the frequency of the maximum of $g(\omega)/\omega^2$ measured by INS. 
Librations of rigid units are also low-lying vibrations in $\beta$-cristobalite, but are located at the zone boundary of a transverse acoustic branch.   
In the glass these modes are characterized by weak bondings (those creating the disorder).
It is therefore natural that they vibrate at low frequency. 
In a connected network they also  hybridize with transverse and longitudinal acoustic phonons of similar frequency. 
This source of scattering combines with the incoherent scattering of acoustic modes to built up the reservoir of boson peak modes.  
The loss of the Brillouin zone in the glass combined with the localization of plane wave makes identification of these vibrations as acoustic- or optic-like useless. 
On heating, the inter-unit bonds gradually break and these libration motions (vibrations) transform into rotation motions (relaxations) in the liquid state.

\subsection{Nuclear inelastic scattering}\label{RHF_sec4_3}

Phonon spectra can be obtained with a M\"{o}ssbauer radioisotope source by analyzing the frequency dependence of its nuclear resonance fluorescence with recoil. 
The principles were formulated soon after the discovery of the M\"{o}ssbauer effect~\cite{VISS1960}, but Nuclear inelastic scattering (NIS) setups have emerged only in the 90's with the development of synchrotron radiation offering highly monochromatic x-rays beams, i.e., compatible with the phonon energy spectra. 
Albeit the technique is limited to elements possessing a M\"{o}ssbauer isotope, this can turn into an important advantage for example when the dynamics of the nuclei must be studied separately.  

\begin{figure}
\includegraphics[width=\textwidth]{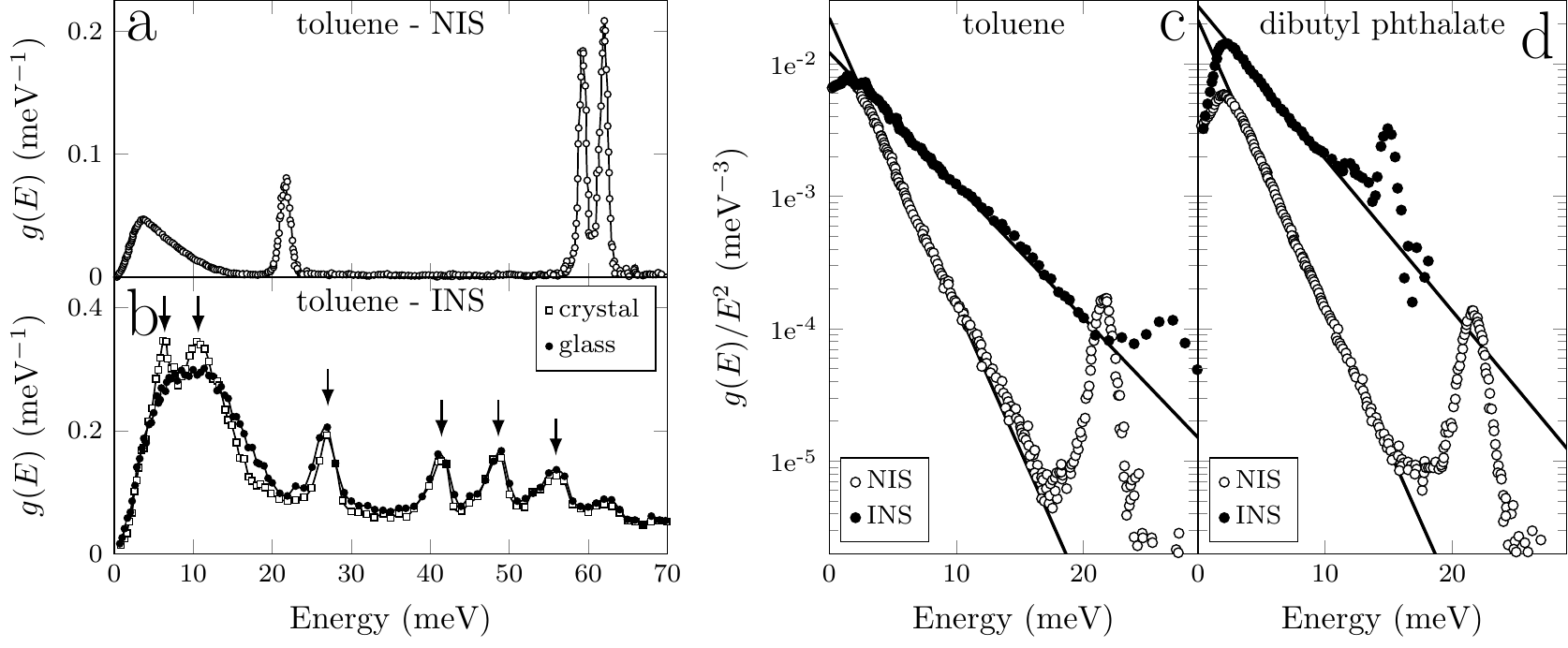}
\caption{Density of states of toluene glass measured by a) nuclear inelastic scattering at \SI{22}{K} from ferrocene molecules and b) inelastic neutron scattering. The arrows indicates local optic modes not seen in the NIS vDOS. Reduced NIS- and INS-vDOS in c) toluene and d) dibutyl phthalate in semi-log plots. Adapted from~\cite{CHUM2004}.}  
\label{IMS}  
\end{figure}

In molecular glasses the probe were neutral ferrocene molecules with the central $^{57}$Fe nucleus as M\"{o}ssbauer isotope~\cite{CHUM2004}. 
In that case, neither rotations of the probe nor intramolecular modes of the glassy matrix are seen. At boson peak frequencies, the probe selects only few modes among the total vDOS, mostly displacements of the rigid probe driven by the translational collective motions of the glass network. 
The coherent length of such vibrations was estimated to be about \SI{20}{\angstrom}.
The net difference between the NIS vDOS and the INS vDOS of toluene is shown in Figs.~\ref{IMS}a,b.
The latter is much more complex due to the participation of rotations and librations of the methyl groups, as well as other local modes at high frequency. 
The comparison of the NIS and INS bosons peaks of toluene and dibutyl phthalate is also very informative (see Figs.~\ref{IMS}c,d). 
The higher number of vibrational states in the total vDOS (INS) as compared to the NIS vDOS (by a factor of 2) in dibutyl phthalate at and below the BP maximum was explained by modes with coherent lengths shorter than the \SI{20}{\angstrom} probed by the technique, i.e., quasi-local or local vibrations. 
%
NIS was also performed in a Na$_2$FeSi$_3$O$_8$ glass with 95\% enrichment of $^{57}$Fe isotope~\cite{MONA2006a, CHUM2011}. 
The structural analysis has shown that iron is mostly in tetrahedral coordination and therefore plays a similar role as silicon.  
A comparative analysis of the NIS vDOS of the glass and of the crystalline polymorph of NaFeSi$_2$O$_6$ revealed that the number of vibrational states of the boson peak is similar to that of the Van--Hove singularity of the TA branch in the crystal. 
It also suggested that the latter are redistributed at lower frequency in the glass and construct the boson peak. 
it is worth  reminding however that in these pioneered NIS campaigns, the M\"{o}ssbauer isotope $^{57}$Fe, positioned at the center of mass of the ferrocene molecule and of the FeO$_4$ tetrahedra, is insensitive to libration motions which are supposed to play an important role at the boson peak frequencies in silicates. 

\subsection{Inelastic x-ray scattering}\label{RHF_sec4_4}

The density of vibrational states of pristine and densified silica glasses have been collected by inelastic x-ray scattering with setups equipped either with a crystal-type or by a M\"{o}ssbauer-type analyzer~\cite{CHUM2014}. 
The data have been compared with those obtained in various crystalline SiO$_2$ polymorphs. 
Calculating the absolute value of the DOSs, the authors have shown that the number of modes in excess relative to the Debye level is similar in the glass and in the crystalline polymorph with a comparable density.
In addition, the shape of $g(\omega)/\omega^2$ in the glass (BP) appears as a smeared out version of $g(\omega)/\omega^2$ in the corresponding crystal which shows the Van--Hove singularity. 
Despite these striking resemblances, it would be misleading to conclude that the BP simply identifies to the Van--Hove singularity of the acoustic branches in the corresponding crystal.
The nature of the vibrations underlying the low frequency vDOS in glasses and  crystals are indeed very different. In the latter, the vibrations are plane waves (acoustic and optic) propagating in an ordered material while in the former the acoustic modes are scattered by the disorder.
These differences produce the shadowed region in $C_\textsc{p}(T)/T^3$ (cf.~Fig.~\ref{fig:thermprop}) and may appear as a \emph{detail} when comparing the total vDOS of crystals and glasses.    
It is however these specific properties which lead to the scattering regime of phonons at the origin, e.g., of the plateau in the thermal conductivity, and  to the boson peak observed in inelastic light spectroscopies.



\bibliographystyle{ws-rv-van}

\begin{thebibliography}{179}
\providecommand{\natexlab}[1]{#1}
\providecommand{\url}[1]{\texttt{#1}}
\expandafter\ifx\csname urlstyle\endcsname\relax
  \providecommand{\doi}[1]{doi: #1}\else
  \providecommand{\doi}{doi: \begingroup \urlstyle{rm}\Url}\fi

\bibitem{BILI1975}
N.~Bilir and W.~A. Phillips, Phonons in {SiO$_2$}: The low-temperature heat
  capacity of cristobalite, \emph{Phil. Mag.} {\bf 32}\penalty0 (1), \penalty0
  113--122  (1975).

\bibitem{DOVE1997}
M.~T. Dove, M.~J. Harris, A.~C. Hannon, J.~M. Parker, I.~P. Swainson, and
  M.~Gambhir, Floppy modes in crystalline and amorphous silicates, \emph{Phys.
  Rev. Lett.} {\bf 78}\penalty0 (6), \penalty0 1070--1073  (1997).

\bibitem{SIGA1999}
V.~N. Sigaev, E.~N. Smelyanskaya, V.~G. Plotnichenko, V.~V. Koltashev, A.~A.
  Volkov, and P.~Pernice, Low-frequency band at 50 cm$^{-1}$ in the raman
  spectrum of cristobalite: identification of similar structural motifs in
  glasses and crystals of similar composition, \emph{J. Non-Cryst. Solids}.
  {\bf 248}\penalty0 (2), \penalty0 141--146  (1999).

\bibitem{WIET1921}
R.~Wietzel, The stability conditions of the glass and crystal phase of silicon
  dioxide, \emph{Z. Anorg. Allg. Chem.} {\bf 116}, \penalty0 71--95  (1921).

\bibitem{SIMO1922}
F.~Simon, Analysis of specific heat capacity in low temperatures, \emph{Annalen
  der Physik}. {\bf 68}\penalty0 (11), \penalty0 241--280  (1922).

\bibitem{SOSM1927}
R.~B. Sosman, \emph{The Properties of Silica: An Introduction to the Properties
  of Substances in the Solid Non-conducting State}. vol.~37, \emph{American
  Chemical Society. Monograph series}, The Chemical Catalog Co, New York, NY
  (1927).
\newblock p. 362.

\bibitem{FLUB1959}
P.~Flubacher, A.~J. Leadbetter, J.~A. Morrison, and B.~P. Stoicheff, The
  low-temperature heat capacity and the raman and brillouin spectra of vitreous
  silica, \emph{J. Phys. Chem. Solids}. {\bf 12}\penalty0 (1), \penalty0 53--65
   (1959).

\bibitem{ZELL1971}
R.~C. Zeller and R.~O. Pohl, Thermal conductivity and specific heat of
  noncrystalline solids, \emph{Phys. Rev. B}. {\bf 4}\penalty0 (6), \penalty0
  2029--2041  (1971).

\bibitem{BUCH1986}
U.~Buchenau, M.~Prager, N.~N\"ucker, A.~J. Dianoux, N.~Ahmad, and W.~A.
  Phillips, Low-frequency modes in vitreous silica, \emph{Phys. Rev. B}. {\bf
  34}\penalty0 (8), \penalty0 5665--5673  (1986).

\bibitem{RAMO1997a}
M.~A. Ramos and U.~Buchenau, Low-temperature thermal conductivity of glasses
  within the soft-potential model, \emph{Phys. Rev. B}. {\bf 55}\penalty0 (9),
  \penalty0 5749--5754  (1997).

\bibitem{LASJ1975}
J.~C. Lasjaunias, A.~Ravex, M.~Vandorpe, and S.~Hunklinger, The density of low
  energy states in vitreous silica: Specific heat and thermal conductivity down
  to 25 mk, \emph{Solid State Commun.} {\bf 17}\penalty0 (9), \penalty0
  1045--1049  (1975).

\bibitem{CAHI1987}
D.~G. Cahill and R.~O. Pohl, Thermal conductivity of amorphous solids above the
  plateau, \emph{Phys. Rev. B}. {\bf 35}\penalty0 (8), \penalty0 4067--4073
  (1987).

\bibitem{EUCK1911}
A.~Eucken, {\"u}ber die temperaturabh{\"a}ngigkeit der
  w{\"a}rmeleit{\"a}higkeit fester nichtmetalle, \emph{Annalen der Physik}.
  {\bf 339}\penalty0 (2), \penalty0 185--221  (1911).

\bibitem{KAYE1926}
G.~W.~C. Kaye, W.~F. Higgins, and J.~E. Petavel, The thermal conductivity of
  vitreous silica, with a note on crystalline quartz, \emph{Proc. R. Soc. Lond.
  Series A}. {\bf 113}\penalty0 (764), \penalty0 335--351  (1926).

\bibitem{BIRC1940}
F.~Birch and H.~Clark, The thermal conductivity of rocks and its dependence
  upon temperature and composition, \emph{Am. Jour. Sc.} {\bf 238}\penalty0
  (8), \penalty0 529--558  (1940).

\bibitem{KUNU1972}
M.~Kunugi, N.~Soga, H.~Sawa, and A.~Konishi, Thermal conductivity of
  cristobalite, \emph{J. Am. Ceram. Soc.} {\bf 55}\penalty0 (11), \penalty0
  580--580  (1972).

\bibitem{SAFA2006}
D.~J. Safarik, R.~B. Schwarz, and M.~F. Hundley, Similarities in the
  ${C}_{p}/{T}^{3}$ peaks in amorphous and crystalline metals, \emph{Phys. Rev.
  Lett.} {\bf 96}\penalty0 (19), \penalty0 195902  (2006).

\bibitem{CHUM2014}
A.~I. Chumakov, G.~Monaco, A.~Fontana, A.~Bosak, R.~P. Hermann, D.~Bessas,
  B.~Wehinger, W.~A. Crichton, M.~Krisch, R.~R\"uffer, G.~Baldi, G.~Carini~Jr.,
  G.~Carini, G.~D'Angelo, E.~Gilioli, G.~Tripodo, M.~Zanatta, B.~Winkler,
  V.~Milman, K.~Refson, M.~T. Dove, N.~Dubrovinskaia, L.~Dubrovinsky,
  R.~Keding, and Y.~Z. Yue, Role of disorder in the thermodynamics and atomic
  dynamics of glasses, \emph{Phys. Rev. Lett.} {\bf 112}\penalty0 (2),
  \penalty0 025502  (2014).

\bibitem{ROTH1983}
M.~Rothenfusser, W.~Dietsche, and H.~Kinder, Linear dispersion of transverse
  high-frequency phonons in vitreous silica, \emph{Phys. Rev. B}. {\bf
  27}\penalty0 (8), \penalty0 5196--5198  (1983).

\bibitem{PHIL1972}
W.~A. Phillips, Tunneling states in amorphous solids, \emph{J. Low Temp. Phys.}
  {\bf 7}\penalty0 (3), \penalty0 351--360  (1972).

\bibitem{ANDE1972}
P.~W. Anderson, H.~B. I, and V.~C. M, Anomalous low-temperature thermal
  properties of glasses and spin glasses, \emph{Philos. Mag.} {\bf 25}\penalty0
  (1), \penalty0 1--9  (1972).

\bibitem{GRAE1986}
J.~E. Graebner, B.~Golding, and L.~C. Allen, Phonon localization in glasses,
  \emph{Phys. Rev. B}. {\bf 34}\penalty0 (8), \penalty0 5696--5701  (1986).

\bibitem{SCHI2006}
W.~Schirmacher, Thermal conductivity of glassy materials and the boson peak,
  \emph{Europhys. Lett.} {\bf 73}\penalty0 (6), \penalty0 892--898  (2006).

\bibitem{BELT2016}
Y.~M. Beltukov, C.~Fusco, D.~A. Parshin, and A.~Tanguy, Boson peak and
  ioffe-regel criterion in amorphous siliconlike materials: The effect of bond
  directionality, \emph{Phys. Rev. E}. {\bf 93}\penalty0 (2), \penalty0 023006
  (2016).

\bibitem{MART1974}
A.~J. Martin and W.~Brenig, Model for {Brillouin}-scattering in amorphous
  solids, \emph{Phys. Status Solidi B}. {\bf 64}\penalty0 (1), \penalty0
  163--172  (1974).

\bibitem{MARI1971}
H.~J. Maris.
\newblock Interaction of sound waves with thermal phonons in dielectric
  crystals.
\newblock In eds. W.~Mason and R.~Thurston, \emph{Principles and Methods},
  vol.~8, \emph{Physical Acoustics}, chapter~6, pp. 279--345. Academic Press,
  New York, NY  (1971).

\bibitem{AKHI1939}
A.~N. Akhiezer, On the absorption of sound in solids, \emph{J. Phys. Acad. Sci.
  USSR}. {\bf 1}, \penalty0 277--287  (1939).

\bibitem{BOMM1960}
H.~E. B{\"o}mmel and K.~Dransfeld, Excitation and attenuation of hypersonic
  waves in quartz, \emph{Phys. Rev.} {\bf 117}\penalty0 (5), \penalty0
  1245--1252  (1960).

\bibitem{VACH2005}
R.~Vacher, E.~Courtens, and M.~Foret, Anharmonic versus relaxational sound
  damping in glasses. ii. vitreous silica, \emph{Phys. Rev. B}. {\bf
  72}\penalty0 (21), \penalty0 214205  (2005).

\bibitem{BLIN1970}
J.~S. Blinick and H.~J. Maris, Velocities of first and zero sound in quartz,
  \emph{Phys. Rev. B}. {\bf 2}\penalty0 (6), \penalty0 2139--2146  (1970).

\bibitem{KRAU1968}
J.~T. Krause and C.~R. Kurkjian, Vibrational anomalies in inorganic glass
  formers, \emph{J. Am. Ceram. Soc.} {\bf 51}\penalty0 (4), \penalty0 226--227
  (1968).

\bibitem{CLAY1978}
T.~N. Claytor and R.~J. Sladek, Ultrasonic velocities in amorphous
  {As$_2$S$_3$} and {As$_2$Se$_3$} between 1.5 {K} and 296 {K}, \emph{Phys.
  Rev. B}. {\bf 18}\penalty0 (10), \penalty0 5842--5850  (1978).

\bibitem{VACH1981}
R.~Vacher, J.~Pelous, F.~Plicque, and A.~Zarembowitch, Ultrasonic and
  {Brillouin}-scattering study of the elastic properties of vitreous silica
  between 10 {K} and 300 {K}, \emph{J. Non-Cryst. Solids}. {\bf 45}\penalty0
  (3), \penalty0 397--410  (1981).

\bibitem{JACK1972}
J.~J\"ackle, Ultrasonic attenuation in glasses at low-temperatures, \emph{Z.
  Phys.} {\bf 257}\penalty0 (3), \penalty0 212--223  (1972).

\bibitem{JACK1976}
J.~J\"ackle, L.~Pich\'e, W.~Arnold, and S.~Hunklinger, Elastic effects of
  structural relaxation in glasses at low temperatures, \emph{J. Non-Cryst.
  Solids}. {\bf 20}\penalty0 (3), \penalty0 365--391  (1976).

\bibitem{GILR1981}
K.~S. Gilroy and W.~A. Phillips, An asymmetric double-well potential model for
  structural relaxation processes in amorphous materials, \emph{Philos. Mag.
  B}. {\bf 43}\penalty0 (5), \penalty0 735--746  (1981).

\bibitem{TIEL1992}
D.~Tielb\"urger, R.~Merz, R.~Ehrenfels, and S.~Hunklinger, Thermally activated
  relaxation processes in vitreous silica - an investigation by
  {Brillouin}-scattering at high-pressures, \emph{Phys. Rev. B}. {\bf
  45}\penalty0 (6), \penalty0 2750--2760  (1992).

\bibitem{HUNK1976}
S.~Hunklinger and W.~Arnold.
\newblock Ultrasonic properties of glasses at low temperatures.
\newblock In eds. W.~Mason and R.~Thurston, \emph{Principles and Methods},
  vol.~12, \emph{Physical Acoustics}, chapter~3, pp. 155--215. Academic Press,
  New York, NY  (1976).

\bibitem{ANDE1955}
O.~L. Anderson and H.~E. B\"ommel, Ultrasonic absorption in fused silica at low
  temperatures and high frequencies, \emph{J. Am. Ceram. Soc.} {\bf
  38}\penalty0 (4), \penalty0 125--131  (1955).

\bibitem{KEIL1993}
R.~Keil, G.~Kasper, and S.~Hunklinger, Distribution of barrier heights in
  a-sio$_2$ and a-se, \emph{J. Non-Cryst. Solids}. {\bf 164--166}, \penalty0
  1183--1186  (1993).

\bibitem{DAMA2018}
T.~Damart and D.~Rodney, Atomistic study of two-level systems in amorphous
  silica, \emph{Phys. Rev. B}. {\bf 97}\penalty0 (1), \penalty0 014201  (2018).

\bibitem{GUIM2012}
G.~Guimbreti\`ere, B.~Ruffl\'e, and R.~Vacher, Acoustic damping and dispersion
  in vitreous germanium oxide, \emph{Phys. Rev. B}. {\bf 86}\penalty0 (9),
  \penalty0 094304  (2012).

\bibitem{CARI2012}
G.~Carini, G.~Carini, G.~Tripodo, G.~Di~Marco, and E.~Gilioli, Elastic and
  anelastic properties of densified vitreous {B$_2$O$_3$}: Relaxations and
  anharmonicity, \emph{Phys. Rev. B}. {\bf 85}\penalty0 (9), \penalty0 094201
  (2012).

\bibitem{CARI2014}
G.~Carini, G.~Carini, G.~D'Angelo, D.~Fioretto, and G.~Tripodo, Identification
  of relaxing structural defects in densified {$B_2O_3$} glasses, \emph{Phys.
  Rev. B}. {\bf 90}\penalty0 (14), \penalty0 140204  (2014).

\bibitem{CARI2009}
G.~Carini, G.~Tripodo, and L.~Borjesson, Hypersonic attenuation in cesium
  borate glasses: Relaxation and anharmonicity, \emph{Materials Sci. Eng. A}.
  {\bf 521--522}, \penalty0 247--250  (2009).

\bibitem{CARI2008}
G.~Carini, Jr., G.~Tripodo, and L.~Borjesson, Thermally activated relaxations
  and vibrational anharmonicity in alkali-borate glasses: {Brillouin}
  scattering study, \emph{Phys. Rev. B}. {\bf 78}\penalty0 (2), \penalty0
  024104  (2008).

\bibitem{LIND1981}
S.~M. Lindsay, M.~W. Anderson, and J.~R. Sandercock, Construction and alignment
  of a high performance multipass vernier tandem {Fabry-Perot} interferometer,
  \emph{Rev. Sci. Instrum.} {\bf 52}\penalty0 (10), \penalty0 1478--1486
  (1981).

\bibitem{VACH1976}
R.~Vacher and J.~Pelous, Behavior of thermal phonons in amorphous media from 4
  to 300 {K}, \emph{Phys. Rev. B}. {\bf 14}\penalty0 (2), \penalty0 823--828
  (1976).

\bibitem{RUFF2010}
B.~Ruffl\'e, S.~Ayrinhac, E.~Courtens, R.~Vacher, M.~Foret, A.~Wischnewski, and
  U.~Buchenau, Scaling the temperature-dependent boson peak of vitreous silica
  with the high-frequency bulk modulus derived from {Brillouin} scattering
  data, \emph{Phys. Rev. Lett.} {\bf 104}\penalty0 (6), \penalty0 067402
  (2010).

\bibitem{CARI2005}
G.~Carini, G.~Carini, G.~D'Angelo, G.~Tripodo, A.~Bartolotta, and G.~Salvato,
  Ultrasonic relaxations, anharmonicity, and fragility in lithium borate
  glasses, \emph{Phys. Rev. B}. {\bf 72}\penalty0 (1), \penalty0 014201
  (2005).

\bibitem{WACH1961}
J.~Wachtman, W.~Tefft, D.~Lam, and C.~Apstein, Exponential temperature
  dependence of {Youngs} modulus for several oxides, \emph{Phys. Rev.} {\bf
  122}\penalty0 (6), \penalty0 1754--1759  (1961).

\bibitem{AYRI2011a}
S.~Ayrinhac, B.~Ruffl\'e, M.~Foret, H.~Tran, S.~Cl\'ement, R.~Vialla,
  R.~Vacher, J.~C. Chervin, P.~Munsch, and A.~Polian, Dynamical origin of
  anomalous temperature hardening of elastic modulus in vitreous silica,
  \emph{Phys. Rev. B}. {\bf 84}, \penalty0 024201  (2011).

\bibitem{CARI2006}
G.~Carini, G.~Carini, G.~Tripodo, A.~Bartolotta, and G.~D. Marco, Effect of
  cation sizes on tunnelling states, relaxations and anharmonicity of alkali
  borate glasses, \emph{J. Phys.-Condens. Matter}. {\bf 18}\penalty0 (12),
  \penalty0 3251--3262  (2006).

\bibitem{FEDE1982}
G.~Federle and S.~Hunklinger, Ultrasonic studies of some polymers at low
  temperatures, \emph{J. Phys. Colloques}. {\bf 43}\penalty0 (C9), \penalty0
  505--508  (1982).

\bibitem{FEDE1982a}
S.~Federle, G.and~Hunklinger.
\newblock Ultrasonic absorption in polymethylmethacrylate at low temperatures.
\newblock In eds. G.~Hartwig and D.~Evans, \emph{Nonmetallic Materials and
  Composites at Low Temperatures}, pp. 49--57. Springer US, Boston, MA  (1982).

\bibitem{SCHM1982}
M.~Schmidt, R.~Vacher, J.~Pelous, and S.~Hunklinger, Brillouin scattering from
  amorphous polymers at low temperatures, \emph{J. Phys. Colloques}. {\bf
  43}\penalty0 (C9), \penalty0 501--504  (1982).

\bibitem{RAT2005}
E.~Rat, M.~Foret, G.~Massiera, R.~Vialla, M.~Arai, R.~Vacher, and E.~Courtens,
  Anharmonic versus relaxational sound damping in glasses. i. brillouin
  scattering from densified silica, \emph{Phys. Rev. B}. {\bf 72}, \penalty0
  214204  (2005).

\bibitem{VACH1985}
{Vacher, R.} and {Pelous, J.}, Rayleigh brillouin scattering and molecular
  disorder in glassy crystals and molecular glasses, \emph{J. Chim. Phys.} {\bf
  82}, \penalty0 311--316  (1985).

\bibitem{RUFF2007a}
B.~Ruffl\'e, unpublished  (2007).

\bibitem{CALI2001}
G.~Caliskan, A.~Kisliuk, V.~N. Novikov, and A.~P. Sokolov, Relaxation spectra
  in poly(methylmethacrylate): Comparison of acoustic attenuation and light
  scattering data, \emph{J. Chem. Phys.} {\bf 114}\penalty0 (22), \penalty0
  10189--10195  (2001).

\bibitem{WEIS1996}
G.~Weiss, A.~Daum, M.~Sohn, and J.~Arndt, Vibrating reed experiments on
  compacted vitreous silica, \emph{Physica B}. {\bf 219-220}, \penalty0
  290--292  (1996).

\bibitem{MASC2004}
C.~Masciovecchio, A.~Gessini, S.~Di~Fonzo, L.~Comez, S.~C. Santucci, and
  D.~Fioretto, Inelastic ultraviolet scattering from high frequency acoustic
  modes in glasses, \emph{Phys. Rev. Lett.} {\bf 92}\penalty0 (24), \penalty0
  247401  (2004).

\bibitem{BENA2005}
P.~Benassi, S.~Caponi, R.~Eramo, A.~Fontana, A.~Giugni, M.~Nardone, M.~Sampoli,
  and G.~Viliani, Sound attenuation in a unexplored frequency region:
  {Brillouin} ultraviolet light scattering measurements in {$v$-SiO$_2$},
  \emph{Phys. Rev. B}. {\bf 71}\penalty0 (17), \penalty0 172201  (2005).

\bibitem{VACH2006}
R.~Vacher, S.~Ayrinhac, M.~Foret, B.~Ruffl\'e, and E.~Courtens, Finite size
  effects in brillouin scattering from silica glass, \emph{Phys. Rev. B}. {\bf
  74}\penalty0 (1), \penalty0 012203  (2006).

\bibitem{GRUB1994}
W.~T. Grubbs and R.~A. MacPhail, Dynamics in supercooled glycerol by high
  resolution stimulated brillouin gain spectroscopy, \emph{J. Chem. Phys.} {\bf
  100}\penalty0 (4), \penalty0 2561--2570  (1994).

\bibitem{MASC2006}
C.~Masciovecchio, G.~Baldi, S.~Caponi, L.~Comez, S.~Di~Fonzo, D.~Fioretto,
  A.~Fontana, A.~Gessini, S.~C. Santucci, F.~Sette, G.~Viliani, P.~Vilmercati,
  and G.~Ruocco, Evidence for a crossover in the frequency dependence of the
  acoustic attenuation in vitreous silica, \emph{Phys. Rev. Lett.} {\bf
  97}\penalty0 (3), \penalty0 035501  (2006).

\bibitem{RUFF2011}
B.~Ruffl\'e, E.~Courtens, and M.~Foret, Inelastic ultraviolet {Brillouin}
  scattering from superpolished vitreous silica, \emph{Phys. Rev. B}. {\bf
  84}\penalty0 (13), \penalty0 132201  (2011).

\bibitem{SAND1972}
J.~Sandercock, Brillouin-scattering measurements on silicon and germanium,
  \emph{Phys. Rev. Lett.} {\bf 28}\penalty0 (4), \penalty0 237--240  (1972).

\bibitem{INAM2010}
F.~Inam, J.~P. Lewis, and D.~A. Drabold, Hidden structure in amorphous solids,
  \emph{Phys. Status Solidi A-Appl. Mat.} {\bf 207}\penalty0 (3), \penalty0
  599--604  (2010).

\bibitem{THOM1984}
C.~Thomsen, J.~Strait, Z.~Vardeny, H.~Maris, J.~Tauc, and J.~Hauser, Coherent
  phonon generation and detection by picosecond light-pulses, \emph{Phys. Rev.
  Lett.} {\bf 53}\penalty0 (10), \penalty0 989--992  (1984).

\bibitem{ZHU1991}
T.~Zhu, H.~Maris, and J.~Tauc, Attenuation of longitudinal-acoustic phonons in
  amorphous {SiO$_2$} at frequencies up to {440 GHz}, \emph{Phys. Rev. B}. {\bf
  44}\penalty0 (9), \penalty0 4281--4289  (1991).

\bibitem{THOM1986}
C.~Thomsen, H.~Grahn, M.~HJ, and J.~Tauc, Surface generation and detection of
  phonons by picosecond light-pulses, \emph{Phys. Rev. B}. {\bf 34}\penalty0
  (6), \penalty0 4129--4138  (1986).

\bibitem{BENA1996}
P.~Benassi, M.~Krisch, C.~Masciovecchio, V.~Mazzacurati, G.~Monaco, G.~Ruocco,
  F.~Sette, and R.~Verbeni, Evidence of high frequency propagating modes in
  vitreous silica, \emph{Phys. Rev. Lett.} {\bf 77}\penalty0 (18), \penalty0
  3835--3838  (1996).

\bibitem{DEVO2008}
A.~Devos, M.~Foret, S.~Ayrinhac, P.~Emery, and B.~Ruffl\'e, Hypersound damping
  in vitreous silica measured by picosecond acoustics, \emph{Phys. Rev. B}.
  {\bf 77}\penalty0 (10), \penalty0 100201  (2008).

\bibitem{AYRI2011}
S.~Ayrinhac, M.~Foret, A.~Devos, B.~Ruffl\'e, E.~Courtens, and R.~Vacher,
  Subterahertz hypersound attenuation in silica glass studied via picosecond
  acoustics, \emph{Phys. Rev. B}. {\bf 83}\penalty0 (1), \penalty0 014204
  (2011).

\bibitem{KLIE2011}
C.~Klieber, E.~Peronne, K.~Katayama, J.~Choi, M.~Yamaguchi, T.~Pezeril, and
  K.~A. Nelson, Narrow-band acoustic attenuation measurements in vitreous
  silica at frequencies between 20 and 400 ghz, \emph{Appl. Phys. Lett.} {\bf
  98}\penalty0 (21), \penalty0 211908  (2011).

\bibitem{WEN2011}
Y.-C. Wen, S.-H. Guol, H.-P. Chen, J.-K. Sheu, and C.-K. Sun, Femtosecond
  ultrasonic spectroscopy using a piezoelectric nanolayer: Hypersound
  attenuation in vitreous silica films, \emph{Appl. Phys. Lett.} {\bf
  99}\penalty0 (5), \penalty0 051913  (2011).

\bibitem{LIN1991}
H.~Lin, R.~Stoner, H.~Maris, and J.~Tauc, Phonon attenuation and
  velcoity-measurements in transparent materials by picosecond acosutic
  interferometry, \emph{J. Appl. Phys.} {\bf 69}\penalty0 (7), \penalty0
  3816--3822  (1991).

\bibitem{MORA1996}
C.~Morath and H.~Maris, Phonon attenuation in amorphous solids studied by
  picosecond ultrasonics, \emph{Phys. Rev. B}. {\bf 54}\penalty0 (1), \penalty0
  203--213  (1996).

\bibitem{MERM1998}
A.~Mermet, A.~Cunsolo, E.~Duval, M.~Krisch, C.~Masciovecchio, S.~Perghem,
  G.~Ruocco, F.~Sette, R.~Verbeni, and G.~Viliani, Pressure-induced in-glass
  structural transformation in the amorphous polymer poly(methylmethacrylate),
  \emph{Phys. Rev. Lett.} {\bf 80}, \penalty0 4205--4208  (1998).

\bibitem{HUYN2017}
A.~Huynh, E.~P\'eronne, C.~Gingreau, X.~Lafosse, A.~Lema\^{\i}tre, B.~Perrin,
  R.~Vacher, B.~Ruffl\'e, and M.~Foret, Temperature dependence of hypersound
  attenuation in silica films via picosecond acoustics, \emph{Phys. Rev. B}.
  {\bf 96}, \penalty0 174206  (2017).

\bibitem{MASC1996}
C.~Masciovecchio, G.~Ruocco, F.~Sette, M.~Krisch, R.~Verbeni, U.~Bergmann, and
  M.~Soltwisch, Observation of large momentum phononlike modes in glasses,
  \emph{Phys. Rev. Lett.} {\bf 76}\penalty0 (18), \penalty0 3356--3359  (1996).

\bibitem{FORE1998}
M.~Foret, B.~Hehlen, G.~Taillades, E.~Courtens, R.~Vacher, H.~Casalta, and
  B.~Dorner, Neutron {Brillouin} and umklapp scattering from glassy selenium,
  \emph{Phys. Rev. Lett.} {\bf 81}\penalty0 (10), \penalty0 2100--2103  (1998).

\bibitem{FORE1996}
M.~Foret, E.~Courtens, R.~Vacher, and J.~Suck, Scattering investigation of
  acoustic localization in fused silica, \emph{Phys. Rev. Lett.} {\bf
  77}\penalty0 (18), \penalty0 3831--3834  (1996).

\bibitem{POLA1988}
G.~Polatsek and O.~Entin-Wohlman, Effective-medium approximation for a
  percolation network: The structure factor and the ioffe-regel criterion,
  \emph{Phys. Rev. B}. {\bf 37}\penalty0 (13), \penalty0 7726--7730  (1988).

\bibitem{BUCH2014}
U.~Buchenau, Evaluation of x-ray brillouin scattering data, \emph{Phys. Rev.
  E}. {\bf 90}\penalty0 (6), \penalty0 062319  (2014).

\bibitem{COUR1988}
E.~Courtens, R.~Vacher, J.~Pelous, and T.~Woignier, Observation of fractons in
  silica aerogels, \emph{Europhysics Lett.} {\bf 6}\penalty0 (3), \penalty0
  245--250  (1988).

\bibitem{MASC1997}
C.~Masciovecchio, G.~Ruocco, F.~Sette, P.~Benassi, A.~Cunsolo, M.~Krisch,
  V.~Mazzacurati, A.~Mermet, G.~Monaco, and R.~Verbeni, High-frequency
  propagating modes in vitreous silica at 295 {K}, \emph{Phys. Rev. B}. {\bf
  55}\penalty0 (13), \penalty0 8049--8051  (1997).

\bibitem{MASC1998}
C.~Masciovecchio, G.~Monaco, G.~Ruocco, F.~Sette, A.~Cunsolo, M.~Krisch,
  A.~Mermet, M.~Soltwisch, and R.~Verbeni, High frequency dynamics of glass
  forming liquids at the glass transition, \emph{Phys. Rev. Lett.} {\bf
  80}\penalty0 (3), \penalty0 544--547  (1998).

\bibitem{SETT1998}
F.~Sette, M.~Krisch, C.~Masciovecchio, G.~Ruocco, and G.~Monaco, Dynamics of
  glasses and glass-forming liquids studied by inelastic x-ray scattering,
  \emph{Science}. {\bf 280}\penalty0 (5369), \penalty0 1550--1555  (1998).

\bibitem{MONA1998}
G.~Monaco, C.~Masciovecchio, G.~Ruocco, and F.~Sette, Determination of the
  infinite frequency sound velocity in the glass former o-terphenyl,
  \emph{Phys. Rev. Lett.} {\bf 80}\penalty0 (10), \penalty0 2161--2164  (1998).

\bibitem{FIOR1999}
D.~Fioretto, U.~Buchenau, L.~Comez, A.~Sokolov, C.~Masciovecchio, A.~Mermet,
  G.~Ruocco, F.~Sette, L.~Willner, B.~Frick, D.~Richter, and L.~Verdini,
  High-frequency dynamics of glass-forming polybutadiene, \emph{Phys. Rev. E}.
  {\bf 59}\penalty0 (4), \penalty0 4470--4475  (1999).

\bibitem{PILL2000}
O.~Pilla, A.~Cunsolo, A.~Fontana, C.~Masciovecchio, G.~Monaco, M.~Montagna,
  G.~Ruocco, T.~Scopigno, and F.~Sette, Nature of the short wavelength
  excitations in vitreous silica: An x-ray {Brillouin} scattering study,
  \emph{Phys. Rev. Lett.} {\bf 85}\penalty0 (10), \penalty0 2136--2139  (2000).

\bibitem{MATI2001}
A.~Matic, L.~Borjesson, G.~Ruocco, C.~Masciovecchio, A.~Mermet, F.~Sette, and
  R.~Verbeni, Contrasting behaviour of acoustic modes in network and
  non-network glasses, \emph{Europhys. Lett.} {\bf 54}\penalty0 (1), \penalty0
  77--83  (2001).

\bibitem{MATT2003}
J.~Mattsson, A.~Matic, G.~Monaco, D.~Engberg, and L.~Borjesson, High-frequency
  collective excitations in a molecular glass-former, \emph{J. Phys.-Condens.
  Matter}. {\bf 15}\penalty0 (11, SI), \penalty0 S1259--S1267  (2003).

\bibitem{SCOP2004}
T.~Scopigno, R.~Di~Leonardo, G.~Ruocco, A.~Baron, S.~Tsutsui, F.~Bossard, and
  S.~Yannopoulos, High frequency dynamics in a monatomic glass, \emph{Phys.
  Rev. Lett.} {\bf 92}\penalty0 (2), \penalty0 025503  (2004).

\bibitem{MATI2004}
A.~Matic, C.~Masciovecchio, D.~Engberg, G.~Monaco, L.~Borjesson, S.~Santucci,
  and R.~Verbeni, Crystal-like nature of acoustic excitations in glassy
  ethanol, \emph{Phys. Rev. Lett.} {\bf 93}\penalty0 (14), \penalty0 145502
  (2004).

\bibitem{BOVE2005}
L.~E. Bove, E.~Fabiani, A.~Fontana, F.~Paoletti, C.~Petrillo, O.~Pilla, and
  I.~C.~V. Bento, Brillouin neutron scattering of {$v$-GeO$_2$},
  \emph{Europhys. Lett.} {\bf 71}\penalty0 (4), \penalty0 563--569  (2005).

\bibitem{RAT1999}
E.~Rat, M.~Foret, E.~Courtens, R.~Vacher, and M.~Arai, Observation of the
  crossover to strong scattering of acoustic phonons in densified silica,
  \emph{Phys. Rev. Lett.} {\bf 83}\penalty0 (7), \penalty0 1355--1358  (1999).

\bibitem{MATI2001a}
A.~Matic, D.~Engberg, C.~Masciovecchio, and L.~Borjesson, Sound wave scattering
  in network glasses, \emph{Phys. Rev. Lett.} {\bf 86}\penalty0 (17), \penalty0
  3803--3806  (2001).

\bibitem{FORE2002}
M.~Foret, R.~Vacher, E.~Courtens, and G.~Monaco, Merging of the acoustic branch
  with the boson peak in densified silica glass, \emph{Phys. Rev. B}. {\bf
  66}\penalty0 (2), \penalty0 024204  (2002).

\bibitem{TARA1999}
S.~N. Taraskin and S.~R. Elliott, Low-frequency vibrational excitations in
  vitreous silica: the ioffe-regel limit, \emph{J. Phys.-Condens. Matter}. {\bf
  11}\penalty0 (10A), \penalty0 A219--A227  (1999).

\bibitem{TARA2000a}
S.~N. Taraskin and S.~R. Elliott, Ioffe-regel crossover for plane-wave
  vibrational excitations in vitreous silica, \emph{Phys. Rev. B}. {\bf
  61}\penalty0 (18), \penalty0 12031--12037  (2000).

\bibitem{RUFF2003}
B.~Ruffl\'e, M.~Foret, E.~Courtens, R.~Vacher, and G.~Monaco, Observation of
  the onset of strong scattering on high frequency acoustic phonons in
  densified silica glass, \emph{Phys. Rev. Lett.} {\bf 90}\penalty0 (9),
  \penalty0 095502  (2003).

\bibitem{RUFF2006}
B.~Ruffl\'e, G.~Guimbreti\`ere, E.~Courtens, R.~Vacher, and G.~Monaco,
  Glass-specific behavior in the damping of acousticlike vibrations,
  \emph{Phys. Rev. Lett.} {\bf 96}\penalty0 (4), \penalty0 045502  (2006).

\bibitem{VACH2008}
R.~Vacher, B.~Ruffl\'e, B.~Hehlen, G.~Guimbreti\`ere, G.~Simon, and
  E.~Courtens, The vibrational excitations of glasses in the boson-peak region:
  application to borates, \emph{Phys. Chem. Glasses-Eur. J. Glass Sci. Technol.
  Part B}. {\bf 49}\penalty0 (1), \penalty0 19--25  (2008).

\bibitem{COUR2006}
E.~Courtens, B.~Rufflé, and R.~Vacher, Low-q brillouin scattering from
  glasses, \emph{J. Neut. Res.} {\bf 14}\penalty0 (4), \penalty0 361--366
  (2006).

\bibitem{VACH1999}
R.~Vacher, E.~Courtens, and M.~Foret, Are high frequency acoustic modes in
  glasses dominated by strong scattering or by lifetime broadening?,
  \emph{Philos. Mag. B}. {\bf 79}\penalty0 (11--12), \penalty0 1763--1774
  (1999).

\bibitem{RUFF2007}
B.~Ruffl\'e, G.~Guimbreti\`ere, E.~Courtens, R.~Vacher, and G.~Monaco, Comment
  on ``glass-specific behavior in the damping of acousticlike vibrations{''} -
  reply, \emph{Phys. Rev. Lett.} {\bf 98}\penalty0 (7), \penalty0 079602
  (2007).

\bibitem{SCHI2007}
W.~Schirmacher, G.~Ruocco, and T.~Scopigno, Acoustic attenuation in glasses and
  its relation with the boson peak, \emph{Phys. Rev. Lett.} {\bf 98}\penalty0
  (2), \penalty0 025501  (2007).

\bibitem{SCHO2004}
H.~R. Schober, Vibrations and relaxations in a soft sphere glass: boson peak
  and structure factors, \emph{J. Phys.: Condens. Matter}. {\bf 16}\penalty0
  (27), \penalty0 S2659--S2670  (2004).

\bibitem{SCOP2006}
T.~Scopigno, J.~Suck, R.~Angelini, F.~Albergamo, and G.~Ruocco, High-frequency
  dynamics in metallic glasses, \emph{Phys. Rev. Lett.} {\bf 96}\penalty0 (13),
  \penalty0 135501  (2006).

\bibitem{COUR2007}
E.~Courtens, M.~Foret, B.~Ruffl\'e, and R.~Vacher, Comment on ''high-frequency
  dynamics in metallic glasses'', \emph{Phys. Rev. Lett.} {\bf 98}\penalty0
  (7), \penalty0 079603  (2007).

\bibitem{BRUN2011}
P.~Bruna, G.~Baldi, E.~Pineda, J.~Serrano, J.~B. Suck, D.~Crespo, and
  G.~Monaco, Communication: Are metallic glasses different from other glasses?
  a closer look at their high frequency dynamics, \emph{J. Chem. Phys.} {\bf
  135}\penalty0 (10), \penalty0 101101  (2011).

\bibitem{KARP1983}
V.~Karpov, M.~Klinger, and F.~Ignatiev, Theory of low-temperature anomalies in
  thermal-properties of amorphic structures, \emph{Zh. Eksp. Teor. Fiz.} {\bf
  84}\penalty0 (2), \penalty0 760--775  (1983).

\bibitem{KARP1985}
V.~Karpov and D.~Parshin, Heat-conductivity of glasses at temperatures below
  the {Debye} temperature, \emph{Zh. Eksp. Teor. Fiz.} {\bf 88}\penalty0 (6),
  \penalty0 2212--2227  (1985).

\bibitem{PARS2007}
D.~A. Parshin, H.~R. Schober, and V.~L. Gurevich, Vibrational instability,
  two-level systems, and the boson peak in glasses, \emph{Phys. Rev. B}. {\bf
  76}\penalty0 (6), \penalty0 064206  (2007).

\bibitem{BUCH1992}
U.~Buchenau, Y.~Galperin, V.~Gurevich, D.~Parshin, M.~Ramos, and H.~Schober,
  Interaction of soft modes and sound-waves in glasses, \emph{Phys. Rev. B}.
  {\bf 46}\penalty0 (5), \penalty0 2798--2808  (1992).

\bibitem{PARS2001}
D.~Parshin and C.~Laermans, Interaction of quasilocal harmonic modes and boson
  peak in glasses, \emph{Phys. Rev. B}. {\bf 63}\penalty0 (13), \penalty0
  132203  (2001).

\bibitem{SCHI1993}
W.~Schirmacher and M.~Wagener, Vibrational anomalies and phonon localization in
  glasses, \emph{Solid State Commun.} {\bf 86}\penalty0 (9), \penalty0 597--603
   (1993).

\bibitem{SCHI1998}
W.~Schirmacher, G.~Diezemann, and C.~Ganter, Harmonic vibrational excitations
  in disordered solids and the boson peak, \emph{Phys. Rev. Lett.} {\bf 81},
  \penalty0 136--139  (1998).

\bibitem{SCHO2011}
H.~R. Schober, Quasi-localized vibrations and phonon damping in glasses,
  \emph{J. Non-Cryst. Solids}. {\bf 357}\penalty0 (2), \penalty0 501--505
  (2011).

\bibitem{RUOC2007}
G.~Ruocco, A.~Matic, T.~Scopigno, and S.~N. Yannopoulos, Comment on
  ``glass-specific behavior in the damping of acousticlike vibrations'',
  \emph{Phys. Rev. Lett.} {\bf 98}, \penalty0 079601  (2007).

\bibitem{WITT2002}
J.~P. Wittmer, A.~Tanguy, J.-L. Barrat, and L.~Lewis, Vibrations of amorphous,
  nanometric structures: When does continuum theory apply?, \emph{Europhysics
  Lett.} {\bf 57}\penalty0 (3), \penalty0 423--429  (2002).

\bibitem{TANG2002}
A.~Tanguy, J.~P. Wittmer, F.~Leonforte, and J.-L. Barrat, Continuum limit of
  amorphous elastic bodies: A finite-size study of low-frequency harmonic
  vibrations, \emph{Phys. Rev. B}. {\bf 66}\penalty0 (17), \penalty0 174205
  (2002).

\bibitem{YOSH2004}
K.~Yoshimoto, T.~S. Jain, K.~V. Workum, P.~F. Nealey, and J.~J. de~Pablo,
  Mechanical heterogeneities in model polymer glasses at small length scales,
  \emph{Phys. Rev. Lett.} {\bf 93}\penalty0 (17), \penalty0 175501  (2004).

\bibitem{LEON2005}
F.~Leonforte, R.~Boissi\`ere, A.~Tanguy, J.~P. Wittmer, and J.-L. Barrat,
  Continuum limit of amorphous elastic bodies. iii. three-dimensional systems,
  \emph{Phys. Rev. B}. {\bf 72}\penalty0 (22), \penalty0 224206  (2005).

\bibitem{MIZU2014}
H.~Mizuno, S.~Mossa, and J.-L. Barrat, Acoustic excitations and elastic
  heterogeneities in disordered solids, \emph{Proc. Natl. Acad. Sci. USA}. {\bf
  11}\penalty0 (33), \penalty0 11949--11954  (2014).

\bibitem{LAIR1991}
B.~B. Laird and H.~R. Schober, Localized low-frequency vibrational modes in a
  simple model glass, \emph{Phys. Rev. Lett.} {\bf 66}\penalty0 (5), \penalty0
  636--639  (1991).

\bibitem{HAFN1994}
J.~Hafner and M.~Krajci, Propagating and localized vibrational modes in {Ni-Zr}
  glasses, \emph{J. Phys.-Condens. Matter}. {\bf 6}\penalty0 (25), \penalty0
  4631--4654  (1994).

\bibitem{TARA1999a}
S.~N. Taraskin and S.~R. Elliott, Anharmonicity and localization of atomic
  vibrations in vitreous silica, \emph{Phys. Rev. B}. {\bf 59}, \penalty0
  8572--8585  (1999).

\bibitem{LERN2016}
E.~Lerner, G.~D\"uring, and E.~Bouchbinder, Statistics and properties of
  low-frequency vibrational modes in structural glasses, \emph{Phys. Rev.
  Lett.} {\bf 117}\penalty0 (3), \penalty0 035501  (2016).

\bibitem{MIZU2017}
H.~Mizuno, H.~Shiba, and A.~Ikeda, Continuum limit of the vibrational
  properties of amorphous solids, \emph{Proc. Natl. Acad. Sci. USA}. {\bf
  114}\penalty0 (46), \penalty0 E9767--E9774  (2017).

\bibitem{MIZU2018}
H.~Mizuno and A.~Ikeda, Phonon transport and vibrational excitations in
  amorphous solids, \emph{Phys. Rev. E}. {\bf 98}, \penalty0 062612  (2018).

\bibitem{SHIM2018}
M.~Shimada, H.~Mizuno, and A.~Ikeda, Anomalous vibrational properties in the
  continuum limit of glasses, \emph{Phys. Rev. E}. {\bf 97}, \penalty0 022609
  (2018).

\bibitem{SHIM2018a}
M.~Shimada, H.~Mizuno, M.~Wyart, and A.~Ikeda, Spatial structure of
  quasilocalized vibrations in nearly jammed amorphous solids, \emph{Phys. Rev.
  E}. {\bf 98}, \penalty0 060901  (2018).

\bibitem{KAPT2018}
G.~Kapteijns, E.~Bouchbinder, and E.~Lerner, Universal nonphononic density of
  states in 2d, 3d, and 4d glasses, \emph{Phys. Rev. Lett.} {\bf 121}\penalty0
  (5), \penalty0 055501  (2018).

\bibitem{BONF2020}
S.~Bonfanti, R.~Guerra, C.~Mondal, I.~Procaccia, and S.~Zapperi, Universal
  low-frequency vibrational modes in silica glasses, \emph{Phys. Rev. Lett.}
  {\bf 125}\penalty0 (8), \penalty0 085501  (2020).

\bibitem{RICH2020}
D.~Richard, K.~Gonz\'alez-L\'opez, G.~Kapteijns, R.~Pater, T.~Vaknin,
  E.~Bouchbinder, and E.~Lerner, Universality of the nonphononic vibrational
  spectrum across different classes of computer glasses, \emph{Phys. Rev.
  Lett.} {\bf 125}\penalty0 (8), \penalty0 085502  (2020).

\bibitem{RUFF2008}
B.~Ruffl\'e, D.~A. Parshin, E.~Courtens, and R.~Vacher, Boson peak and its
  relation to acoustic attenuation in glasses, \emph{Phys. Rev. Lett.} {\bf
  100}\penalty0 (1), \penalty0 015501  (JAN 11, 2008).

\bibitem{SCHI2008}
W.~Schirmacher, B.~Schmid, C.~Tomaras, G.~Viliani, G.~Baldi, G.~Ruocco, and
  T.~Scopigno, Vibrational excitations in systems with correlated disorder,
  \emph{phys. stat. sol. (c)}. {\bf 5}\penalty0 (3), \penalty0 862--866
  (2008).

\bibitem{KOHL2013}
S.~K\"ohler, G.~Ruocco, and W.~Schirmacher, Coherent potential approximation
  for diffusion and wave propagation in topologically disordered systems,
  \emph{Phys. Rev. B}. {\bf 88}\penalty0 (6), \penalty0 064203  (2013).

\bibitem{MONA2009}
G.~Monaco and V.~M. Giordano, Breakdown of the debye approximation for the
  acoustic modes with nanometric wavelengths in glasses, \emph{Proc. Natl.
  Acad. Sci. USA}. {\bf 106}\penalty0 (10), \penalty0 3659--3663  (2009).

\bibitem{SCHI2013}
W.~Schirmacher, The boson peak, \emph{phys. stat. sol. (b)}. {\bf 250}\penalty0
  (5), \penalty0 937--943  (2013).

\bibitem{BALD2010}
G.~Baldi, V.~M. Giordano, G.~Monaco, and B.~Ruta, Sound attenuation at
  terahertz frequencies and the boson peak of vitreous silica, \emph{Phys. Rev.
  Lett.} {\bf 104}\penalty0 (19), \penalty0 195501  (2010).

\bibitem{RUTA2010}
B.~Ruta, G.~Baldi, V.~M. Giordano, L.~Orsingher, S.~Rols, F.~Scarponi, and
  G.~Monaco, Communication: {H}igh-frequency acoustic excitations and boson
  peak in glasses: {A} study of their temperature dependence, \emph{J. Chem.
  Phys.} {\bf 133}\penalty0 (4), \penalty0 041101  (2010).

\bibitem{RUTA2012}
B.~Ruta, G.~Baldi, F.~Scarponi, D.~Fioretto, V.~M. Giordano, and G.~Monaco,
  Acoustic excitations in glassy sorbitol and their relation with the fragility
  and the boson peak, \emph{J. Chem. Phys.} {\bf 137}\penalty0 (21), \penalty0
  214502  (2012).

\bibitem{BALD2014}
G.~Baldi, V.~M. Giordano, B.~Ruta, R.~Dal~Maschio, A.~Fontana, and G.~Monaco,
  Anharmonic damping of terahertz acoustic waves in a network glass and its
  effect on the density of vibrational states, \emph{Phys. Rev. Lett.} {\bf
  112}  (2014).

\bibitem{BALD2011}
G.~Baldi, V.~M. Giordano, and G.~Monaco, Elastic anomalies at terahertz
  frequencies and excess density of vibrational states in silica glass,
  \emph{Phys. Rev. B}. {\bf 83}\penalty0 (17), \penalty0 174203  (2011).

\bibitem{MONA2006}
A.~Monaco, A.~I. Chumakov, Y.~Z. Yue, G.~Monaco, L.~Comez, D.~Fioretto, W.~A.
  Crichton, and R.~Ruffer, Density of vibrational states of a hyperquenched
  glass, \emph{Phys. Rev. Lett.} {\bf 96}\penalty0 (20), \penalty0 205502
  (2006).

\bibitem{MONA2006a}
A.~Monaco, A.~I. Chumakov, G.~Monaco, W.~A. Crichton, A.~Meyer, L.~Comez,
  D.~Fioretto, J.~Korecki, and R.~Ruffer, Effect of densification on the
  density of vibrational states of glasses, \emph{Phys. Rev. Lett.} {\bf
  97}\penalty0 (13), \penalty0 135501  (2006).

\bibitem{BALD2009}
G.~Baldi, A.~Fontana, G.~Monaco, L.~Orsingher, S.~Rols, F.~Rossi, and B.~Ruta,
  Connection between boson peak and elastic properties in silicate glasses,
  \emph{Phys. Rev. Lett.} {\bf 102}\penalty0 (19), \penalty0 195502  (2009).

\bibitem{NISS2007}
K.~Niss, B.~Begen, B.~Frick, J.~Ollivier, A.~Beraud, A.~Sokolov, V.~N. Novikov,
  and C.~Alba-Simionesco, Influence of pressure on the boson peak: Stronger
  than elastic medium transformation, \emph{Phys. Rev. Lett.} {\bf 99}\penalty0
  (5), \penalty0 055502  (AUG 3, 2007).

\bibitem{HONG2008}
L.~Hong, B.~Begen, A.~Kisliuk, C.~Alba-Simionesco, V.~N. Novikov, and A.~P.
  Sokolov, Pressure and density dependence of the boson peak in polymers,
  \emph{Phys. Rev. B}. {\bf 78}\penalty0 (13), \penalty0 134201  (OCT, 2008).

\bibitem{ORSI2012}
L.~Orsingher, A.~Fontana, E.~Gilioli, G.~Carini, G.~Carini, G.~Tripodo,
  T.~Unruh, and U.~Buchenau, Vibrational dynamics of permanently densified
  {GeO$_2$} glasses: Densification-induced changes in the boson peak, \emph{J.
  Chem. Phys.} {\bf 132}\penalty0 (12), \penalty0 124508  (2010).

\bibitem{DEGI2014}
E.~DeGiuli, A.~Laversanne-Finot, G.~Düring, E.~Lerner, and M.~Wyart, Effects
  of coordination and pressure on sound attenuation{,} boson peak and
  elasticity in amorphous solids, \emph{Soft Matter}. {\bf 10}, \penalty0
  5628--5644  (2014).

\bibitem{FABI2008}
E.~Fabiani, A.~Fontana, and U.~Buchenau, Neutron scattering study of the
  vibrations in vitreous silica and germania, \emph{J. Chem. Phys.} {\bf
  128}\penalty0 (24), \penalty0 244507  (2008).

\bibitem{ENGB1999}
D.~Engberg, A.~Wischnewski, U.~Buchenau, L.~B\"orjesson, A.~J. Dianoux, A.~P.
  Sokolov, and L.~M. Torell, Origin of the boson peak in a network glass
  {B$_2$O$_3$}, \emph{Phys. Rev. B}. {\bf 59}\penalty0 (6), \penalty0
  4053--4057  (1999).

\bibitem{BUCH1996}
U.~Buchenau, A.~Wischnewski, D.~Richter, and B.~Frick, Is the fast process at
  the glass transition mainly due to long wavelength excitations?, \emph{Phys.
  Rev. Lett.} {\bf 77}\penalty0 (19), \penalty0 4035--4038  (1996).

\bibitem{TEZU1991}
Y.~Tezuka, S.~Shin, and M.~Ishigame, Observation of the silent soft phonon in
  {$\beta$}-quartz by means of hyper-raman scattering, \emph{Phys. Rev. Lett.}
  {\bf 66}\penalty0 (18), \penalty0 2356--2359  (1991).

\bibitem{SIMO2006}
G.~Simon, B.~Hehlen, E.~Courtens, E.~Longueteau, and R.~Vacher, Hyper-{Raman}
  scattering from vitreous boron oxide: Coherent enhancement of the boson peak,
  \emph{Phys. Rev. Lett.} {\bf 96}\penalty0 (10), \penalty0 105502  (2006).

\bibitem{HEHL2020}
B.~Hehlen and B.~Rufflé.
\newblock Atomic vibrations in glasses.
\newblock In ed. P.~Richet, \emph{Encyclopedia of Glass Science, Technology,
  History, and Culture}. in press, John Wiley \& Sons, Inc.  (2020).

\bibitem{SHUK1970}
R.~Shuker and R.~Gammon, Raman-scattering selection-rule breaking and density
  of states in amorphous materials, \emph{Phys. Rev. Lett.} {\bf 25}\penalty0
  (4), \penalty0 222--225  (1970).

\bibitem{SOKO1992}
A.~P. Sokolov, A.~Kisliuk, M.~Soltwisch, and D.~Quitmann, Medium-range order in
  glasses: Comparison of raman and diffraction measurements, \emph{Phys. Rev.
  Lett.} {\bf 69}\penalty0 (10), \penalty0 1540--1543  (1992).

\bibitem{DUVA1990}
E.~Duval, A.~Boukenter, and T.~Achibat, Vibrational dynamics and the structure
  of glasses, \emph{J. Phys.-Condens. Matter}. {\bf 2}\penalty0 (51), \penalty0
  10227--10234  (1990).

\bibitem{SURO2002}
N.~V. Surovtsev and A.~P. Sokolov, Frequency behavior of raman coupling
  coefficient in glasses, \emph{Phys. Rev. B}. {\bf 66}\penalty0 (5), \penalty0
  054205  (2002).

\bibitem{SIMO2007}
G.~Simon, B.~Hehlen, R.~Vacher, and E.~Courtens, Hyper-raman scattering
  analysis of the vibrations in vitreous boron oxide, \emph{Phys. Rev. B}. {\bf
  76}\penalty0 (5), \penalty0 054210  (2007).

\bibitem{PROT2008}
I.~Prots, V.~Malinovsky, and N.~Surovtsev, Investigation of the fast relaxation
  in glass-forming selenium by low-frequency raman spectroscopy, \emph{Glass
  Phys. and Chem.} {\bf 34}\penalty0 (1), \penalty0 30--36  (2008).

\bibitem{FONT2006}
A.~Fontana, F.~Rossi, and E.~Fabiani, The raman coupling function in
  {$v$-GeO$_2$} and {$v$-SiO$_2$}: A new light and neutron scattering study,
  \emph{J. Non-Cryst. Solids}. {\bf 352}\penalty0 (42--49), \penalty0
  4601--4605  (2006).

\bibitem{SIBI2014}
J.~Sibik, S.~R. Elliott, and J.~A. Zeitler, Thermal decoupling of
  molecular-relaxation processes from the vibrational density of states at
  teraherz frequencies in supercooled hydrogen-bonded liquids, \emph{J. Phys.
  Chem. Lett.} {\bf 5}\penalty0 (11), \penalty0 1968--1972  (2014).

\bibitem{TARA2006}
S.~N. Taraskin, S.~I. Simdyankin, S.~R. Elliott, J.~R. Neilson, and T.~Lo,
  Universal features of terahertz absorption in disordered materials,
  \emph{Phys. Rev. Lett.} {\bf 97}\penalty0 (5), \penalty0 055504  (2006).

\bibitem{KABE2016}
M.~Kabeya, T.~Mori, Y.~Fujii, A.~Koreeda, B.~Wan~Lee, J.-H. Ko, and S.~Kojima,
  Boson peak dynamics of glassy glucose studied by integrated terahertz-band
  spectroscopy, \emph{Phys. Rev. B}. {\bf 94}\penalty0 (22), \penalty0 224204
  (2016).

\bibitem{DENI1987}
V.~Denisov, B.~Mavrin, and V.~Podobedov, Hyper-{Raman} scattering by
  vibrational excitations in crystals, glasses and liquids, \emph{Phys. Rep.}
  {\bf 151}\penalty0 (1), \penalty0 1--92  (1987).

\bibitem{IIJI2018}
Y.~{Iijima}, T.~{Mori}, S.~{Kojima}, Y.~{Fujii}, A.~{Koreeda}, S.~{Kitani},
  H.~{Kawaji}, and J.~{Ko}.
\newblock Terahertz time-domain spectroscopy and low-frequency raman scattering
  of boson peak dynamics of lithium borate glasses.
\newblock In \emph{43$^{rd}$ International Conference on Infrared, Millimeter,
  and Terahertz Waves}, pp. 957--958, Institute of Electrical and Electronics
  Engineers  (2018).

\bibitem{HEHL2012}
B.~Hehlen and G.~Simon, The vibrations of vitreous silica observed in
  hyper-raman scattering, \emph{J. Raman Spectrosc.} {\bf 43}\penalty0 (12),
  \penalty0 1941--1950  (2012).

\bibitem{DOLI1992}
G.~Dolino, B.~Berge, M.~Vallade, and F.~Moussa, Origin of the incommensurate
  phase of quartz .1. inelastic neutron-scattering study of the
  high-temperature beta-phase of quartz, \emph{J. Phys. I}. {\bf 2}\penalty0
  (7), \penalty0 1461--1480  (1992).

\bibitem{VISS1960}
W.~M. Visscher, Study of lattice vibrations by resonance absorption of nuclear
  gamma rays, \emph{Annals of Physics}. {\bf 9}\penalty0 (2), \penalty0
  194--210  (1960).

\bibitem{CHUM2004}
A.~I. Chumakov, I.~Sergueev, U.~van B\"urck, W.~Schirmacher, T.~Asthalter,
  R.~R\"uffer, O.~Leupold, and W.~Petry, Collective nature of the boson peak
  and universal transboson dynamics of glasses, \emph{Phys. Rev. Lett.} {\bf
  92}\penalty0 (24), \penalty0 245508  (2004).

\bibitem{CHUM2011}
A.~I. Chumakov, G.~Monaco, A.~Monaco, W.~A. Crichton, A.~Bosak, R.~R\"uffer,
  A.~Meyer, F.~Kargl, L.~Comez, D.~Fioretto, H.~Giefers, S.~Roitsch,
  G.~Wortmann, M.~H. Manghnani, A.~Hushur, Q.~Williams, J.~Balogh,
  K.~Parli\ifmmode~\acute{n}\else \'{n}\fi{}ski, P.~Jochym, and P.~Piekarz,
  Equivalence of the boson peak in glasses to the transverse acoustic van hove
  singularity in crystals, \emph{Phys. Rev. Lett.} {\bf 106}\penalty0 (22),
  \penalty0 225501  (2011).

\end{thebibliography}

\end{document}